\documentclass[%
 reprint,
 amsmath,amssymb,
 aps,
floatfix,longbibliography
]{revtex4-2}

\usepackage{graphicx}
\usepackage{dcolumn}
\usepackage{bm}
\usepackage[normalem]{ulem}
\usepackage{braket}

\usepackage{color}
\usepackage[dvipsnames]{xcolor}

\begin{document}

\preprint{APS/123-QED}

\title{Variational quantum algorithm for
ergotropy estimation in quantum many-body batteries}

\author{Duc Tuan Hoang\textsuperscript{1}}
\author{Friederike Metz\textsuperscript{1,2,3}}
\author{Andreas Thomasen\textsuperscript{4}}
\author{Tran Duong Anh-Tai\textsuperscript{1}}
\author{Thomas Busch\textsuperscript{1}}
\author{Thom\'as Fogarty\textsuperscript{1}}
\email{thomas.fogarty@oist.jp}
\affiliation{\textsuperscript{1}Quantum Systems Unit, OIST Graduate University, Onna, Okinawa 904-0495, Japan}
\affiliation{\textsuperscript{2}Institute of Physics, École Polytechnique Fédérale de Lausanne (EPFL), CH-1015 Lausanne, Switzerland}
\affiliation{\textsuperscript{3}Center for Quantum Science and Engineering, École Polytechnique Fédérale de Lausanne (EPFL), CH-1015 Lausanne, Switzerland}
\affiliation{\textsuperscript{4}QunaSys Inc., Aqua Hakusan Building 9F, 1-13-7 Hakusan, Bunkyo, Tokyo 113-0001, Japan}

\date{\today}

\begin{abstract}
Quantum batteries are predicted to have the potential to outperform their classical counterparts and are therefore an important element in the development of quantum technologies. Of particular interest is the role of correlations in many-body quantum batteries and how these can affect the maximal work extraction, quantified by the ergotropy.
In this work we simulate the charging process and work extraction of many-body quantum batteries on noisy-intermediate scale quantum (NISQ) devices, and devise the Variational Quantum Ergotropy (VQErgo) algorithm which finds the optimal unitary operation that maximises work extraction from the battery. We test VQErgo by calculating the ergotropy of a many-body quantum battery undergoing transverse field Ising dynamics following a sudden quench.
We investigate the battery for different system sizes and charging times, and analyze the minimum number of ansatz circuit repetitions needed for the variational optimization using both ideal and noisy simulators. We also discuss how the growth of long-range correlations can hamper the accuracy of VQErgo in larger systems, requiring increased repetitions of the ansatz circuit to reduce error. Finally, we optimize part of the VQErgo algorithm and calculate the ergotropy on one of IBM's quantum devices.
\end{abstract}

\maketitle

\section{\label{sec:level1}Introduction}

The allure of modern quantum technologies relies on leveraging quantum effects such as coherence and entanglement to out-perform their classical counterparts. In recent years this has been motivated by rapid experimental advances which has increased control over quantum states and has allowed to explore fundamental concepts in these devices. In particular, quantum thermal machines allow to explore the foundations of quantum thermodynamics, with devices such as quantum heat engines and refrigerators designed to control work output and heat flow with quantum media \cite{Kosloff2014,Mark2019,Myers2022}. Energy can also be stored in quantum batteries to be extracted at a later time \cite{alicki2013entanglement,hovhannisyan2013entanglement,rossini2019many,andolina2019extractable,Hu_2022,konar2022quantum,barra2022quantum,arjmandi2022performance,Binder2023}, which have the potential to outperform their classical counterparts in terms of total stored energy \cite{zhang2019powerful, andolina2018charger}, charging speed \cite{andolina2019quantum, ghosh2020enhancement,Gao2022,gyhm2022quantum,Salvia2023,Hu_2022,Rodriguez2023} and energy extraction \cite{ferraro2018high,le2018spin}. The maximum amount of energy that can be extracted from quantum systems through unitary processes is given by the \textit{ergotropy} \cite{allahverdyan2004maximal} which relies on finding the optimal unitary operation which transforms the system to its lowest energy state, known as its \textit{passive} state. This can be a difficult task as the ergotropy can be sensitive to correlations which can also affect device performance, notably improving efficiency in quantum heat engines coupled to squeezed baths \cite{Rossnagel2014,Klaers2017,Niedenzu2018,Biswas2022extractionof}, while impairing energy extraction from many-body batteries \cite{Goold_2016,Bera2017,Manzano2018,Vitagliano2018,andolina2019extractable,rossini2019many,Dou2021,Gemme2022,Dou2022,Dou2022b,Dou2023}. Simulation of the latter problem will be the focus of our work. 

Simulating the dynamics of many-body quantum systems in itself can be a complex problem due to the non-negligible role of quantum correlations which arise from finite couplings between particles. Furthermore, by today numerical calculations carried out on classical hardware are limited to small numbers of particles. This is in contrast to algorithms based on quantum hardware, which promise to alleviate some of this complexity by simulating quantum wave-functions in the Hilbert space of quantum bits, rather than numerically in classical registers. In addition to quantum physics and other fundamental sciences \cite{nielsen2002quantum,lloyd2014quantum,arute2019quantum,sakurai2022hybrid,cain2023quantum,yonezu2023time,shtanko2023uncovering}, quantum computers promise applications in various technological sectors including chemistry \cite{o2019calculating,mizukami2020orbital,fujii2022deep,simon2023improved,ko2023implementation,senjean2023toward} and materials design and research \cite{ma2020quantum,kanno2022resource,zini2023quantum,rubin2023fault,westermayr2023high}. It is for this reason that they have seen unprecedented growth in recent years: reported milestones include simulation of dynamics and calculations of accurate expectation values on a 127 qubit device \cite{Kim2023}, demonstration of fast converging quantum-enhanced Markov chain Monte-Carlo simulations \cite{layden2023quantum} and generation of large-scale cluster states on superconducting qubit devices \cite{cao2023generation}. While fault tolerant quantum computation (FTQC) based on error corrected qubits is still not technically possible, currently noisy intermediate-scale quantum (NISQ) processors are available. However, they only have short-lived qubits which are not protected from decoherence \cite{preskill2018quantum,bauer2020quantum,wurtz2023aquila}. In this NISQ era, quantum algorithms rely on shallow circuits where qubits are measured quickly \cite{peruzzo2014variational, kandala2017hardware}. There is therefore a need for quantum algorithms which can simulate quantum systems within a limited time-span while still solving problems which exceed the capabilities of their classical counterparts.

Variational quantum algorithms \cite{mcclean2016theory,Cerezo2021,Bharti2022} (VQA) have been deemed particularly promising for NISQ devices. These are a class of algorithms which can be used to find variational approximate solutions to problems of interest. A famous example is the variational quantum eigensolver (VQE) \cite{peruzzo2014variational} which is used to determine the ground state of a Hamiltonian through repeated sampling of an ansatz wave-function in the eigenbases of a set of observables. 
Amongst others, VQAs have been developed to solve the max-cut problem via the quantum approximate optimization algorithm \cite{farhi2014quantum}, to find numerous chemical properties of molecules \cite{omiya2022analytical,ibe2022calculating,nakagawa2023analytical}, or to perform machine learning tasks like the classification of symmetry protected topological phases \cite{cong2019quantum}. While the performance of these algorithms is limited by barren plateaus \cite{McClean2018}, i.e., the problem of exponentially vanishing gradients with the system size, in recent years tools have been developed to study and mitigate this phenomenon \cite{zhao2021analyzing,sack2022avoiding}. In addition, error mitigation has been shown to offer significant improvements when noise is an issue \cite{endo2021hybrid,suzuki2022quantum,cai2022quantum}, and approaches inspired by FTQC have resulted in partial error correction schemes developed for NISQ devices \cite{akahoshi2023partially,bultrini2023battle}.


In this work we propose a VQA called variational quantum ergotropy (VQErgo) to calculate the ergotropy of a quantum battery on NISQ computers. We use the transverse field Ising spin-chain model to benchmark our algorithm whereby the battery is charged by a sudden quench of the interaction among nearest-neighbor spins. The interactions will create correlations between the spins which results in non-trivial dynamics of the ergotropy.
In order to simulate the dynamics we use projected - Variational Quantum Dynamics (p-VQD) \cite{barison2021efficient} to find the time-evolved quantum state
and then a variational optimization is carried out to obtain the optimal unitary which prepares the passive state. Although VQErgo has some elements in common with other variational quantum algorithms such as VQE, VQErgo does not find a variational ground state and its energy or other static quantities relating to a Hamiltonian. VQErgo time-evolves the input wavefunction to then evaluate the ergotropy by considering a reduced state of the full wave-function. In contrast to VQE and its variants which variationally prepares the pure eigenstates of a system, VQErgo performs its state optimization on a mixed state which can also be strongly correlated with the rest of the system.

The performance of the algorithm is analyzed for different system sizes and number of ansatz circuit repetitions, and its accuracy is assessed by comparison with exact results. We also analyze how the creation of correlations between spins can negatively affect the ergotropy estimation, requiring an {increase in the number of repetitions of the variational ansatz circuit.} Finally, we evaluate the effectiveness of our scheme in the presence of noise using a noisy simulator as well as  real hardware. Our work represents one of the first NISQ algorithms designed specifically to calculate the ergotropy of quantum systems, expanding the tools already available to describe quantum thermodynamics and associated devices on quantum computers \cite{Solfanelli2021,Melo2022,Solfanelli2022,Consiglio2023}.

The manuscript is organized as follows. In Sec.~\ref{method}, we briefly review the operation of quantum batteries and the concept of ergotropy. We also present the VQErgo algorithm which is used to calculate the ergotropy on quantum hardware, separated into four steps: initialization, time evolution/charging, mean energy calculation and passive energy optimization. Then, the transverse field Ising spin-chain Hamiltonian and the charging protocol for our quantum battery are introduced in Sec.~\ref{model}. We describe our main results in Sec.~\ref{results}, including the dynamics of the system, the measurement of the total energy and the ergotropy from noise-free (state-vector simulations), noisy simulations, and from calculations run on IBM quantum devices \cite{IBMQuantum}. Finally in Sec.~\ref{conclusion}, we draw our conclusions and discuss future prospects for this algorithm.

\section{METHODS} \label{method}
\subsection{The maximal extractable work - ergotropy}
We describe a quantum battery made from $N$ identical quantum cells which are charged through unitary dynamics by suddenly switching on an external field $V$.
Initially the battery is prepared in the ground state  $| \Psi\left(t=0\right)\rangle$ of a local Hamiltonian $H_0$, and during charging it evolves according to 
\begin{equation}
    {H}_{1} = {H}_{0} +V\,.
\end{equation}
The state of the charged battery is therefore time-dependent, $|\Psi\left(t\right)\rangle$, and energy is discharged by removing the external field $V$ with the total work stored in the battery at time $t$ then given by
\begin{equation} \label{total_work}
    W(t) = \langle \Psi\left(t\right)|{H_{0}}|\Psi\left(t\right) \rangle - \langle \Psi\left(0\right)|{H_{0}}|\Psi\left(0\right) \rangle\;.
\end{equation}
While this is the total energy that is stored in the entire battery after the time $t$, it is not necessarily all extractable, especially when only considering subsystems of the device. This would correspond to extracting energy from $M\leq N$ cells of the battery, which could be required due to a restriction on accessing the full state of the system or in order to only partially discharge the battery. In this scenario energy can be locked in correlations between the $M$ and $N-M$ cells thereby reducing the amount of energy that can be extracted \cite{alicki2013entanglement,rossini2019many,andolina2019extractable}. The maximum amount of work that can be extracted from the $M$-cell state $\rho^M=\mathrm{tr}_{{N-M}}\{\ket{\Psi\left(t\right)}\bra{\Psi\left(t\right)}\}=\sum_{j=1}\lambda_{j}\left|\varphi_{j}\right>\left<\varphi_{j}\right|$ (with $\lambda_{j}\geq \lambda_{j+1}$) through unitary transformations is given by the ergotropy \cite{allahverdyan2004maximal}, which is found by optimizing over all possible unitaries such that the resulting state has the minimum energy with respect to the {$M$-cell} Hamiltonian ${H^M_0}=\sum_{i=1}\varepsilon_{i}\left|\psi_{i}\right>\left<\psi_{i}\right|$ (with $\varepsilon_{i}\leq\varepsilon_{i+1}$) 
\begin{eqnarray} \label{ergotropy1}
    \mathcal{E}&&=\text{tr}\{{H^M_0}\rho^M\}- \text{min}_{U} \{ \text{tr}\{{H^M_0} U\rho^{M} U^{\dagger}\} \} \nonumber \\
    & &=\text{tr}\{{H^M_0}\left(\rho^M-P_{\rho}\right)\}\,.
\end{eqnarray}
This state is known as the passive state $P_{\rho}=\sum_{i}\lambda_{i}\left|\psi_{i}\right>\left<\psi_{i}\right|$ and no further work can be extracted from it by unitary transformations. The ergotropy can then be expressed  in the well-known form \cite{allahverdyan2004maximal}
\begin{equation} 
\label{ergotropy2}    
\mathcal{E}=\sum_i \left( p_i -\lambda_i  \right)\varepsilon_i,
\end{equation} 
where $p_i=\sum\limits_{j}\lambda_{j}{\left| \left\langle \varphi_{j} \right|\left.\psi_{i} \right\rangle  \right|}^{2}$ is the projection of $\rho^M$ on the eigenstates of {$H^M_0$}. To extract energy from the battery we therefore require that $p_i\neq \lambda_i$. If the reduced state $\rho^M$ is mixed there can be a difference between the work 
\begin{equation}
\label{Eq:reduced_work}
W(t)=\text{tr}\{{H^M_0}\rho^M(t)\}-\text{tr}\{{H^M_0}\rho^M(0)\}    
\end{equation}
and the ergotropy, $W(t)\geq\mathcal{E}(t)$, which becomes an equality if $\rho^M$ is pure.  

In classical simulations of quantum batteries, the ergotropy is conventionally calculated by solving Eq.~\eqref{ergotropy2}, i.e., by diagonalizing the sub-system Hamiltonian and the reduced density matrix of the battery state and by computing the relevant overlaps. The analogous way to run this sequence of calculations using NISQ hardware would be to first obtain the full spectrum of the Hamiltonian and its eigenstates \cite{Nakanishi2019,Higgott2019,Jones2019}. Then estimates of the overlaps with the time-evolved wavefunction can be obtained by measurement in the Hamiltonian eigenbasis.
However, we can instead consider the optimization problem in Eq.~\eqref{ergotropy1} which is naturally expressed in terms of expectation values that can be efficiently computed on a quantum computer. Furthermore, the optimization over unitary operators for the passive state can be naturally phrased in the language of variational quantum algorithms \cite{Cerezo2021,mcclean2016theory,Bharti2022} as we detail in the following section. Hence, current state of the art quantum devices allow us to readily simulate and investigate the ergotropy and other properties of many-body quantum batteries.

\subsection{The Variational Quantum Ergotropy (VQErgo) algorithm}
\begin{figure*} 
\centering
\includegraphics[width=\textwidth]{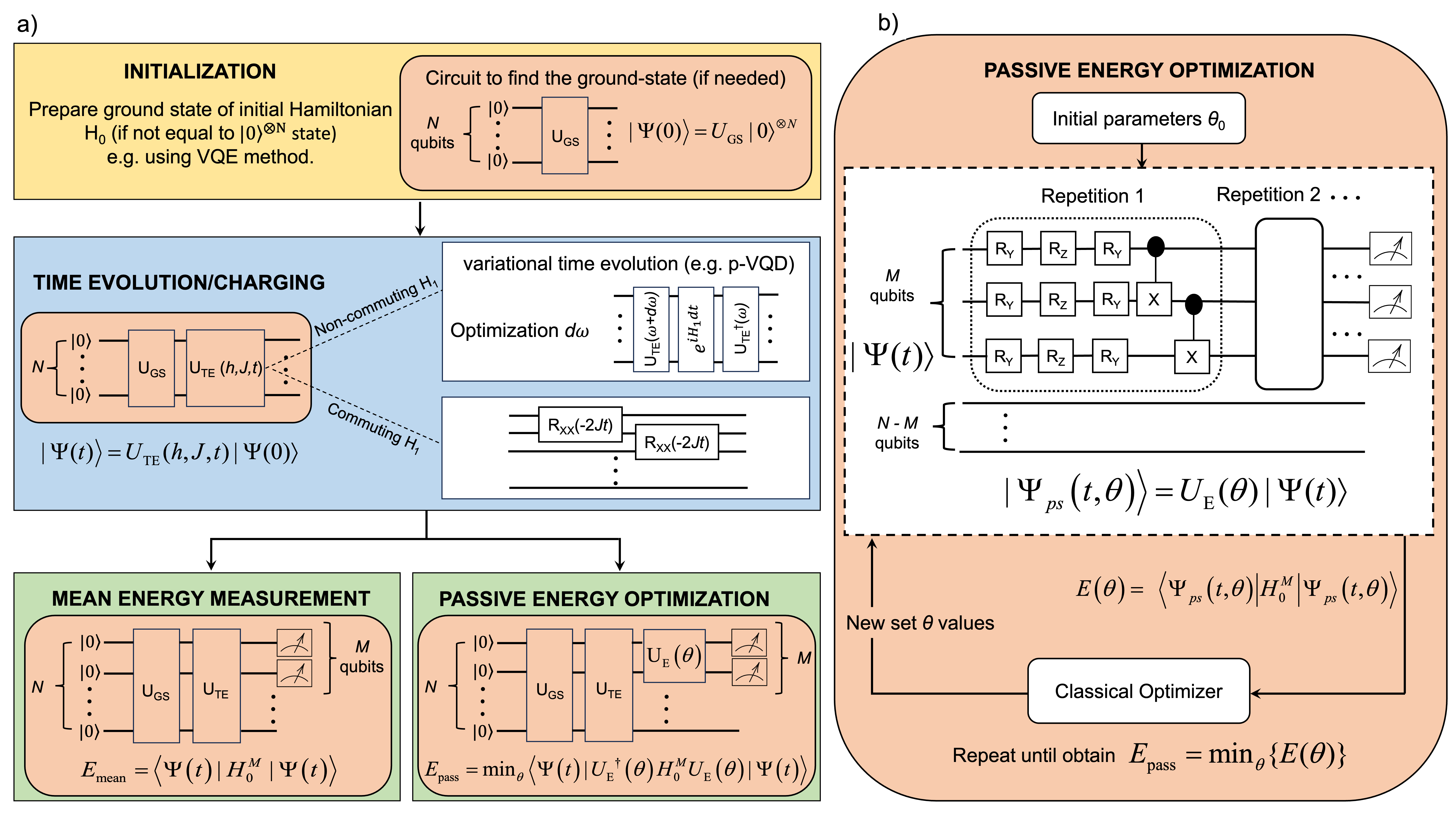}
\caption{\label{fig:diagram} (a) Schematic depiction of the Variational Quantum Ergotropy (VQErgo) algorithm. \textit{Initialization:}~The uncharged battery state $\ket{\psi(0)}$ given by the ground state of the local Hamiltonian $H_0$ is prepared on the quantum device e.g.~using the variational quantum eigensolver (VQE). \textit{Time evolution/charging:}~The battery is charged by time evolving the state with another Hamiltonian $H_1$. As an example, in this work we consider quenches with the transverse field Ising Hamiltonian $H_1=-J\sum_i^{N-1} \sigma^x_i \sigma^x_{i+1} - h \sum_i^{N}\sigma^z_i$. We approximate the time evolution unitary $U_{\text{TE}}=\exp(-iH_1t)$ via a variational circuit $U_{\text{TE}}(\omega)$ optimized using the projected-variational quantum dynamics (p-VQD) algorithm. We also consider the case in which the magnetic field is switched off ($h=0$) and the time evolution unitary can be exactly decomposed into a single layer of two-qubit gates. \textit{Mean energy measurement:}~The ergotropy is the difference between the mean and passive energy (c.f.~Eq.~\eqref{ergotropy1}). The former can be measured as an expectation value of $H^M_0$ on $M<N$ qubits of the time evolved state $\ket{\Psi(t)}$. (b) \textit{Passive energy optimization:}~The passive energy is defined as the minimum attainable expectation value of $H^M_0$ over all unitary transformations $U_{\mathcal{E}}$ acting on the time evolved state. We express $U_{\mathcal{E}}(\theta)$ in terms of a variational circuit with parameters $\theta$ which are optimized using a typical classical-quantum feedback loop.
}
\end{figure*}

In the following we describe our framework for simulating quantum batteries on quantum hardware and how to extract interesting properties, in particular the ergotropy. We therefore refer to the overall algorithm as the Variational Quantum Ergotropy (VQErgo) algorithm. VQErgo can be divided into 4 subroutines which are (i) battery initialization, (ii) battery charging, (iii) mean energy calculation and (iv) passive energy optimization as shown in Fig.~\ref{fig:diagram}.

\textbf{Battery initialization.~}The battery starts off in the uncharged state, corresponding to the ground state of the local Hamiltonian $H_0$. Any ground state preparation routine (e.g.~ VQE) can be employed for this task. Note that in the rest of this work we choose the local Hamiltonian to be of the form $H_0 = -\sum_i^N \sigma^z_i$. Thus, the ground state $\ket{0}^{\otimes N}$ naturally coincides with the initial computational basis state of digital-based quantum computers allowing us to omit a state preparation circuit.

\textbf{Battery charging.}~The battery is charged by time evolving the initial state with the Hamiltonian $H_1$ for a total time $t$. On digital quantum computers time evolution can be achieved by a Trotter-Suzuki decomposition of the global time evolution unitary into local gates. However, the number of Trotter steps (the number of gates) grows with time $t$ 
and hence, generally, only short evolution times can be simulated on noisy hardware. To overcome this limitation, there have been several proposals for performing the time evolution variationally using parameterized circuits {with a fixed, small number of gates} \cite{Crstoiu2020,Yuan2019,li2017efficient,Bharti2021}. Here, we employ the projected-variational quantum dynamics (p-VQD) algorithm due to its efficiency \cite{barison2021efficient}. p-VQD iteratively evolves the parameters $w(t+\delta t) = w(t) + dw$ of a state ansatz $|\psi_{w(t)}\rangle=U(w(t))|0\rangle$ in short time increments $\delta t$ by minimizing the infidelity between the ansatz state $\ket{\psi_{w(t)+d w}} = U(w(t)+dw)|0\rangle$ and the true time evolved state $|\phi(t+\delta t)\rangle=e^{-i H_1 \delta t}|\psi_{w(t)}\rangle$
\begin{equation}\label{eq:pvqd}
    d w = \text{arg}\min_{d w}\left[\frac{1-\left|\left\langle\phi(t+\delta t) \mid \psi_{w(t)+d w}\right\rangle\right|^2}{\delta t^2}\right].
\end{equation}
The unitary $e^{-i H_1 \delta t}$ is typically approximated using the Trotter-Suzuki decomposition with a single Trotter step given that the time step size $\delta t$ is chosen sufficiently small. Importantly, the {repetitions} of the state ansatz circuit and of the circuits used for evaluating Eq.~\eqref{eq:pvqd} do not grow with time $t$. For further details regarding the p-VQD optimization, we refer to Appendix~\ref{app:pvqd}.

Let us note here that our quantum circuit framework for quantum battery simulation is highly modular and, in principle, any of the subroutine algorithms can be exchanged with other viable quantum algorithms for the respective tasks. Specifically, for time evolution, we mention the time-dependent variational algorithm (TDVA) \cite{li2017efficient,Yuan2019}, and subspace variational quantum simulation (SVQS) \cite{heya2019subspace}, which may be used in cases where the number of excited states populated during time-evolution does not exceed the number of qubits. Additionally, one could restrict to quenches with Hamiltonians composed of only commuting terms, e.g., $H_1 = -\sum_i^N \sigma^x_i \sigma^x_{i+1}$. In this case, the time evolution operator is trivially decomposed into a single layer of two-qubit gates and eliminates the need for involved, approximate time evolution algorithms {(see Appendix~\ref{appendix-turning_off_magnetic})}. Finally, the recent advances in analog quantum computing provide yet another promising architecture for quantum battery simulations since time evolution is naturally implemented via global Hamiltonians \cite{Bernien2017,Labuhn2016}.

\textbf{Mean energy calculation.}~The ergotropy of the quantum battery is calculated as the difference of the mean and passive energies of the charged state $\ket{\psi(t)}$ after tracing over $N-M$ sites (c.f.~Eq.~\ref{ergotropy1}). Specifically, the mean energy can be expressed as an expectation value of the local Hamiltonian acting only on the subsystem of $M<N$ qubits
\begin{equation}\label{eq:mean}
    E_{\text{mean}}= \langle \Psi\left(t\right)|H_{0}^M\otimes \mathbb{I}^{\otimes (N-M)}|\Psi\left(t\right) \rangle\,.
\end{equation}

\textbf{Passive energy optimization.}~The computation of the passive energy requires us to find the optimal unitary transformation $U_{\mathcal{E}}$ acting on $M$ qubits of the charged state that minimizes the expectation value of the local Hamiltonian within the subsystem
\begin{equation}\label{eq:passive}
    E_{\text{pass}}= \min_{U_{\mathcal{E}}}\left[\langle \Psi\left(t\right)|(U_{\mathcal{E}}^\dagger H_{0}^M U_{\mathcal{E}})\otimes \mathbb{I}^{\otimes (N-M)}|\Psi\left(t\right) \rangle\right]\,.
\end{equation}
We can efficiently perform the optimization over unitaries on current quantum hardware using the tools of variational quantum algorithms. In particular, we define a circuit ansatz $U_{\mathcal{E}}(\theta)$ composed of two-qubit and single-qubit gates with a set of parameters $\theta$. The optimization of the passive state then amounts to finding the optimal parameters that minimize the expectation value in Eq.~\eqref{eq:passive} which can be iteratively achieved by using a classical optimizer like gradient descent and the parameter-shift rule for evaluating gradients \cite{mitarai2018,schuld2019}. 

In order to limit the amount of noise during the quantum simulation to a minimum, we employ a hardware-efficient ansatz for the p-VQD and the passive state variational circuits. This means that the ansatz circuits are composed of several layers each containing arbitrary, parameterized single-qubit rotations (decomposed into $R_YR_ZR_Y$ gates) followed by a series of CNOT gates applied only to neighboring qubits (see Fig.~\ref{fig:diagram}(b)). Note that the passive state circuit $U_{\mathcal{E}}$ is only defined on the $M<N$ subsystem qubits. Appendix~\ref{app:vqergo} contains additional details about the circuit optimization.

\section{MODEL} \label{model}
To model the quantum battery we consider the paradigmatic transverse field Ising spin-chain \cite{le2018spin,rossini2019many,Mondal2022,Catalano2023}. The competition between the nearest-neighbour interactions and an external field can result in strongly correlated reduced states $\rho^M$ which give rise to a non-trivial dependence of the ergotropy on the charging time $t$ and subsystem size $M$ as we show further below. Moreover, the system is amenable to simulations on currently available NISQ devices for a couple of qubits. The discharged battery is described by the non-interacting Hamiltonian
\begin{equation} \label{noninteracting-hamiltonian}
    H_0 = -h\sum_{i=1}^{N}\sigma_i^z,       
\end{equation}
where $h>0$ is the external magnetic field, $N$ is the number of spins (cells) in the battery, and $\sigma_i^{k}$ with $k = x,y,z$ denotes the spin-1/2 Pauli matrices. At $t=0$ the battery is initialized in the spin polarized ground state $\ket{\Psi(0)} = \ket{\uparrow}^{\otimes N}\equiv \ket{0}^{\otimes N}$ and thus, naturally coincides with the initial state of the quantum computer.

In order to charge the battery we implement a sudden quench $H_0\rightarrow H_1$ for $t>0$ which switches on the nearest-neighbor interaction
\begin{equation} \label{pvqdhamiltonian}
    H_1 =  -h\sum_{i=1}^{N}\sigma_i^z - J\sum_{i=1}^{N-1} \sigma_i^x\sigma_{i+1}^x,
\end{equation}
where $J$ is the coupling strength and we consider open boundary conditions. We simulate the time evolved state $\ket{\Psi(t)} = \exp(-iH_1 t)\ket{\Psi(0)}$ on a quantum computer using the aforementioned p-VQD algorithm. In Appendix~\ref{appendix-turning_off_magnetic} we provide results obtained with an alternative charging protocol in which the external field is turned off during the charging time ($h=0$). In this case, the time evolution operator can be trivially decomposed into a single layer of local two-qubit gates and thus, be implemented without the need for a variational optimization. Throughout the remainder of this work we set $\hbar=1$ and consider the transverse field Ising model with fixed parameters $h=0.6$ and $J=2$.

\begin{figure} [tb]
\centering
\includegraphics[width=85mm]{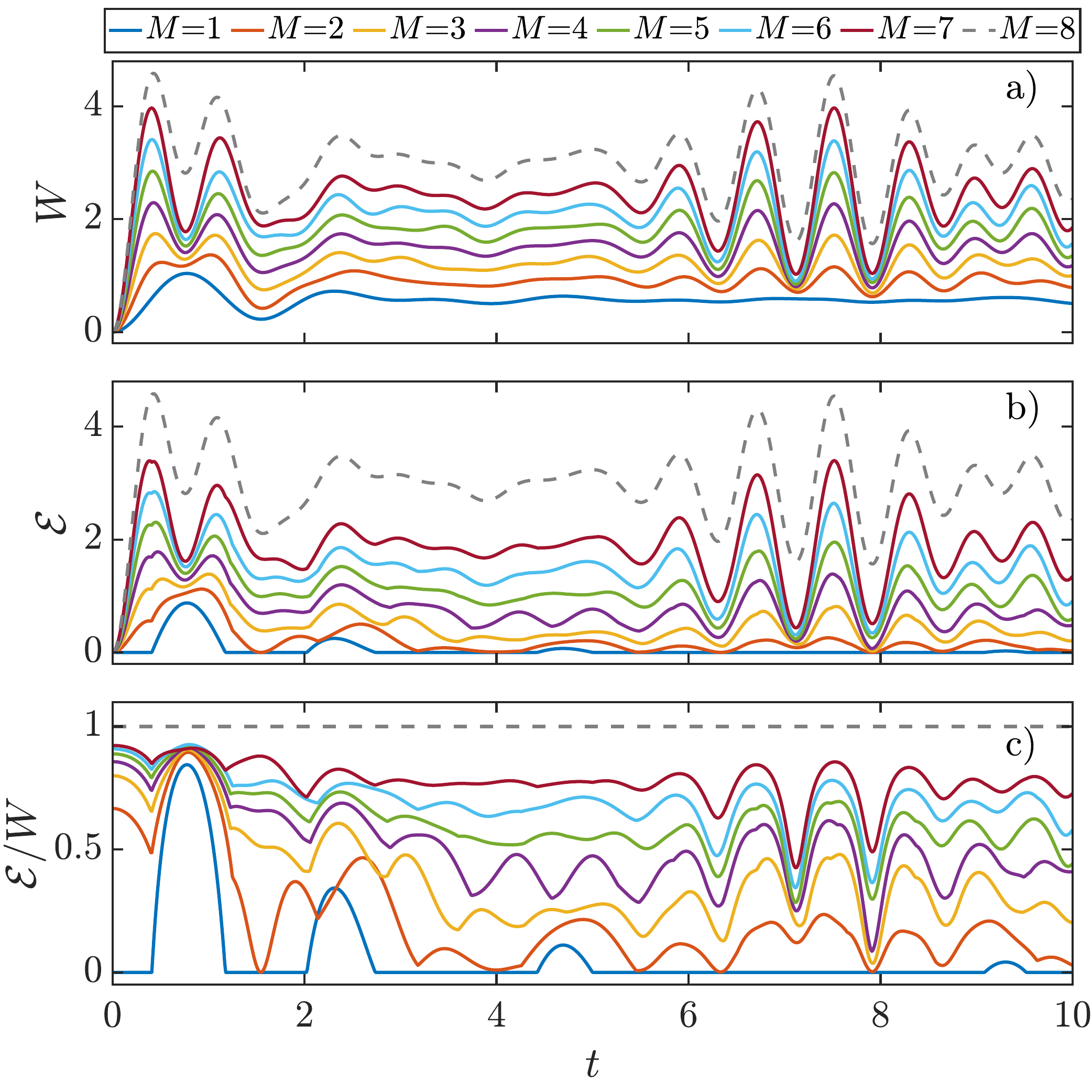}
\caption{\label{fig:dynamics_N8} (a) The total injected work $W$, (b) the ergotropy $\mathcal{E}$  and (c) the efficiency of the battery $\mathcal{E}/W$ as a function of the charging time $t$. The total system is comprised of $N=8$ spins while each line corresponds to a different sub-system size $M$ from which energy is extracted.
}
\end{figure}

The quench dynamics of the state can be computed directly through
\begin{equation} \label{time evolution}
    \left|\Psi\left(t\right)\right\rangle =\sum\limits_{j}\langle \Phi^F_{j} | \Psi\left(0\right)\rangle \exp \left( -\frac{iE_j^{F}t}{\hbar} \right)|\Phi^F_{j}\rangle,
\end{equation}
where $|\Phi_j^F\rangle$ and $E_j^F$ are the eigenstates and eigenvalues of $H_{1}$. The stored work and ergotropy are then calculated using Eqs.~\eqref{Eq:reduced_work} and \eqref{ergotropy2}, respectively. Fig.~\ref{fig:dynamics_N8} shows the work $W$ and ergotropy $\mathcal{E}$ stored in $M$ cells as a function of charging time for a total system size of $N=8$ spins. We also plot their ratio $\mathcal{E}/W$ describing how efficiently the battery can be discharged which saturates for $M=N$ as expected. One can see that immediately following the quench  the work and the ergotropy rapidly increase as the quench drives the system far out-of-equilibrium, while subsequent oscillations are the result of finite size effects. 
We note that the maximum ergotropy and injected work can be achieved at short charging times $t\sim 0.4$ for $M>2$ and the charging process subsequently stabilizes in the region $2\lesssim t \lesssim 6$ after which revivals induce further oscillations.

In general, the larger the sub-system $M$, the more work, ergotropy and efficiency can be achieved. However, for smaller cell size $M$ the efficiency necessarily suffers as correlated cells in the rest of the battery ($N-M$) are discarded. This is apparent for times $t>2$ when the reduced state $\rho^M$ is sufficiently mixed. We also note that, for long intervals for the $M=1$ system the ergotropy is exactly zero although the total injected work is non-zero (see panel (b) and (c)). This is related to the equivalence of the reduced and passive states when $M=1$, and will be discussed in detail in the next section.

\section{RESULTS} \label{results}

\subsection{VQErgo state-vector simulation}\label{res:state-vector}

\begin{figure*}[tb] 
\centering
\includegraphics[width=\textwidth]{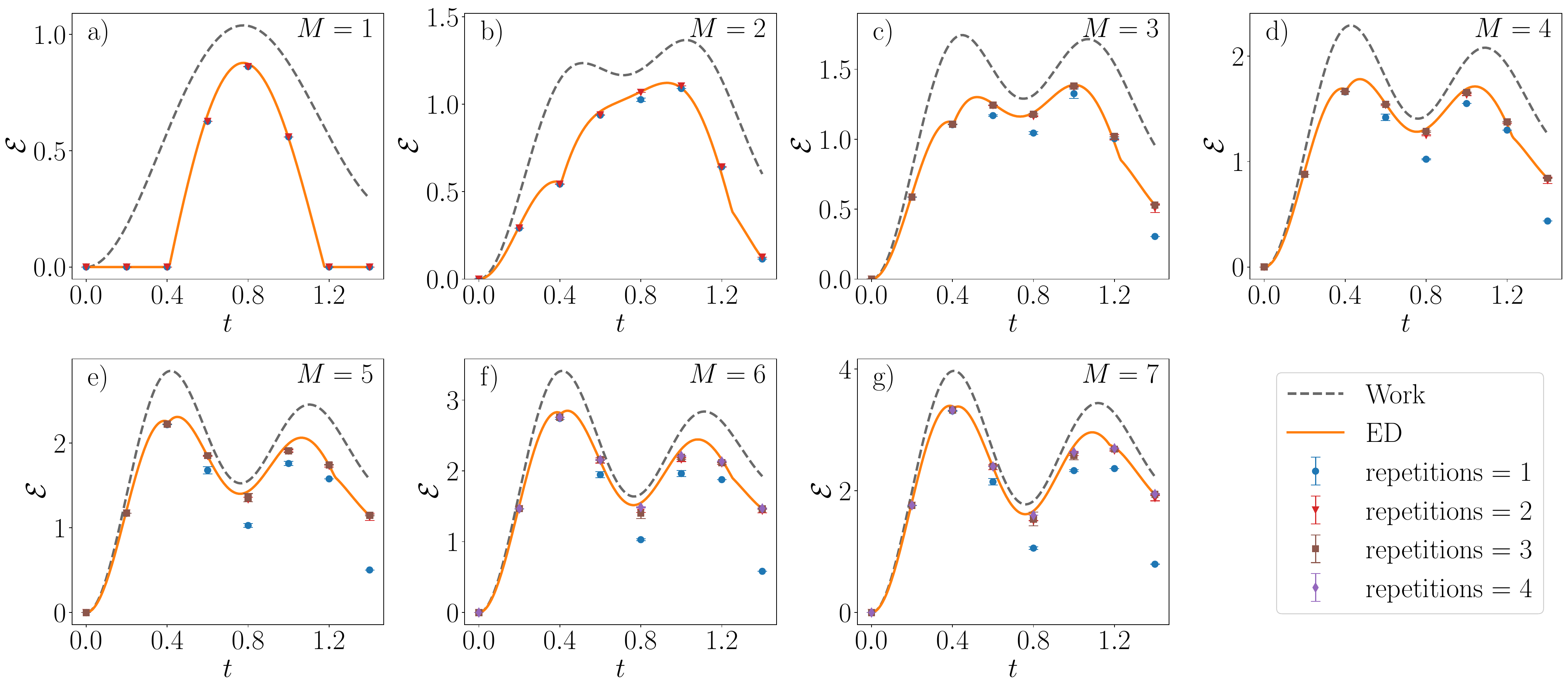}
\caption{\label{fig:pvqd_statevector_depthscale} The ergotropy $\mathcal{E}$ as a function of the charging time $t$ for different subsystem sizes $M$ and a total system size of $N=8$. The grey dashed line denotes the work $W$ stored in the battery cells. The orange line corresponds to the value of the ergotropy computed numerically from exact diagonalization (ED) calculations while the markers show the variationally obtained ergotropies for different {repetitions} of the passive state ansatz. Each point is an average over 100 optimization runs using a statevector simulation (the standard deviation is indicated by the error bars).
}
\end{figure*}

First we simulate the quantum battery using our proposed VQErgo algorithm and analyze the variational optimization in an ideal, noise-free setting via statevector simulation. We restrict ourselves to charging times $0<t<1.4$ which include the first two maxima of the work and ergotropy curves (see Fig.~\ref{fig:dynamics_N8}). We simulate the time evolution starting from the polarized product state using p-VQD which optimizes the variational circuit parameters iteratively in small time increments and hence we automatically obtain the evolved states at all intermediate times as well. {The ansatz used for the time-evolution and the passive state optimization was the hardware efficient ansatz.} All the details regarding the optimization {and choice of ansatz} are reported in Appendix~\ref{app:pvqd}. In particular, for a given number of spins $N$ we repeat the optimization with different {ansatz circuit repetitions}, {resulting in a} different number of variational parameters and compare the p-VQD states to the exact time evolved states in Fig.~\ref{fig:pvqd_infidelity} of Appendix~\ref{app:pvqd}. For each simulated system size we choose a final p-VQD {ansatz circuit repetition number} that gives rise to small errors with respect to the exact state.

Once the optimized time evolution circuit is obtained we can measure the mean energy on the circuit (see~Eq.~\eqref{eq:mean}) which allows us to calculate the stored work $W$. Next, we perform the VQErgo optimization to prepare the passive state on a subsystem of $M<N$ qubits and measure the passive energy from which we can extract the ergotropy $\mathcal{E}$.
In Fig.~\ref{fig:pvqd_statevector_depthscale} we show the results obtained for a total system size of $N=8$ qubits and subsystem sizes $M=1,\dots, 7$. We compare the exact ergotropy (orange line) to the ergotropies evaluated on optimized circuits $U_{\mathcal{E}}$ of different {repetition numbers}. Each point represents an average over 100 different runs of the algorithm (i.e., using different random seeds for the circuit initialization). Overall, we find a good agreement of the variationally obtained ergotropies with their exact values given that the {number of repetitions} is chosen large enough. Note that some of the observed discrepancies have to be attributed to the preceding p-VQD optimization, which also introduces an error in the state.

\begin{figure}[tb]
\centering
\includegraphics[width=85 mm]{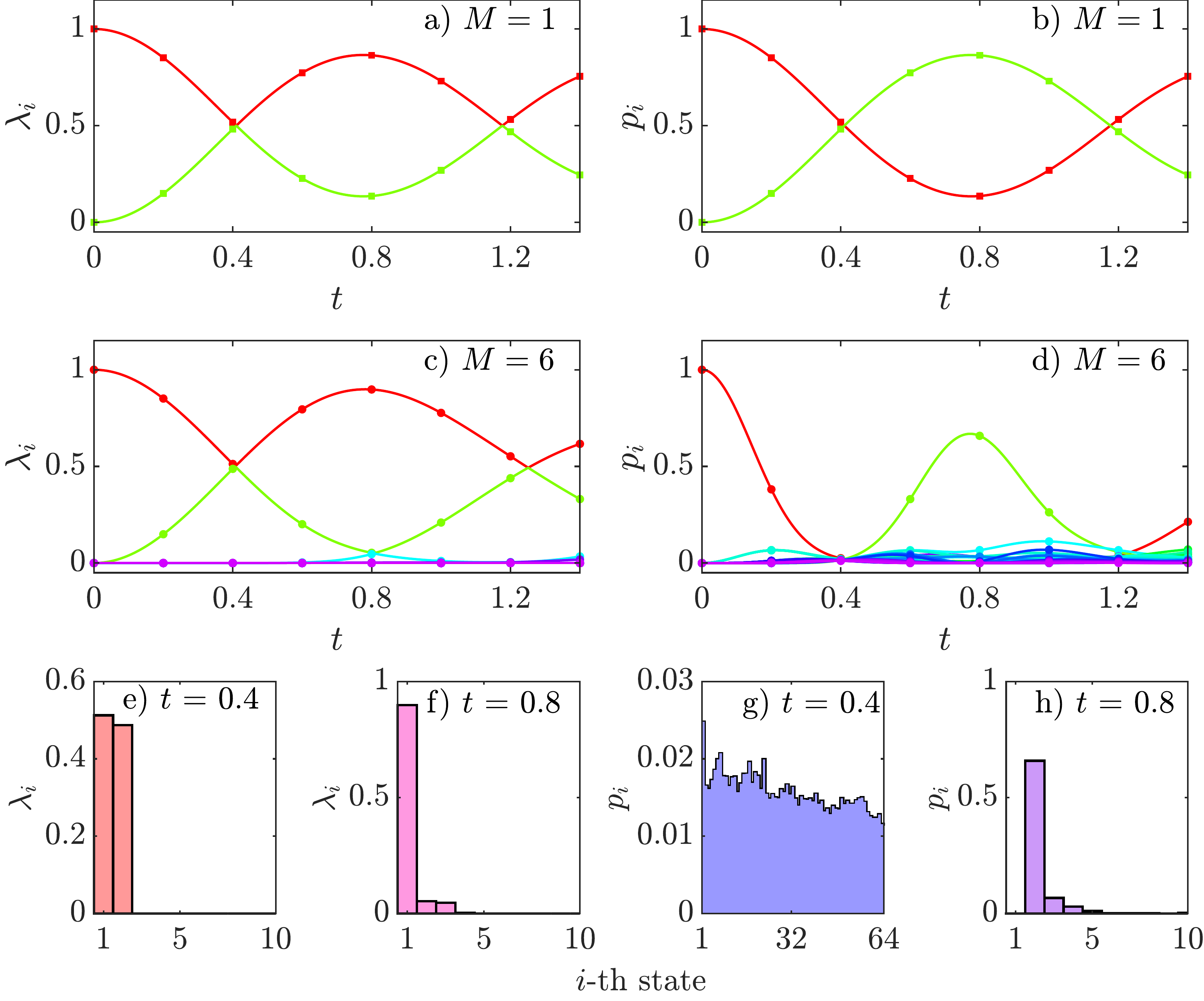}
\caption{\label{fig:occupation_probability} (a), (c) The passive state probabilities $\lambda_i$ as a function of $t$ for $M=1$ and $M=6$, respectively. In (b), (d) we show the corresponding reduced state probabilities $p_i$. For $M=6$ we plot $\lambda_i$ at (e) $t=0.4$ and (f) $t=0.8$, and $p_i$ at (g) $t=0.4$ and (h) $t=0.8$. Data in all figures are obtained numerically via exact diagonalization, with the specific times $t=\{0, 0.2 , 0.4, 0.6, 0.8 ,1 , 1.2 , 1.4\}$ denoted by square markers.}
\end{figure}

To further understand the dynamics of the ergotropy in Fig.~\ref{fig:pvqd_statevector_depthscale} we examine its constituent time dependent parts, namely the exact probabilities $p_i$ of the reduced state and the $\lambda_i$ of its passive state. These are shown in Fig.~\ref{fig:occupation_probability} for both $M=1$ and $M=6$. The simplest case is $M=1$ as its dynamics is that of a two level system (there are only two accessible states), with $\lambda_i=p_i$ for times $t<0.4$ and therefore the ergotropy is zero. At $t=0.4$ there is a crossing in the probability distribution with $p_2>p_1$ and finite energy can now be extracted from the battery through reordering of these occupancies (see Fig.\ref{fig:pvqd_statevector_depthscale}(a)). For $t\geq 1.2$ a subsequent crossing restores the original ordering of the probabilties and thus the ergotropy again vanishes.

This behaviour is echoed in the dynamics of larger subsystems, albeit with more complexity, as the number of occupied states $p_i$ and $\lambda_i$ is increased. For instance, the dynamics of $\lambda_i$ for $M=6$ possess a similar structure (see Fig.~\ref{fig:occupation_probability}(c)) with contributions mainly from the two lowest energy eigenstates. On the contrary, the dynamics of $p_i$ is more complex and includes contributions from higher energy eigenstates of $H^M_0$. This results in a non-zero ergotropy at all times $t>0$. In Figs.~\ref{fig:occupation_probability}(e) and (g) we show $\lambda_i$ and $p_i$ at $t=0.4$ which corresponds to the maximum ergotropy for $M=6$. The large difference between the reduced state and its passive state is readily apparent, as $p_i$ is distributed over all possible states, while $\lambda_i$ is again concentrated around the two lowest eigenstates. However, at $t=0.8$ the ergotropy has a local minimum as the $p_i$'s occupy lower energy states owing to less work stored in the battery (see Fig.~\ref{fig:occupation_probability}(h)).

\begin{figure}[tb]
\centering
\includegraphics[width=85mm]{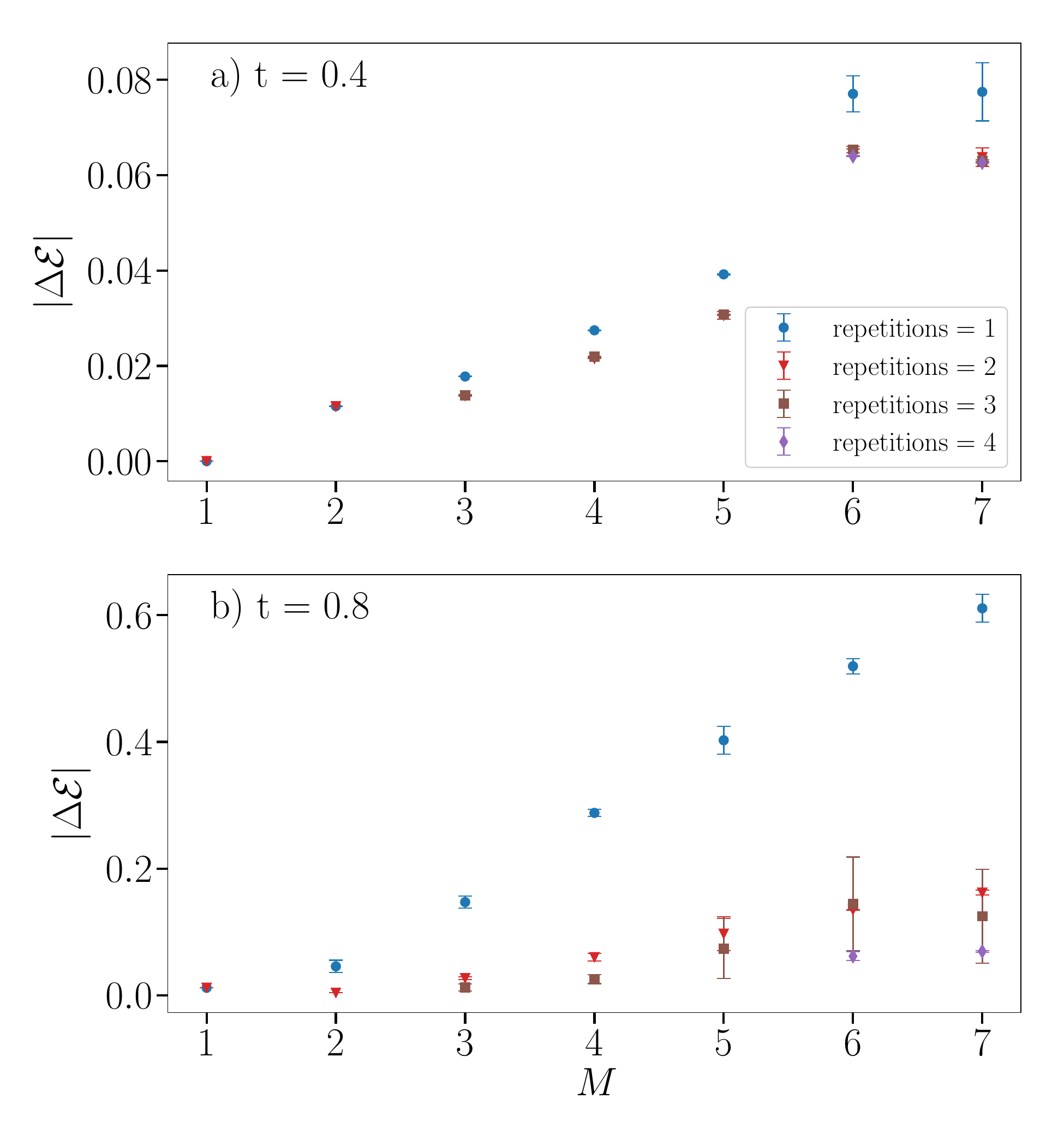}
\caption{\label{fig:pvqd_statevector_abs_error} The absolute error between the ergotropy calculated variationally via a statevector simulation and its exact value versus the battery subsystem size $M$. We show the error for two distinct charging times $t=0.4$ (a) and $t=0.8$ (b) for different {numbers of repetitions} of the passive state ansatz. Note the different order of magnitudes in the error for the two considered times which indicates that the required {number of repetitions} depends not only on the cell size $M$, but also on the charging time $t$.
}
\end{figure}

Similarly to other variational quantum algorithms, the {number of repetitions} of the ansatz {circuit} is an important hyper-parameter of the optimization. Fig.~\ref{fig:pvqd_statevector_depthscale} suggests that for VQErgo the required {repetitions} depends both on the subsystem size $M$ and charging time $t$. For better visualization, we plot the error in the measured ergotropies as a function of the subsystem size in Fig.~\ref{fig:pvqd_statevector_abs_error} at $t=0.4$ and $t=0.8$. In the case of a single quantum cell $M=1$, we always only require one general single-qubit rotation to prepare the passive state. However, with increasing subsystem size $M>1$ more layers of single-and two-qubit gates are needed to reduce errors. This is due to correlations that are spread over larger distances within the system which can be quantified through the $C_{XX}$ and $C_{ZZ}$ correlations between qubit $i$ and a second qubit at $i+l$
\begin{equation}
    C_{XX/ZZ}(i,\ell) = |\langle \sigma^{x/z}_{i+\ell} \sigma^{x/z}_{i} \rangle - \langle\sigma^{x/z}_{i+\ell}\rangle\langle\sigma^{x/z}_{i}\rangle|^2,
\end{equation}
where $\langle\cdot \rangle = \langle \Psi(t) |\cdot| \Psi(t)\rangle$ denotes an expectation value calculated with the time evolved state. 

\begin{figure}[tb] 
\centering
\includegraphics[width=85mm]{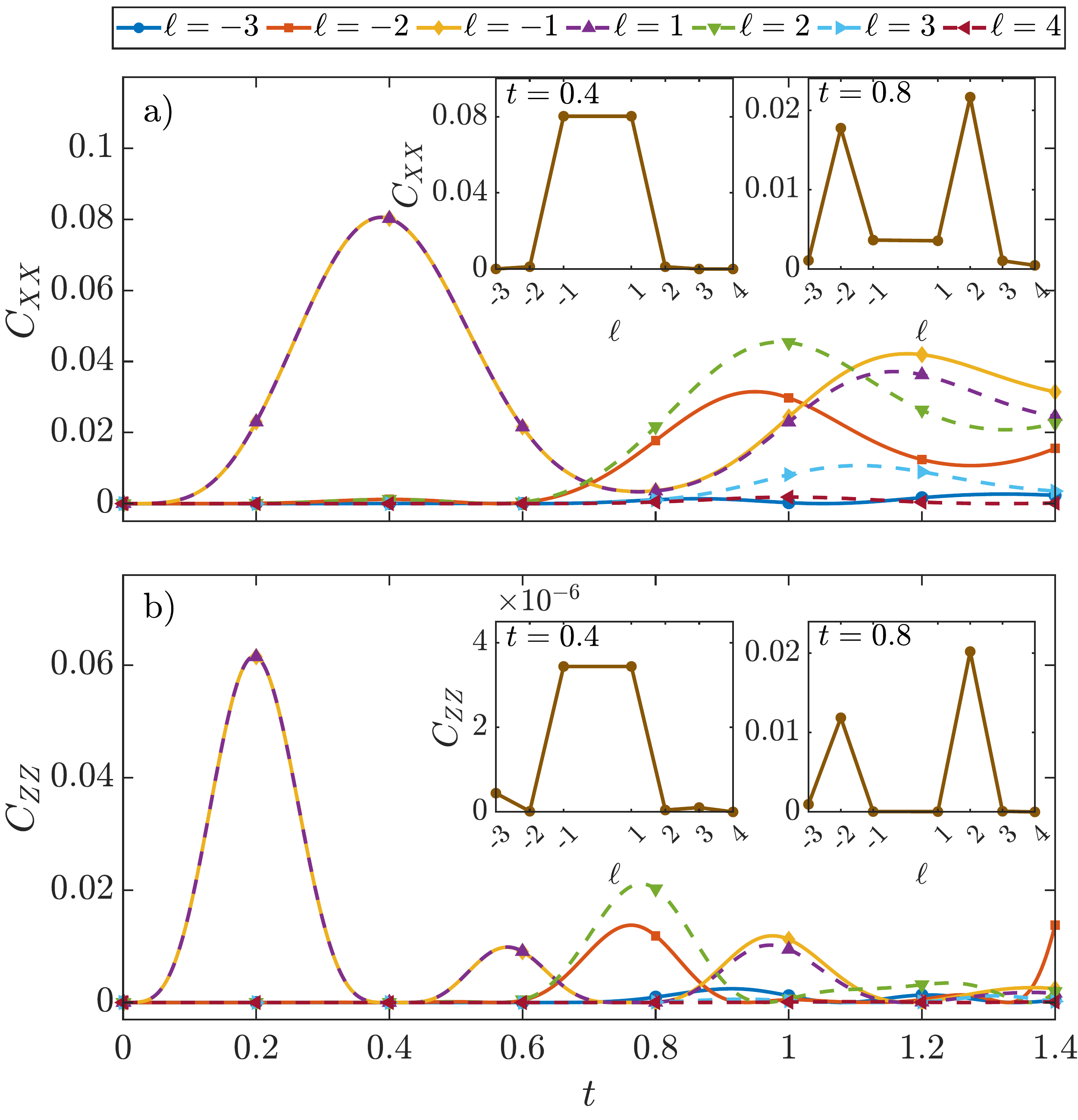}
\caption{\label{fig:CXX_CZZ_i=4} 
The Pauli-$XX$ (a) and $ZZ$ (b) correlations in the charged state between the 4th qubit at the center of the chain and a qubit $\ell$ sites apart as a function of the charging time $t$. The insets show the respective correlations as a function of the distance $\ell$ for two specific times $t=0.4$ and $t=0.8$. For charging times $t\lesssim 0.6$ nearest neighbor correlations dominate while at later times $t > 0.6$ also long-range correlations appear. All data is from exact diagonalization calculations. 
}
\end{figure}

We take qubit $i=4$ at the center of the $N=8$ spin chain as an example and plot its correlations with the other qubits as a function of time in Figs.~\ref{fig:CXX_CZZ_i=4} (a) and (b). The maximum ergotropy coincides with the maximum correlations in the $x$-directions, while correlations in the $z$-direction vanish. Furthermore, up to times $t\lesssim 0.6$ the qubit is correlated only with its nearest neighbors at $\ell=\pm 1$. We observe a lack of long-range correlations also for the other spins in the chain (not shown here) and can thus infer that a single layer of two-qubit gates (paired with parameterized single-qubit rotations) is sufficient to disentangle all qubits of the subsystem and prepare the exact passive state. However, this is not the case for times $t> 0.6$ as long-range correlations and entanglement are built up. We therefore require multiple layers of two-qubit gates to rotate the reduced state into the uncorrelated basis set of the passive state and hence, to increase the accuracy of the ergotropy estimation. This is apparent in Fig.~\ref{fig:pvqd_statevector_abs_error}(b) which shows a significant increase in error at $t=0.8$ (note the difference in order of magnitudes between Figs.~\ref{fig:pvqd_statevector_abs_error} (a) and (b)). However, we have found that for the particular Ising system considered here, the errors quickly decrease with {the number of repetitions and that two repetitions} are already often sufficient. Any extra {repetitions} provide only a small additional advantage which suggests that the {number of repetitions of the circuit} scales sub-linear with the battery cell size $M$, making the optimization less prone to barren plateaus \cite{McClean2018,Cerezo2021b}.

\subsection{VQErgo quantum device experiments}\label{res:hardware}

Following the analysis of VQErgo under ideal, noise-free conditions, we now evaluate its performance on a real quantum device. To that end, we perform VQErgo on one of the freely accessible quantum computers provided through the IBM Quantum cloud. While the most recent state-of-the-art quantum computers operate on more than 100 qubits and feature small error rates \cite{Kim2023}, the freely available quantum devices are still small in size and very noisy. Hence, we restrict ourselves to quantum batteries comprised of only a handful of spins that can be simulated with shallow-{repetition} circuits and can be mapped to the device qubit layout without the need for long-range gates (or SWAP gates).  
We also substitute our real hardware experiments with noisy, classical simulations that mimic the device noise model. All our experiments are performed on the 7 qubit \texttt{ibm\_perth} device and its classical simulator analog FakePerth. 
\begin{figure} 
\centering
\includegraphics[width=85mm]{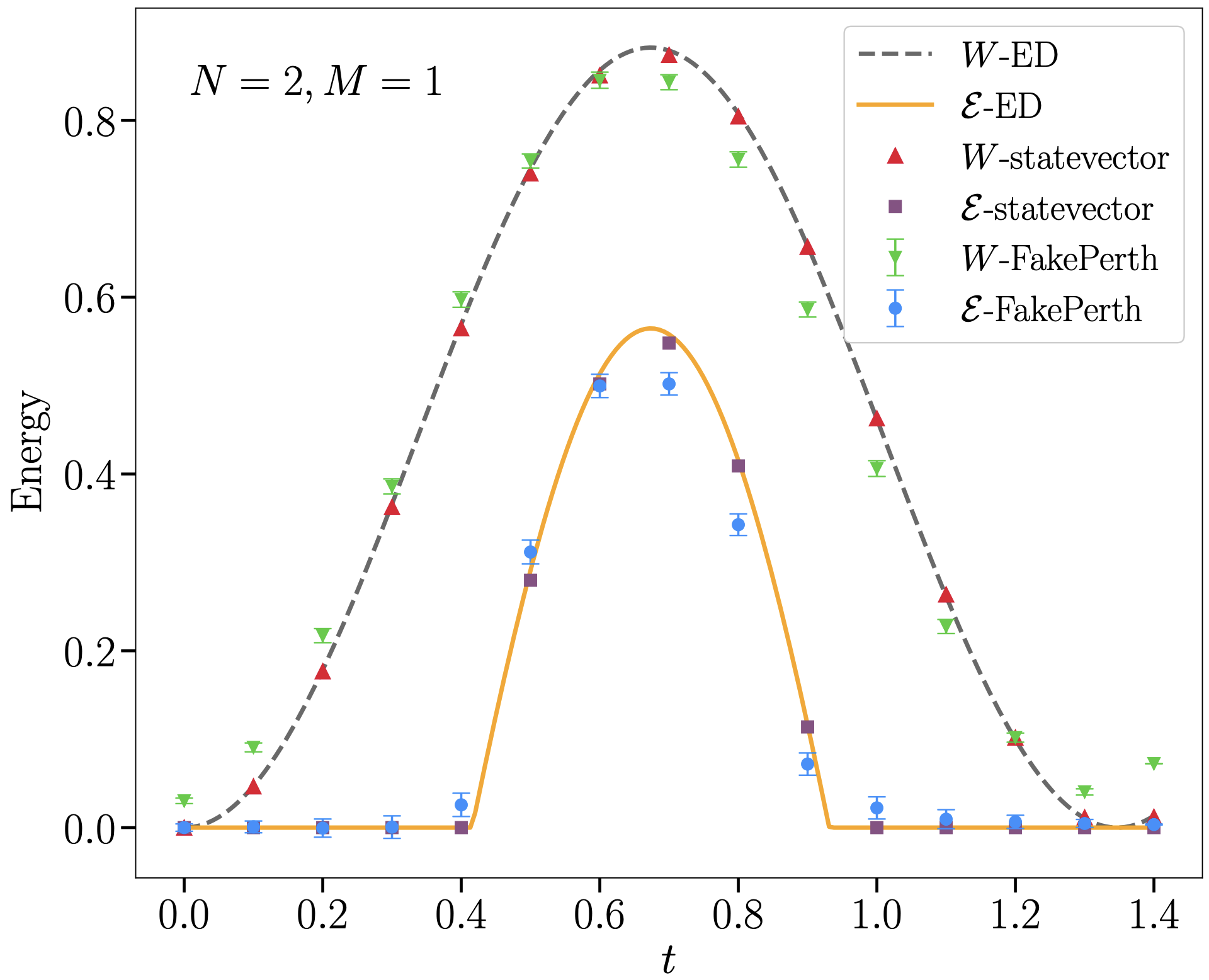}
\caption{\label{fig:pvqd_N2_M1} The total amount of stored work $W$ (dashed line) and the ergotropy $\mathcal{E}$ (solid line) as a function of the battery charging time $t$ computed via exact techniques. We also show their values extracted via VQErgo using an ideal statevector simulation and a noisy classical simulation with FakePerth which mimics the \texttt{ibm\_perth} quantum device. The results are an average over 100 independently run optimizations. The error bars indicate the corresponding standard deviation.
}
\end{figure}

\textbf{Full VQErgo results} We start by running the full VQErgo pipeline (including the p-VQD and passive state optimization) on the noisy classical simulator using the SPSA optimizer with 250 optimization steps and 2048 shots per measurement. The training curves and any further technical details can be found in Appendix~\ref{app:vqergo}. We report the final measured work and ergotropy as a function of the charging time for a system with $N=2$ and $M=1$ battery cells in Fig.~\ref{fig:pvqd_N2_M1}. Each point is again an average over 100 independent runs of the algorithm. For this small battery system, the ergotropies are in good agreement with their exact values. Any discrepancies and the increased standard deviation compared to the statevector simulation can be attributed to various error sources, such as shot noise, state preparation and measurement (SPAM) errors, coherent and incoherent noise. Note that the observed error in the work and ergotropy increases slightly with time which is to be expected since p-VQD iteratively evolves the ansatz state in time and hence, errors naturally build up in the charged state.

\textbf{Noise free p-VQD optimization} Using the noisy simulator has not allowed us to obtain converged results for p-VQD with $N=4$. This can be understood from the fact that p-VQD carries out a variational optimization for each time-step that is simulated. Therefore any error resulting from the noisy hardware or a simulation of it compounds with every iteration. As explained in Appendix~\ref{app:pvqd}, when $N=4$, the {repetition numbers of the p-VQD ansatz circuit} must be increased to twice what it is when $N=2$ while still needing 14 iterations. Although circuits of this {number of repetitions} can be run without the resulting quantum state decohering totally, the accumulated error due to noise is too great to result in accurate time-evolution. However, with the optimized p-VQD parameters from state-vector simulations we are still able to show convergence of the passive state optimization using the noisy simulator and the actual device. In the remainder of this work we perform the p-VQD optimization using the classical statevector simulation and only run the optimized time evolution circuit on the quantum device followed by the variational passive state optimization. Note that in Appendix~\ref{appendix-turning_off_magnetic} we consider a simplified charging protocol that does not require variational time evolution and thus, can be executed end-to-end on current hardware.

\begin{figure}[tb]
\centering
\includegraphics[width=85mm]{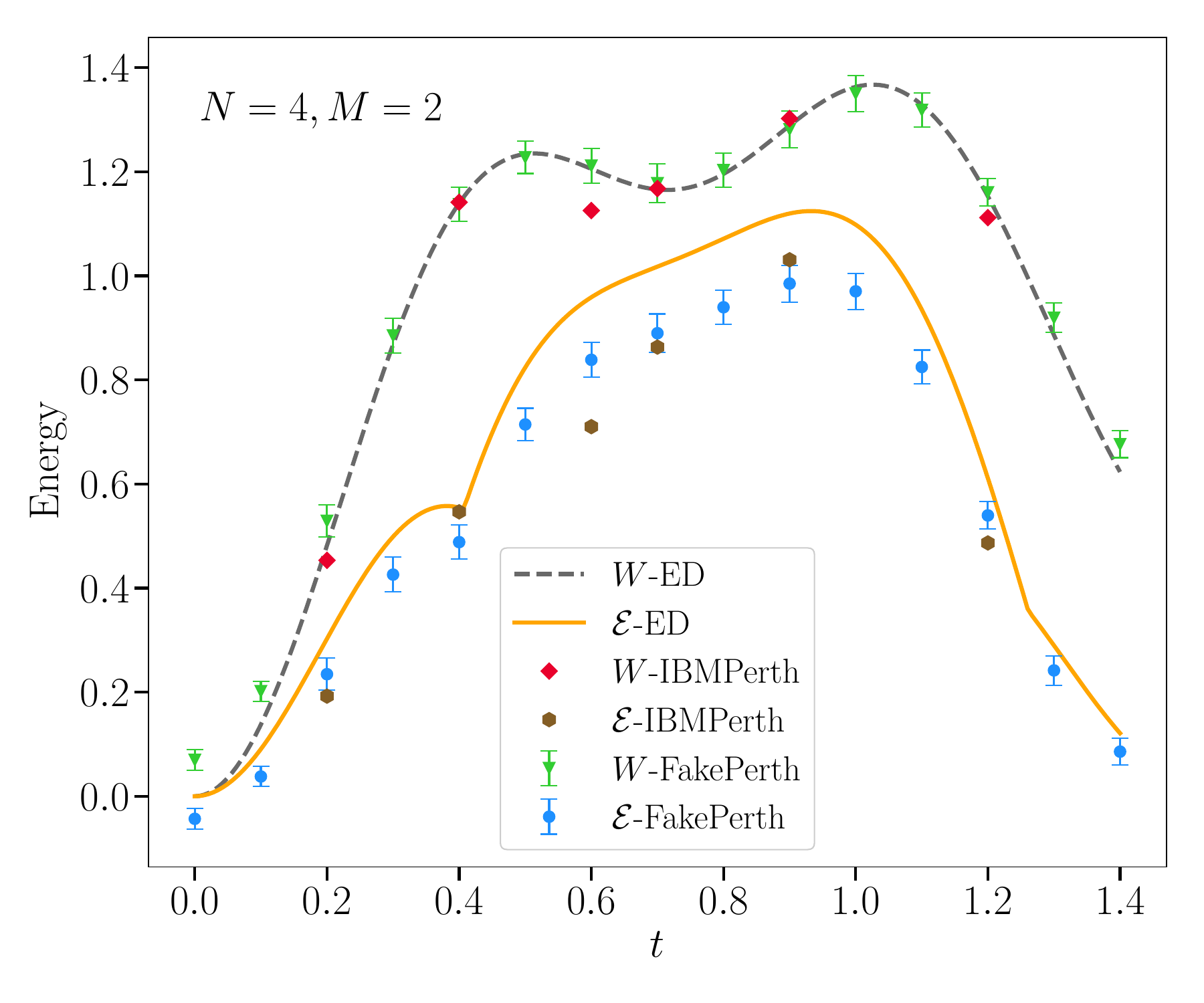}
\caption{\label{fig:pvqd_runtime_N4_M2} The exact stored work $W$ (dashed line) and ergotropy $\mathcal{E}$ (solid line) are plotted against the charging time $t$ for a system with $N=4, M=2$. The results were obtained using state-vector optimized p-VQD parameters together with a noisy simulation of the \texttt{ibm\_perth} backend as well as actual device results. For the real-device experiments only a single run of the passive state optimization is carried out, while we display the average and standard deviation of 100 runs in the case of the noisy classical simulator. Due to the limited available quantum computing time, we performed the VQErgo optimization on \texttt{ibm\_perth} only on a subset of the shown times.
}
\end{figure}

In Fig.~\ref{fig:pvqd_runtime_N4_M2} we display the real-device and noisy simulator results obtained for $N=4$ and subsystem size $M=2$. The measured injected work is in nearly perfect agreement with their theoretically computed values with the exception of the point at time $t=0.6$ which is slightly lower. We believe that this outlier can be attributed to naturally occurring fluctuations in the experimental hardware over time (our simulations have been performed over several weeks). On the other hand, the evaluated ergotropies are consistently lower than their exact values for the noisy simulator and the real-device experiments. We expect this to be the result of decoherence as the passive state circuit contains two additional layers of CNOT gates compared to the circuit the work was measured on. It would be interesting to investigate whether quantum error mitigation such as zero-noise extrapolation can improve these results \cite{Temme2017,Kandala2019,majumdar2023best}. However, despite the small discrepancies, the qualitative dependence of the ergotropy with time can be successfully inferred. Importantly, VQErgo allows us to determine the time at which the ergotropy becomes maximal which is crucial for designing many-body quantum batteries that perform optimally.

\section{CONCLUSIONS} \label{conclusion}

In this work, we have studied the ergotropy - the maximal extractable work - of quantum batteries. We have shown that the calculation of the ergotropy can be naturally phrased in terms of a variational quantum algorithm and thus, the ergotropy is readily amenable to current NISQ device computations. We have embedded the ergotropy calculation in an end-to-end variational simulation routine for quantum batteries called VQErgo that includes battery initialization, charging, and the ergotropy estimation. Note that due to the modularity of the presented framework, different algorithms for any of its subroutines like initial state preparation or time evolution can be chosen and adapted to the system at hand.

We tested VQErgo on a battery undergoing transverse field Ising dynamics. To that end, we investigated the passive state optimization and the required {number of repetitions} of the variational circuit with the battery cell size and charging time. In particular, we showed that {more than one repetition} is necessary beyond a critical charging time after which long-range correlations set in. Subsequently, we demonstrated VQErgo using a noisy classical simulator with noise characteristics from a current IBM quantum device, and demonstrated that the passive state optimization can be carried out on the actual physical device. In both cases, we were able to successfully measure the injected work and ergotropy for different charging times. While the estimated ergotropies were slightly below their true exact values, the qualitative dependence of the ergotropy with time still matched the theoretical predictions. In particular, the results allow us to infer the optimal charging time of the quantum battery that leads to the maximal ergotropy. 

{While we only consider quench dynamics in this work, our algorithm could also be expanded to also optimize the charging protocol for a specific charging time, allowing to time-dependently tune coupling terms in the Hamiltonian to maximize the ergotropy on short times.} Our algorithm is also not restricted to the simulation of quantum batteries, but is also amenable to other thermodynamic devices {where correlations can affect work statistics and therefore depend on accurate estimations of the ergotropy to account for all energy contributions. For instance, in quantum heat engines coupled to non-classical environments, such as squeezed baths \cite{Rossnagel2014,Klaers2017,Niedenzu2018,Biswas2022extractionof}, both passive thermal energy and ergotropy can be transferred to the system. Work can also be extracted from quantum heat engines by coupling to quantum batteries or so called quantum flywheels \cite{Levy2016,flywheel2019}, where again it is important to distinguish between the amount of useful work quantified by the ergotropy and thermal fluctuations that can degrade efficiency. Finally, due to its sensitivity to correlations the ergotropy can also be used to measure genuine multipartite entanglement \cite{Puliyil2022}, as the mixed state $\rho^M$ is entangled with the remaining $N-M$ cells leading a non-trivial gap between the ergotropy and the average work. As the passive state is diagonal in $H^M_0$ the entanglement spectrum $\lambda_i$ can be recovered, provided that the eigenbasis can be resolved on quantum hardware
\cite{cerezo2022variational}.} 

In this work we have shown a viable path towards studying many-body quantum batteries using quantum hardware. It is remarkable that the non-trivial dynamics of the transverse field Ising model can be probed even on the relatively noisy 7 qubit \texttt{ibm\_perth} device as this is a device with a quantum volume of only 32 \cite{cross2019validating,IBMQuantum}. In contrast, the state of the art Falcon r10 device is reported to have a quantum volume of 256 \cite{IBMQuantumHas2021}. The complexity of the types of systems that can currently be interrogated with VQErgo is therefore expected to exceed the capabilities we have shown here. As a concrete example, it is feasible that the p-VQD optimization, which we had to carry out using a state-vector simulator for $N=4$, could be carried out on a quantum volume 256 device.

\begin{acknowledgments}
This work is supported by the Okinawa Institute of Science and Technology Graduate School (OIST). The classical simulations were performed on the high-performance computing cluster (Deigo) provided by the Scientific Computing and Data Analysis section at OIST. We acknowledge the use of IBM Quantum services for this work. The views expressed are those of the authors, and do not reflect the official policy or position of IBM or the IBM Quantum team. FM acknowledges support by the NCCR MARVEL, a National Centre of Competence in Research, funded by the Swiss National Science Foundation (grant number 205602).  TF acknowledges support from JSPS KAKENHI Grant Number JP23K03290. TF and TB are also supported by JST Grant
Number JPMJPF2221. 
\end{acknowledgments}

\appendix
\section{Technical details of the optimization}
All quantum simulations were performed using the Qiskit python library \cite{Qiskit} and the Qiskit Runtime Estimator primitive. In all shot-based simulations, expectation values were estimated using 2048 shots. Furthermore, we employed readout error mitigation implemented in Qiskit in all simulations that were subject to noise. {The calibration data of the \texttt{ibm\_perth} device that was used for the hardware experiments is shown in Table~\ref{tab:perth}.} {The ansatz used for the time-evolved state in the p-VQD algorithm was the hardware efficient ansatz. This circuit is comprised of repeated layers of single qubit rotations followed by entangling CNOT gates. The CNOT gates are applied successively on pairs of neighboring qubits (see Fig.~\ref{fig:diagram}(b) for a single repetition of the ansatz circuit in the case of $M=3$ qubits). 
The hardware efficient ansatz is used for the passive state optimization as well.
}



\begin{table*}[!ht]
    \centering
    \begin{tabular}{|l|l|l|l|l|l|l|l|l|l|l|}
    \hline
        Qubit & $T_1$ ($\mu$s) & $T_2$ ($\mu$s) & $f$ (GHz) & $\delta$ (GHz) & $\epsilon_{\text{readout}}$   & $p(0|1)$  & $p(1|0)$  & Single-qubit error  & CNOT error   & $t_{\text{gate}}$ (ns) \\ \hline
        0 & 188.6 & 82.5 & 5.2 & -0.34 & 0.39 & 0.070 & 0.701 & 3.8$\times 10^{-4}$ & 6.3$\times 10^{-3}$ & 391 \\ \hline
        1 & 24.6 & 44.7 & 5.0 & -0.34 & 0.03 & 0.030 & 0.029 & 7.5$\times 10^{-4}$ & 1.1$\times 10^{-2}$ & 479 \\ \hline
        2 & 131.9 & 93.5 & 4.9 & -0.35 & 0.02 & 0.023 & 0.026 & 2.6$\times 10^{-4}$ & 2.1$\times 10^{-2}$ & 604 \\ \hline
        3 & 158.5 & 201.4 & 5.1 & -0.34 & 0.02 & 0.017 & 0.018 & 2.4$\times 10^{-4}$ & 6.0$\times 10^{-3}$ & 309 \\ \hline
        4 & 146.4 & 160.2 & 5.2 & -0.33 & 0.12 & 0.123 & 0.122 & 2.9$\times 10^{-4}$ & 7.2$\times 10^{-3}$ & 590 \\ \hline
        5 & 156.7 & 151.0 & 5.0 & -0.35 & 0.03 & 0.031 & 0.031 & 2.5$\times 10^{-4}$ & 8.1$\times 10^{-3}$ & 529 \\ \hline
        6 & 133.2 & 236.4 & 5.2 & -0.34 & 0.01 & 0.018 & 0.011 & 3.3$\times 10^{-4}$ & 1.1$\times 10^{-2}$ & 604 \\ \hline
        Median & 146.4 & 151.0 & 5.1 & -0.34 & 0.03 & 0.030 & 0.029 & 2.9$\times 10^{-4}$ & 8.1$\times 10^{-3}$ & 529 \\ \hline
    \end{tabular}
    \label{tab:perth}
    \caption{{Further information about the \texttt{ibm\_perth} device used in this work. The table shows from left to right column: relaxation time $T_1$, dephasing time $T_2$, frequency $f$, anharmonicity $\delta$, readout assignment error $\epsilon_{\text{readout}}$, the probability $p(0|1)$ of measuring 0 if the qubit was prepared in the  $\ket{1}$ state, the probability $p(1|0)$ of measuring 1 if the qubit was prepared in the  $\ket{0}$ state, single-qubit gate error, average CNOT gate error, average gate times $t_{\text{gate}}$. The final row shows the median computed over all qubits. Note that our experiments were performed over the course of several weeks, while the calibration data is only a snapshot of the device properties at a single point in time (data from Nov 13, 2023).}}
\end{table*}

\begin{figure*}
\centering
\includegraphics[width=\textwidth]{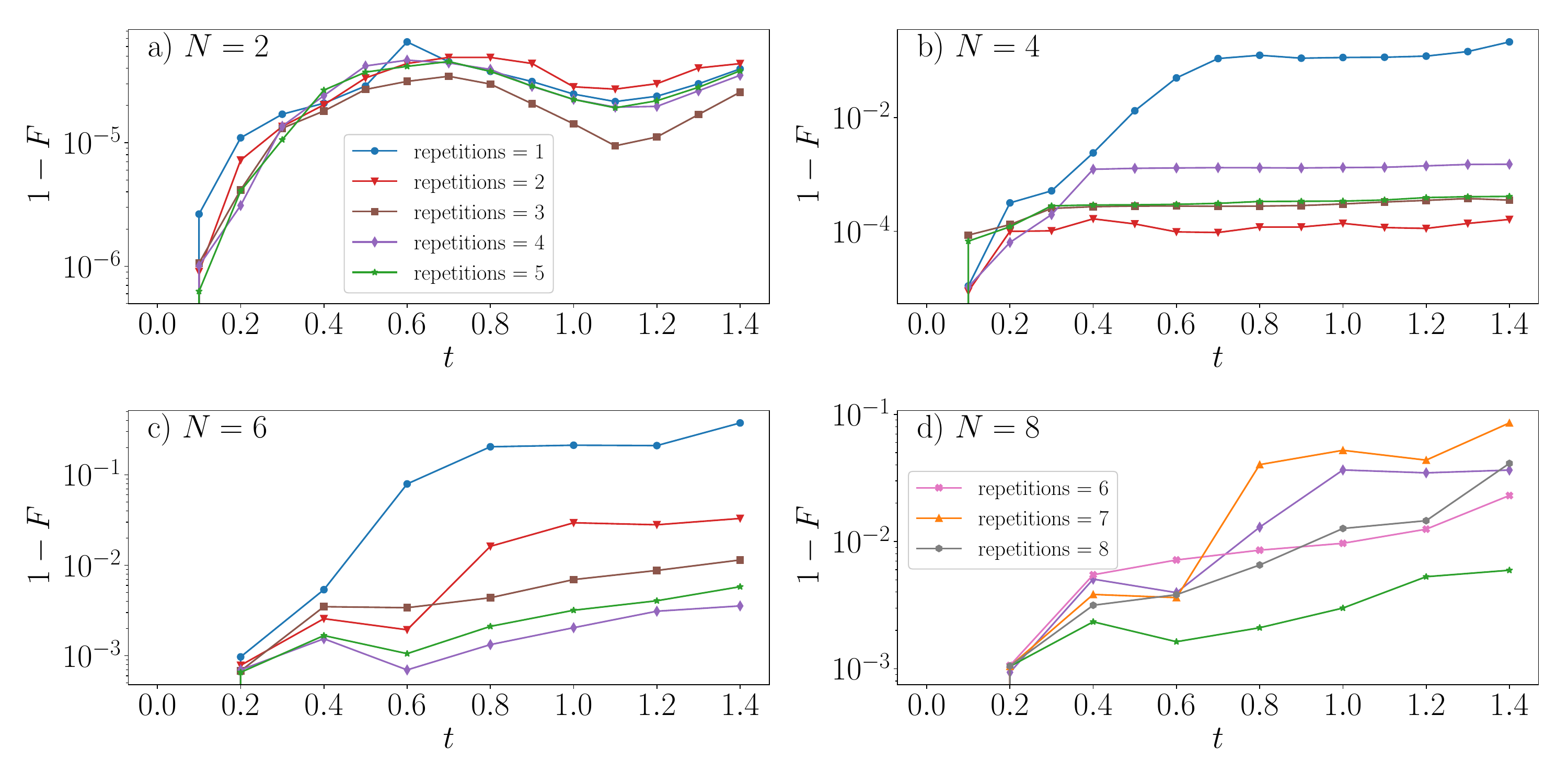}
\caption{\label{fig:pvqd_infidelity} Infidelity between the p-VQD optimized state (via ideal, noise-free statevector simulations) and the exact time-evolved state for system sizes $N=2$ (a), $N=4$ (b), $N=6$ (c), $N=8$ (d), and different {number of repetitions} of the variational circuit. The optimizations have been performed using BFGS for the Ising chain dynamics defined in the main text.}
\end{figure*}

\subsection{p-VQD}\label{app:pvqd}

We optimize p-VQD using the BFGS optimizer for the state-vector simulations and SPSA \cite{spall1992multivariate} for the noisy simulations. While SPSA only performs approximate gradient descent, it also only requires two circuit evaluations per optimization step independent of the number of parameters and can thus be efficiently executed on quantum devices. Moreover, the stochasticity in the perturbation directions make it robust to noise. The fidelity in Eq.~\eqref{eq:pvqd} is evaluated via sampling and can be replaced by a local cost function to make the optimization less prone to barren plateaus \cite{barison2021efficient,Cerezo2021b}. As a termination condition for the BFGS optimizer we set a precision goal of $10^{-6}$ in the cost function, i.e., the infidelity in Eq.~\eqref{eq:pvqd}. When using SPSA we set the number of optimization steps per time step to 1000 instead. We run several state-vector simulations of the evolution up to a total time $t=1.4$ with different time increments $\delta t$, {ansatz repetitions}, and 3 random seeds to determine the optimal hyper-parameters that minimize the infidelity with respect to the exact state computed using the QuSpin package \cite{quspin}. In Fig.~\ref{fig:pvqd_infidelity} we plot the infidelity as a function of time for 4 different system sizes showing only the best out of each run. As expected, we find that the infidelity on average increases with time as errors build up in the time evolved state. Furthermore, the infidelity also grows with system size and as such we require {more ansatz repetitions} to faithfully represent the increasingly correlated quantum states. For the $N=8$ state-vector simulations of Section~\ref{res:state-vector} we use {5 repetitions of the hardware efficient gate-fabric}, for the $N=2$ noisy simulations of Fig.~\ref{fig:pvqd_N2_M1} we set the {number of repetitions} to 1 and for the $N=4$ quantum device experiments of Fig.~\ref{fig:pvqd_runtime_N4_M2} the {number of repetitions} is equal to 2. Note that in the latter case, we used the pre-optimized p-VQD parameters from the state-vector simulation to time-evolve the state on the noisy hardware. Additionally, we experimented with different numbers of time-steps and found that the optimal number of time-steps needed is 7 for $N=6$ and $N=8$ while it is 14 for $N=2$ and $N=4$.

\subsection{Passive state optimization}\label{app:vqergo}

\begin{figure}[tb]
\centering
\includegraphics[width=85mm]{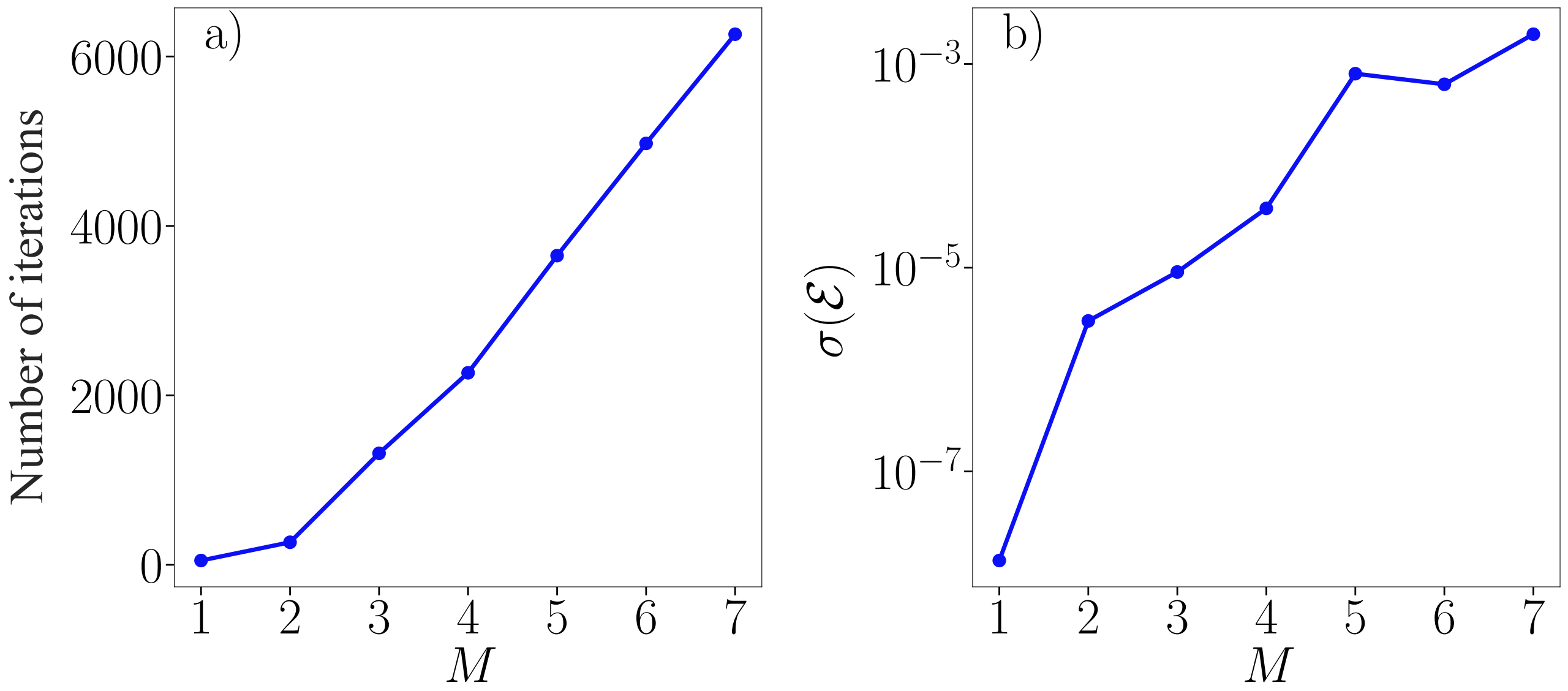}
\caption{\label{fig:pvqd_statevector_Mscale} (a) The average number of BFGS iterations required to achieve a final precision of $10^{-6}$ in the cost of the passive state optimization as a function of the subsystem size $M$. (b) The corresponding standard deviation in the ergotropy over 100 runs. The optimizations were performed using the noise-free statevector simulator, a total system size $N=8$, {number of repetitions} of 2, and a charging time $t=0.4$.
}
\end{figure}

Analogous to p-VQD, we use the BFGS optimizer for the statevector simulations and SPSA for the noisy and real-device simulations. With the exception of the hardware experiments for which only a single data point per time is collected, we repeat each classical simulation with 100 random seeds and take the average. Fig.~\ref{fig:pvqd_statevector_Mscale}(a) shows the average number of required BFGS optimization steps to reach a precision of $10^{-6}$ in the cost as a function of the subsystem size $M$ for an example of a charged state at $t=0.4$. The {number of repetitions} was fixed to 2. Note that the number of parameters in the circuit ansatz grow linearly in $M$. We also display the observed standard deviation of the ergotropy versus the subsystem size (see Fig.~\ref{fig:pvqd_statevector_Mscale}(b)).

\begin{figure}[tb] 
\centering
\includegraphics[width=85mm]{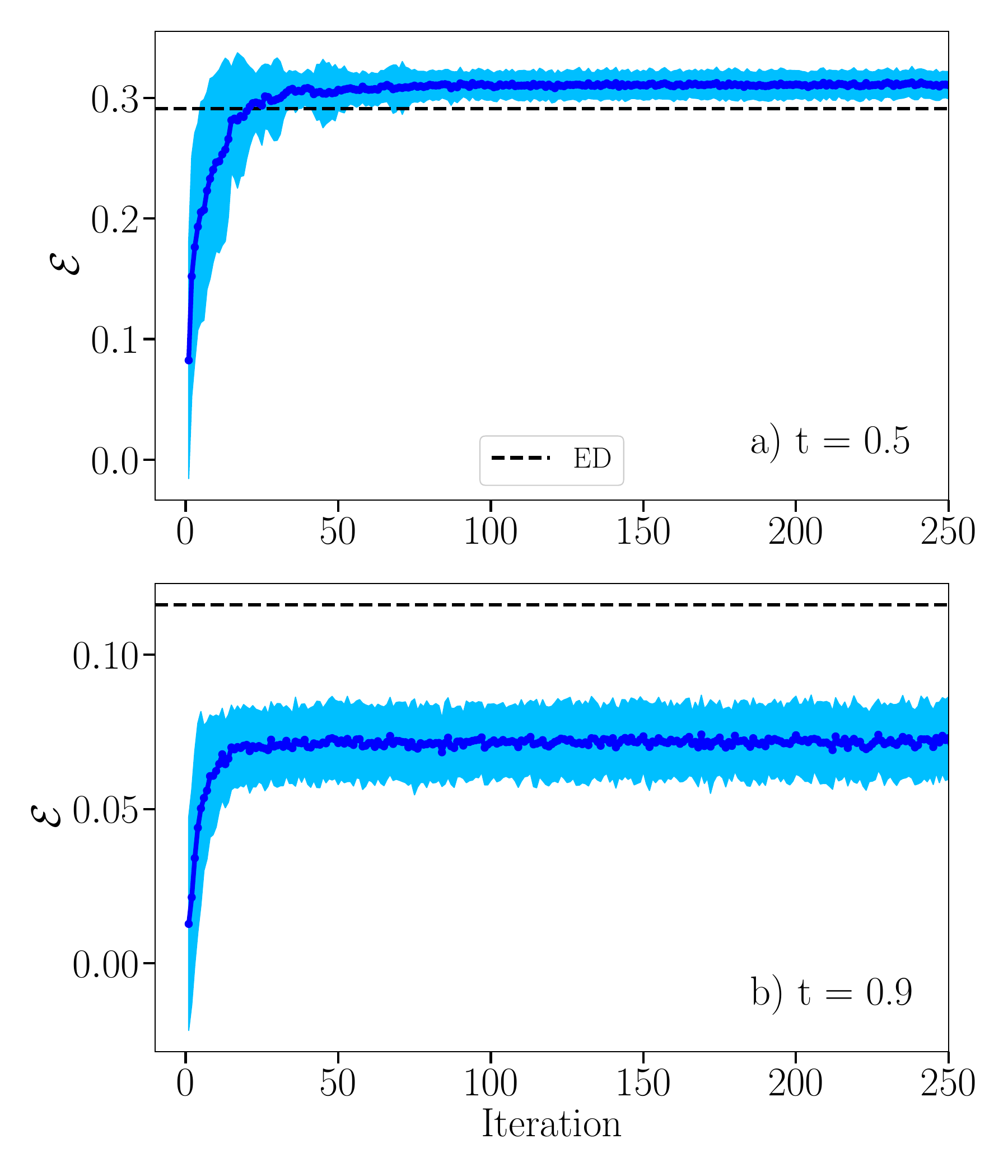}
\caption{\label{fig:pvqd_noisy_N2_M1_200iteration}The average ergotropy $\mathcal{E}$ over 100 runs at each SPSA iteration for charging times $t=0.5$ (a) and $t=0.9$ (b) using noisy simulation with FakePerth (see also Fig.~\ref{fig:pvqd_N2_M1} in the main text). The shaded region indicates the standard deviation. The black dashed line corresponds to the exact result calculated using ED methods. $N=2,M=1$.}
\end{figure}

In Fig.~\ref{fig:pvqd_noisy_N2_M1_200iteration} we show two training curves of the passive state optimization using SPSA that were collected for the two exemplary charging times $t=0.5$ and $t=0.9$ of Fig.~\ref{fig:pvqd_N2_M1} in the main text. The dark (bright) color corresponds to the mean (std) over 100 independent noisy simulations using the FakePerth backend while the dashed line indicates the theoretically exact ergotropy. The optimization usually converged within the first $50\sim 100$ steps. However, the final value can deviate from its exact prediction due to noise coming from the mean and passive energy measurements as well as errors arising in the p-VQD time evolution.

\begin{figure}[tb]
\centering
\includegraphics[width=85mm]{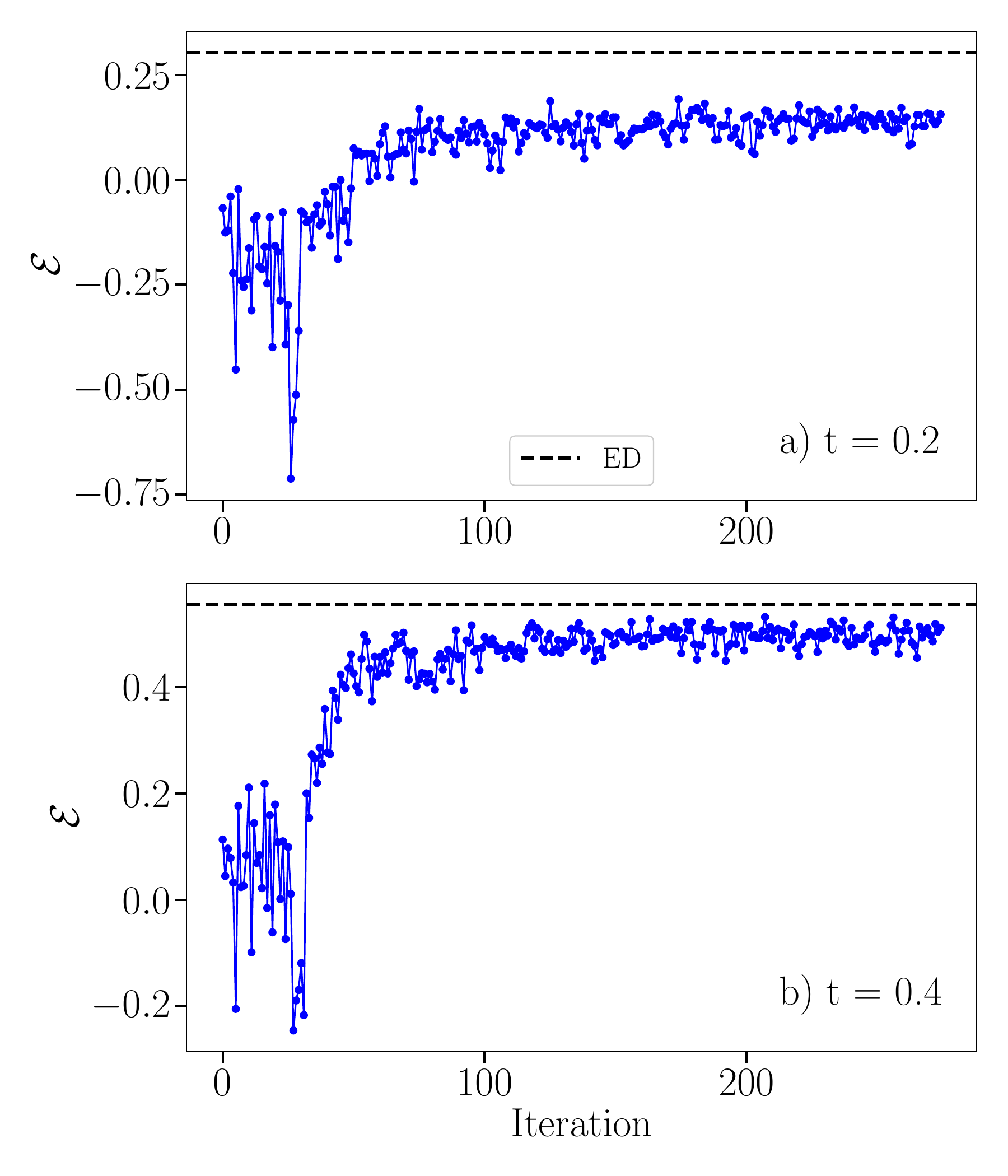}
\caption{\label{fig:runtime_iteration}The estimated ergotropy $\mathcal{E}$ at each SPSA iteration for charging times $t=0.2$ (a) and $t=0.4$ (b) measured on the \texttt{ibm\_perth} quantum hardware (see also Fig.~\ref{fig:pvqd_runtime_N4_M2} in the main text). The black dashed line corresponds to the exact result calculated using ED methods. $N=4,M=2$.}
\end{figure}

We also show two typical training curves for the real-device experiment results (see Fig.~\ref{fig:pvqd_runtime_N4_M2} in the main text) performed on \texttt{ibm\_perth} in Fig.~\ref{fig:runtime_iteration}. Rather than measuring the passive energy after each optimization step which would require additional circuit evaluations, we instead estimate its value by averaging the two expectation values used by SPSA at each iteration.

\section{Alternative charging protocol}\label{appendix-turning_off_magnetic}

In this section we provide results obtained with a simplified battery charging scheme that does not require trotterization or variational optimization and thus can be easily implemented on NISQ hardware. Instead of evolving the system with the transverse field Ising Hamiltonian of Eq.~\eqref{pvqdhamiltonian} we turn off the magenetic field during the quench and only evolve with the term containing the nearest-neighbor coupling
\begin{equation} \label{rxxhamiltonian}
    H_1 = -J\sum_{i=1}^{N-1} \sigma_i^x\sigma_{i+1}^x.
\end{equation}
Note that $H_1$ is composed of only commuting terms. Therefore, the time evolution operator $e^{-iH_1 t}$ can be exactly decomposed into a single layer of two-qubit gates acting only on neighboring spins leading to
\begin{equation} \label{rxxquenchedstate}
    |\Psi\left(t\right)\rangle = e^{-iH_1 t} |0\rangle= \prod_{i=1}^{N-1} R_{XX}^{i,i+1}\left(\theta\right) |0\rangle,
\end{equation}
where $R_{XX}^{i,i+1}\left(\theta\right) = \text{exp}\left(-i\dfrac{\theta}{2}\sigma_x^i\otimes \sigma_x^{i+1} \right)$ and $\theta=-2Jt$. {Note that the charging time $t$ is encoded as the angle of the $R_{XX}$ gates.}

\subsection{Statevector simulation}

\begin{figure*}[tb] 
\centering
\includegraphics[width=\textwidth]{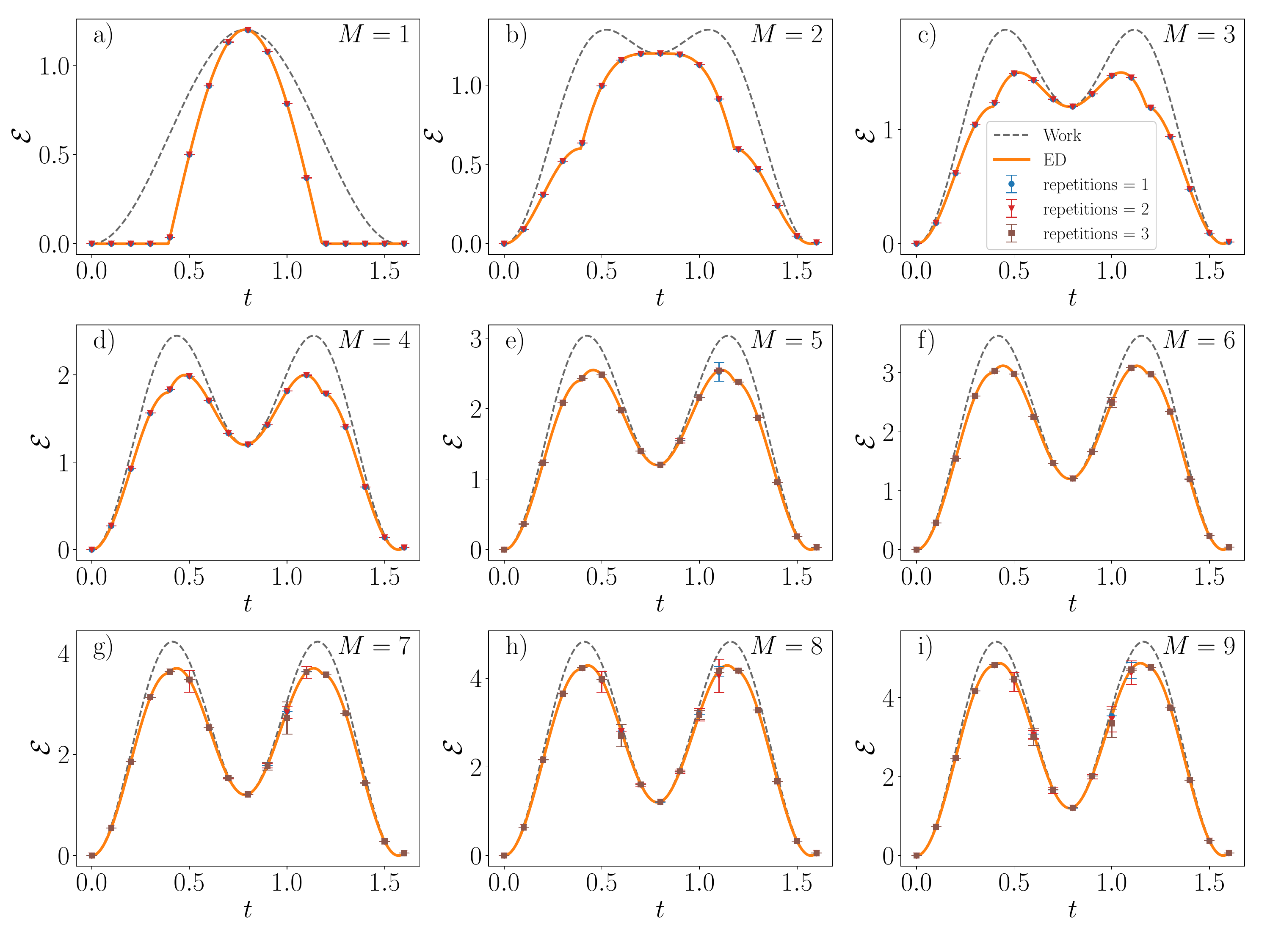}
\caption{\label{fig:rxx_statevector_depthscale} The ergotropy $\mathcal{E}$ as a function of the charging time $t$ for different subsystem sizes $M$ (total system size $N=10$). The transverse magnetic field is turned off during the charging quench. The orange line indicates the exact values while the markers denote average values of 100 optimization runs using a statevector simulation with different {repetition numbers}. The total injected work is shown as a gray dashed line.}
\end{figure*}

We run VQErgo on charged states of an $N=10$ spin system and show the achieved final ergotropies for different subsystem sizes $M$ in Fig.~\ref{fig:rxx_statevector_depthscale}. Interestingly, we find that a {single repetition} is sufficient to achieve a high accuracy with the exactly computed values (orange line) for all considered cell sizes and charging times. This suggests that the evolved state only contains nearest neighbor correlations irrespective of the charging time $t$. Moreover, we observe that the stored work and ergotropy coincide at times $t=(k+1/2)\pi/J$, with $k=0,1,2,\ldots{}$ reaching an energy $W = \mathcal{E} = 2h$. At these times the charged state is in a fully disentangled product state. The ergotropy in this case depends solely on the number of qubits $M<N$ from which we want to extract the energy instead of the full system size $N$. This is shown in Fig.~\ref{fig:rxx_noisy_depth1} for $N=6$ whereby the ergotropy and work are the same as in the $N=10$ case in Fig.~\ref{fig:rxx_statevector_depthscale}. Note that for the case of  $M=1$, it is also possible to obtain an analytical expression for the ergotropy
\begin{equation}
    \mathcal{E}(t)=\begin{cases}
			0, & \text{if $\tan^2(Jt) \leq 1$}\\
            2h\left[\sin^2(Jt)-\cos^2(Jt)\right], & \text{if $\tan^2(Jt) > 1$}
		 \end{cases}.
\end{equation}
Overall, and unsurprisingly, the dynamics in this case is more trivial than the one generated by the full transverse field Ising Hamiltonian. The time evolved state is entangled only over short distances and thus, the passive state can be prepared with at most a single layer of nearest-neighbor two-qubit gates. It would be interesting to study the quantum battery with charging protocols interpolating the simplified case discussed here and the Ising dynamics from the main text by applying a small number of quenches with alternating non-commuting generators.

\subsection{Noisy simulations}

\begin{figure*} 
\centering
\includegraphics[width=\textwidth]{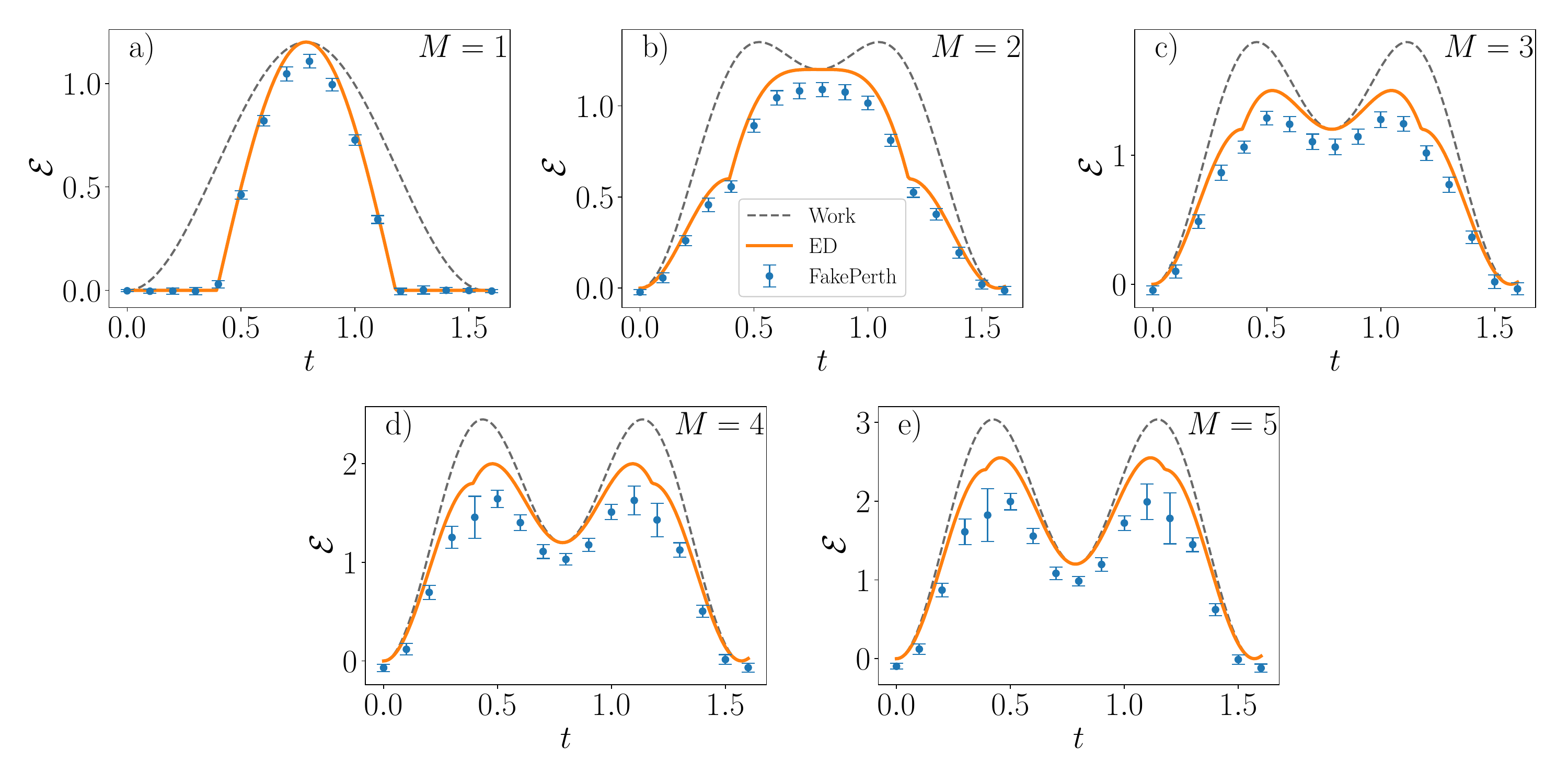}
\caption{\label{fig:rxx_noisy_depth1} Same as Fig. \ref{fig:rxx_statevector_depthscale} but for a total system size of $N=6$ and using a noisy simulator (FakePerth) instead. The {number of repetitions} is fixed to 1.}
\end{figure*}

We have also tested VQErgo with the simplified charging protocol on noisy simulators and hardware. We choose 6 qubits of the 7 qubit \texttt{ibm\_perth} device and plot an average of the measured ergotropies for different subsystem sizes $M$ obtained on the noisy classical simulator in Fig.~\ref{fig:rxx_noisy_depth1}. The {number of repetitions} of the passive state ansatz circuit is 1. However, extra SWAP gates are required to map the full circuit to the underlying topology of the real device which ultimately introduces more noise. The error between the exact and variationally obtained ergotropies grows with the battery cell size. On the other hand, the error is independent of the charging time which is in contrast to the p-VQD based simulation where errors naturally built up over time. Despite the noisy values, we can successfully infer the qualitative dependence of the ergotropy with the charging time.

\begin{figure}
\centering
\includegraphics[width=85 mm]{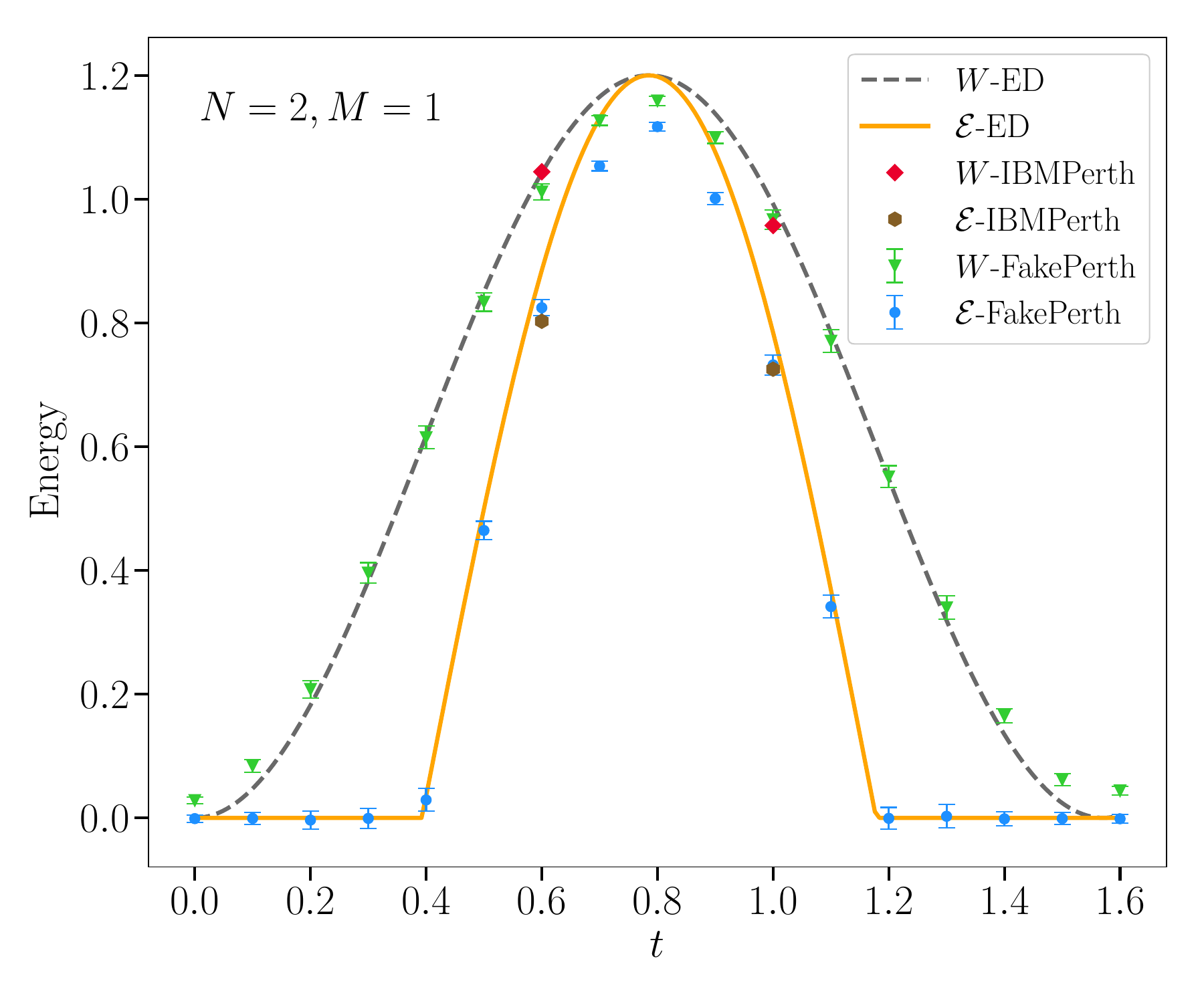}
\caption{\label{fig:rxx_runtime_N2_M1} The ergotropy (orange solid line) and total work (gray dashed line) computed via exact techniques versus the charging time $t$ for a system with $N=2,M=1$ undergoing the simplified dynamics. The markers indicate the measured ergotropies obtained either from a classical noisy VQErgo simulation or real-hardware experiments performed on \texttt{ibm\_perth}. In the case of the former, the points (error bars) denote the average (standard deviation) over 100 independent runs of the optimization.
}
\end{figure}

Finally, we also report two results obtained on the \texttt{ibm\_perth} quantum device for a system with $N=2,M=1$ in Fig. \ref{fig:rxx_runtime_N2_M1}. Again, we find an overall good agreement between the measured values and their exact prediction. 

\bibliography{apssamp}

\providecommand{\noopsort}[1]{}\providecommand{\singleletter}[1]{#1}%
\begin{thebibliography}{111}%
\makeatletter
\providecommand \@ifxundefined [1]{%
 \@ifx{#1\undefined}
}%
\providecommand \@ifnum [1]{%
 \ifnum #1\expandafter \@firstoftwo
 \else \expandafter \@secondoftwo
 \fi
}%
\providecommand \@ifx [1]{%
 \ifx #1\expandafter \@firstoftwo
 \else \expandafter \@secondoftwo
 \fi
}%
\providecommand \natexlab [1]{#1}%
\providecommand \enquote  [1]{``#1''}%
\providecommand \bibnamefont  [1]{#1}%
\providecommand \bibfnamefont [1]{#1}%
\providecommand \citenamefont [1]{#1}%
\providecommand \href@noop [0]{\@secondoftwo}%
\providecommand \href [0]{\begingroup \@sanitize@url \@href}%
\providecommand \@href[1]{\@@startlink{#1}\@@href}%
\providecommand \@@href[1]{\endgroup#1\@@endlink}%
\providecommand \@sanitize@url [0]{\catcode `\\12\catcode `\$12\catcode
  `\&12\catcode `\#12\catcode `\^12\catcode `\_12\catcode `\%12\relax}%
\providecommand \@@startlink[1]{}%
\providecommand \@@endlink[0]{}%
\providecommand \url  [0]{\begingroup\@sanitize@url \@url }%
\providecommand \@url [1]{\endgroup\@href {#1}{\urlprefix }}%
\providecommand \urlprefix  [0]{URL }%
\providecommand \Eprint [0]{\href }%
\providecommand \doibase [0]{https://doi.org/}%
\providecommand \selectlanguage [0]{\@gobble}%
\providecommand \bibinfo  [0]{\@secondoftwo}%
\providecommand \bibfield  [0]{\@secondoftwo}%
\providecommand \translation [1]{[#1]}%
\providecommand \BibitemOpen [0]{}%
\providecommand \bibitemStop [0]{}%
\providecommand \bibitemNoStop [0]{.\EOS\space}%
\providecommand \EOS [0]{\spacefactor3000\relax}%
\providecommand \BibitemShut  [1]{\csname bibitem#1\endcsname}%
\let\auto@bib@innerbib\@empty
\bibitem [{\citenamefont {Kosloff}\ and\ \citenamefont
  {Levy}(2014)}]{Kosloff2014}%
  \BibitemOpen
  \bibfield  {author} {\bibinfo {author} {\bibfnamefont {R.}~\bibnamefont
  {Kosloff}}\ and\ \bibinfo {author} {\bibfnamefont {A.}~\bibnamefont {Levy}},\
  }\bibfield  {title} {\bibinfo {title} {Quantum heat engines and
  refrigerators: Continuous devices},\ }\href
  {https://doi.org/10.1146/annurev-physchem-040513-103724} {\bibfield
  {journal} {\bibinfo  {journal} {Annual Review of Physical Chemistry}\
  }\textbf {\bibinfo {volume} {65}},\ \bibinfo {pages} {365} (\bibinfo {year}
  {2014})},\ \bibinfo {note} {pMID: 24689798},\ \Eprint
  {https://arxiv.org/abs/https://doi.org/10.1146/annurev-physchem-040513-103724}
  {https://doi.org/10.1146/annurev-physchem-040513-103724} \BibitemShut
  {NoStop}%
\bibitem [{\citenamefont {Mitchison}(2019)}]{Mark2019}%
  \BibitemOpen
  \bibfield  {author} {\bibinfo {author} {\bibfnamefont {M.~T.}\ \bibnamefont
  {Mitchison}},\ }\bibfield  {title} {\bibinfo {title} {Quantum thermal
  absorption machines: refrigerators, engines and clocks},\ }\href
  {https://doi.org/10.1080/00107514.2019.1631555} {\bibfield  {journal}
  {\bibinfo  {journal} {Contemporary Physics}\ }\textbf {\bibinfo {volume}
  {60}},\ \bibinfo {pages} {164} (\bibinfo {year} {2019})},\ \Eprint
  {https://arxiv.org/abs/https://doi.org/10.1080/00107514.2019.1631555}
  {https://doi.org/10.1080/00107514.2019.1631555} \BibitemShut {NoStop}%
\bibitem [{\citenamefont {Myers}\ \emph {et~al.}(2022)\citenamefont {Myers},
  \citenamefont {Abah},\ and\ \citenamefont {Deffner}}]{Myers2022}%
  \BibitemOpen
  \bibfield  {author} {\bibinfo {author} {\bibfnamefont {N.~M.}\ \bibnamefont
  {Myers}}, \bibinfo {author} {\bibfnamefont {O.}~\bibnamefont {Abah}},\ and\
  \bibinfo {author} {\bibfnamefont {S.}~\bibnamefont {Deffner}},\ }\bibfield
  {title} {\bibinfo {title} {{Quantum thermodynamic devices: From theoretical
  proposals to experimental reality}},\ }\href
  {https://doi.org/10.1116/5.0083192} {\bibfield  {journal} {\bibinfo
  {journal} {AVS Quantum Science}\ }\textbf {\bibinfo {volume} {4}},\ \bibinfo
  {pages} {027101} (\bibinfo {year} {2022})}\BibitemShut {NoStop}%
\bibitem [{\citenamefont {Alicki}\ and\ \citenamefont
  {Fannes}(2013)}]{alicki2013entanglement}%
  \BibitemOpen
  \bibfield  {author} {\bibinfo {author} {\bibfnamefont {R.}~\bibnamefont
  {Alicki}}\ and\ \bibinfo {author} {\bibfnamefont {M.}~\bibnamefont
  {Fannes}},\ }\bibfield  {title} {\bibinfo {title} {Entanglement boost for
  extractable work from ensembles of quantum batteries},\ }\href
  {https://doi.org/https://doi.org/10.1103/PhysRevE.87.042123} {\bibfield
  {journal} {\bibinfo  {journal} {Physical Review E}\ }\textbf {\bibinfo
  {volume} {87}},\ \bibinfo {pages} {042123} (\bibinfo {year}
  {2013})}\BibitemShut {NoStop}%
\bibitem [{\citenamefont {Hovhannisyan}\ \emph {et~al.}(2013)\citenamefont
  {Hovhannisyan}, \citenamefont {Perarnau-Llobet}, \citenamefont {Huber},\ and\
  \citenamefont {Ac{\'\i}n}}]{hovhannisyan2013entanglement}%
  \BibitemOpen
  \bibfield  {author} {\bibinfo {author} {\bibfnamefont {K.~V.}\ \bibnamefont
  {Hovhannisyan}}, \bibinfo {author} {\bibfnamefont {M.}~\bibnamefont
  {Perarnau-Llobet}}, \bibinfo {author} {\bibfnamefont {M.}~\bibnamefont
  {Huber}},\ and\ \bibinfo {author} {\bibfnamefont {A.}~\bibnamefont
  {Ac{\'\i}n}},\ }\bibfield  {title} {\bibinfo {title} {Entanglement generation
  is not necessary for optimal work extraction},\ }\href
  {https://doi.org/https://doi.org/10.1103/PhysRevLett.111.240401} {\bibfield
  {journal} {\bibinfo  {journal} {Physical Review Letters}\ }\textbf {\bibinfo
  {volume} {111}},\ \bibinfo {pages} {240401} (\bibinfo {year}
  {2013})}\BibitemShut {NoStop}%
\bibitem [{\citenamefont {Rossini}\ \emph {et~al.}(2019)\citenamefont
  {Rossini}, \citenamefont {Andolina},\ and\ \citenamefont
  {Polini}}]{rossini2019many}%
  \BibitemOpen
  \bibfield  {author} {\bibinfo {author} {\bibfnamefont {D.}~\bibnamefont
  {Rossini}}, \bibinfo {author} {\bibfnamefont {G.~M.}\ \bibnamefont
  {Andolina}},\ and\ \bibinfo {author} {\bibfnamefont {M.}~\bibnamefont
  {Polini}},\ }\bibfield  {title} {\bibinfo {title} {Many-body localized
  quantum batteries},\ }\href
  {https://doi.org/https://doi.org/10.1103/PhysRevB.100.115142} {\bibfield
  {journal} {\bibinfo  {journal} {Physical Review B}\ }\textbf {\bibinfo
  {volume} {100}},\ \bibinfo {pages} {115142} (\bibinfo {year}
  {2019})}\BibitemShut {NoStop}%
\bibitem [{\citenamefont {Andolina}\ \emph
  {et~al.}(2019{\natexlab{a}})\citenamefont {Andolina}, \citenamefont {Keck},
  \citenamefont {Mari}, \citenamefont {Campisi}, \citenamefont {Giovannetti},\
  and\ \citenamefont {Polini}}]{andolina2019extractable}%
  \BibitemOpen
  \bibfield  {author} {\bibinfo {author} {\bibfnamefont {G.~M.}\ \bibnamefont
  {Andolina}}, \bibinfo {author} {\bibfnamefont {M.}~\bibnamefont {Keck}},
  \bibinfo {author} {\bibfnamefont {A.}~\bibnamefont {Mari}}, \bibinfo {author}
  {\bibfnamefont {M.}~\bibnamefont {Campisi}}, \bibinfo {author} {\bibfnamefont
  {V.}~\bibnamefont {Giovannetti}},\ and\ \bibinfo {author} {\bibfnamefont
  {M.}~\bibnamefont {Polini}},\ }\bibfield  {title} {\bibinfo {title}
  {Extractable work, the role of correlations, and asymptotic freedom in
  quantum batteries},\ }\href
  {https://doi.org/https://doi.org/10.1103/PhysRevLett.122.047702} {\bibfield
  {journal} {\bibinfo  {journal} {Physical Review Letters}\ }\textbf {\bibinfo
  {volume} {122}},\ \bibinfo {pages} {047702} (\bibinfo {year}
  {2019}{\natexlab{a}})}\BibitemShut {NoStop}%
\bibitem [{\citenamefont {Hu}\ \emph {et~al.}(2022)\citenamefont {Hu},
  \citenamefont {Qiu}, \citenamefont {Souza}, \citenamefont {Yuan},
  \citenamefont {Zhou}, \citenamefont {Zhang}, \citenamefont {Chu},
  \citenamefont {Pan}, \citenamefont {Hu}, \citenamefont {Li}, \citenamefont
  {Xu}, \citenamefont {Zhong}, \citenamefont {Liu}, \citenamefont {Yan},
  \citenamefont {Tan}, \citenamefont {Bachelard}, \citenamefont {Villas-Boas},
  \citenamefont {Santos},\ and\ \citenamefont {Yu}}]{Hu_2022}%
  \BibitemOpen
  \bibfield  {author} {\bibinfo {author} {\bibfnamefont {C.-K.}\ \bibnamefont
  {Hu}}, \bibinfo {author} {\bibfnamefont {J.}~\bibnamefont {Qiu}}, \bibinfo
  {author} {\bibfnamefont {P.~J.~P.}\ \bibnamefont {Souza}}, \bibinfo {author}
  {\bibfnamefont {J.}~\bibnamefont {Yuan}}, \bibinfo {author} {\bibfnamefont
  {Y.}~\bibnamefont {Zhou}}, \bibinfo {author} {\bibfnamefont {L.}~\bibnamefont
  {Zhang}}, \bibinfo {author} {\bibfnamefont {J.}~\bibnamefont {Chu}}, \bibinfo
  {author} {\bibfnamefont {X.}~\bibnamefont {Pan}}, \bibinfo {author}
  {\bibfnamefont {L.}~\bibnamefont {Hu}}, \bibinfo {author} {\bibfnamefont
  {J.}~\bibnamefont {Li}}, \bibinfo {author} {\bibfnamefont {Y.}~\bibnamefont
  {Xu}}, \bibinfo {author} {\bibfnamefont {Y.}~\bibnamefont {Zhong}}, \bibinfo
  {author} {\bibfnamefont {S.}~\bibnamefont {Liu}}, \bibinfo {author}
  {\bibfnamefont {F.}~\bibnamefont {Yan}}, \bibinfo {author} {\bibfnamefont
  {D.}~\bibnamefont {Tan}}, \bibinfo {author} {\bibfnamefont {R.}~\bibnamefont
  {Bachelard}}, \bibinfo {author} {\bibfnamefont {C.~J.}\ \bibnamefont
  {Villas-Boas}}, \bibinfo {author} {\bibfnamefont {A.~C.}\ \bibnamefont
  {Santos}},\ and\ \bibinfo {author} {\bibfnamefont {D.}~\bibnamefont {Yu}},\
  }\bibfield  {title} {\bibinfo {title} {Optimal charging of a superconducting
  quantum battery},\ }\href {https://doi.org/10.1088/2058-9565/ac8444}
  {\bibfield  {journal} {\bibinfo  {journal} {Quantum Science and Technology}\
  }\textbf {\bibinfo {volume} {7}},\ \bibinfo {pages} {045018} (\bibinfo {year}
  {2022})}\BibitemShut {NoStop}%
\bibitem [{\citenamefont {Konar}\ \emph {et~al.}(2022)\citenamefont {Konar},
  \citenamefont {Lakkaraju}, \citenamefont {Ghosh}, \citenamefont {Sen} \emph
  {et~al.}}]{konar2022quantum}%
  \BibitemOpen
  \bibfield  {author} {\bibinfo {author} {\bibfnamefont {T.~K.}\ \bibnamefont
  {Konar}}, \bibinfo {author} {\bibfnamefont {L.~G.~C.}\ \bibnamefont
  {Lakkaraju}}, \bibinfo {author} {\bibfnamefont {S.}~\bibnamefont {Ghosh}},
  \bibinfo {author} {\bibfnamefont {A.}~\bibnamefont {Sen}}, \emph {et~al.},\
  }\bibfield  {title} {\bibinfo {title} {Quantum battery with ultracold atoms:
  Bosons versus fermions},\ }\href
  {https://doi.org/https://doi.org/10.1103/PhysRevA.106.022618} {\bibfield
  {journal} {\bibinfo  {journal} {Physical Review A}\ }\textbf {\bibinfo
  {volume} {106}},\ \bibinfo {pages} {022618} (\bibinfo {year}
  {2022})}\BibitemShut {NoStop}%
\bibitem [{\citenamefont {Barra}\ \emph {et~al.}(2022)\citenamefont {Barra},
  \citenamefont {Hovhannisyan},\ and\ \citenamefont
  {Imparato}}]{barra2022quantum}%
  \BibitemOpen
  \bibfield  {author} {\bibinfo {author} {\bibfnamefont {F.}~\bibnamefont
  {Barra}}, \bibinfo {author} {\bibfnamefont {K.~V.}\ \bibnamefont
  {Hovhannisyan}},\ and\ \bibinfo {author} {\bibfnamefont {A.}~\bibnamefont
  {Imparato}},\ }\bibfield  {title} {\bibinfo {title} {Quantum batteries at the
  verge of a phase transition},\ }\href
  {https://doi.org/10.1088/1367-2630/ac43ed} {\bibfield  {journal} {\bibinfo
  {journal} {New Journal of Physics}\ }\textbf {\bibinfo {volume} {24}},\
  \bibinfo {pages} {015003} (\bibinfo {year} {2022})}\BibitemShut {NoStop}%
\bibitem [{\citenamefont {Arjmandi}\ \emph {et~al.}(2022)\citenamefont
  {Arjmandi}, \citenamefont {Shokri}, \citenamefont {Faizi},\ and\
  \citenamefont {Mohammadi}}]{arjmandi2022performance}%
  \BibitemOpen
  \bibfield  {author} {\bibinfo {author} {\bibfnamefont {M.~B.}\ \bibnamefont
  {Arjmandi}}, \bibinfo {author} {\bibfnamefont {A.}~\bibnamefont {Shokri}},
  \bibinfo {author} {\bibfnamefont {E.}~\bibnamefont {Faizi}},\ and\ \bibinfo
  {author} {\bibfnamefont {H.}~\bibnamefont {Mohammadi}},\ }\bibfield  {title}
  {\bibinfo {title} {Performance of quantum batteries with correlated and
  uncorrelated chargers},\ }\href
  {https://doi.org/https://doi.org/10.1103/PhysRevA.106.062609} {\bibfield
  {journal} {\bibinfo  {journal} {Physical Review A}\ }\textbf {\bibinfo
  {volume} {106}},\ \bibinfo {pages} {062609} (\bibinfo {year}
  {2022})}\BibitemShut {NoStop}%
\bibitem [{\citenamefont {\ifmmode~\check{S}\else \v{S}\fi{}afr\'anek}\ \emph
  {et~al.}(2023)\citenamefont {\ifmmode~\check{S}\else \v{S}\fi{}afr\'anek},
  \citenamefont {Rosa},\ and\ \citenamefont {Binder}}]{Binder2023}%
  \BibitemOpen
  \bibfield  {author} {\bibinfo {author} {\bibfnamefont {D.}~\bibnamefont
  {\ifmmode~\check{S}\else \v{S}\fi{}afr\'anek}}, \bibinfo {author}
  {\bibfnamefont {D.}~\bibnamefont {Rosa}},\ and\ \bibinfo {author}
  {\bibfnamefont {F.~C.}\ \bibnamefont {Binder}},\ }\bibfield  {title}
  {\bibinfo {title} {Work extraction from unknown quantum sources},\ }\href
  {https://doi.org/10.1103/PhysRevLett.130.210401} {\bibfield  {journal}
  {\bibinfo  {journal} {Physical Review Letters}\ }\textbf {\bibinfo {volume}
  {130}},\ \bibinfo {pages} {210401} (\bibinfo {year} {2023})}\BibitemShut
  {NoStop}%
\bibitem [{\citenamefont {Zhang}\ \emph {et~al.}(2019)\citenamefont {Zhang},
  \citenamefont {Yang}, \citenamefont {Fu},\ and\ \citenamefont
  {Wang}}]{zhang2019powerful}%
  \BibitemOpen
  \bibfield  {author} {\bibinfo {author} {\bibfnamefont {Y.-Y.}\ \bibnamefont
  {Zhang}}, \bibinfo {author} {\bibfnamefont {T.-R.}\ \bibnamefont {Yang}},
  \bibinfo {author} {\bibfnamefont {L.}~\bibnamefont {Fu}},\ and\ \bibinfo
  {author} {\bibfnamefont {X.}~\bibnamefont {Wang}},\ }\bibfield  {title}
  {\bibinfo {title} {Powerful harmonic charging in a quantum battery},\ }\href
  {https://doi.org/https://doi.org/10.1103/PhysRevE.99.052106} {\bibfield
  {journal} {\bibinfo  {journal} {Physical Review E}\ }\textbf {\bibinfo
  {volume} {99}},\ \bibinfo {pages} {052106} (\bibinfo {year}
  {2019})}\BibitemShut {NoStop}%
\bibitem [{\citenamefont {Andolina}\ \emph {et~al.}(2018)\citenamefont
  {Andolina}, \citenamefont {Farina}, \citenamefont {Mari}, \citenamefont
  {Pellegrini}, \citenamefont {Giovannetti},\ and\ \citenamefont
  {Polini}}]{andolina2018charger}%
  \BibitemOpen
  \bibfield  {author} {\bibinfo {author} {\bibfnamefont {G.~M.}\ \bibnamefont
  {Andolina}}, \bibinfo {author} {\bibfnamefont {D.}~\bibnamefont {Farina}},
  \bibinfo {author} {\bibfnamefont {A.}~\bibnamefont {Mari}}, \bibinfo {author}
  {\bibfnamefont {V.}~\bibnamefont {Pellegrini}}, \bibinfo {author}
  {\bibfnamefont {V.}~\bibnamefont {Giovannetti}},\ and\ \bibinfo {author}
  {\bibfnamefont {M.}~\bibnamefont {Polini}},\ }\bibfield  {title} {\bibinfo
  {title} {Charger-mediated energy transfer in exactly solvable models for
  quantum batteries},\ }\href
  {https://doi.org/https://doi.org/10.1103/PhysRevB.98.205423} {\bibfield
  {journal} {\bibinfo  {journal} {Physical Review B}\ }\textbf {\bibinfo
  {volume} {98}},\ \bibinfo {pages} {205423} (\bibinfo {year}
  {2018})}\BibitemShut {NoStop}%
\bibitem [{\citenamefont {Andolina}\ \emph
  {et~al.}(2019{\natexlab{b}})\citenamefont {Andolina}, \citenamefont {Keck},
  \citenamefont {Mari}, \citenamefont {Giovannetti},\ and\ \citenamefont
  {Polini}}]{andolina2019quantum}%
  \BibitemOpen
  \bibfield  {author} {\bibinfo {author} {\bibfnamefont {G.~M.}\ \bibnamefont
  {Andolina}}, \bibinfo {author} {\bibfnamefont {M.}~\bibnamefont {Keck}},
  \bibinfo {author} {\bibfnamefont {A.}~\bibnamefont {Mari}}, \bibinfo {author}
  {\bibfnamefont {V.}~\bibnamefont {Giovannetti}},\ and\ \bibinfo {author}
  {\bibfnamefont {M.}~\bibnamefont {Polini}},\ }\bibfield  {title} {\bibinfo
  {title} {Quantum versus classical many-body batteries},\ }\href
  {https://doi.org/https://doi.org/10.1103/PhysRevB.99.205437} {\bibfield
  {journal} {\bibinfo  {journal} {Physical Review B}\ }\textbf {\bibinfo
  {volume} {99}},\ \bibinfo {pages} {205437} (\bibinfo {year}
  {2019}{\natexlab{b}})}\BibitemShut {NoStop}%
\bibitem [{\citenamefont {Ghosh}\ \emph {et~al.}(2020)\citenamefont {Ghosh},
  \citenamefont {Chanda}, \citenamefont {Sen} \emph
  {et~al.}}]{ghosh2020enhancement}%
  \BibitemOpen
  \bibfield  {author} {\bibinfo {author} {\bibfnamefont {S.}~\bibnamefont
  {Ghosh}}, \bibinfo {author} {\bibfnamefont {T.}~\bibnamefont {Chanda}},
  \bibinfo {author} {\bibfnamefont {A.}~\bibnamefont {Sen}}, \emph {et~al.},\
  }\bibfield  {title} {\bibinfo {title} {Enhancement in the performance of a
  quantum battery by ordered and disordered interactions},\ }\href
  {https://doi.org/https://doi.org/10.1103/PhysRevA.101.032115} {\bibfield
  {journal} {\bibinfo  {journal} {Physical Review A}\ }\textbf {\bibinfo
  {volume} {101}},\ \bibinfo {pages} {032115} (\bibinfo {year}
  {2020})}\BibitemShut {NoStop}%
\bibitem [{\citenamefont {Gao}\ \emph {et~al.}(2022)\citenamefont {Gao},
  \citenamefont {Cheng}, \citenamefont {He}, \citenamefont {Mondaini},
  \citenamefont {Guan},\ and\ \citenamefont {Lin}}]{Gao2022}%
  \BibitemOpen
  \bibfield  {author} {\bibinfo {author} {\bibfnamefont {L.}~\bibnamefont
  {Gao}}, \bibinfo {author} {\bibfnamefont {C.}~\bibnamefont {Cheng}}, \bibinfo
  {author} {\bibfnamefont {W.-B.}\ \bibnamefont {He}}, \bibinfo {author}
  {\bibfnamefont {R.}~\bibnamefont {Mondaini}}, \bibinfo {author}
  {\bibfnamefont {X.-W.}\ \bibnamefont {Guan}},\ and\ \bibinfo {author}
  {\bibfnamefont {H.-Q.}\ \bibnamefont {Lin}},\ }\bibfield  {title} {\bibinfo
  {title} {Scaling of energy and power in a large quantum battery-charger
  model},\ }\href {https://doi.org/10.1103/PhysRevResearch.4.043150} {\bibfield
   {journal} {\bibinfo  {journal} {Physical Review Research}\ }\textbf
  {\bibinfo {volume} {4}},\ \bibinfo {pages} {043150} (\bibinfo {year}
  {2022})}\BibitemShut {NoStop}%
\bibitem [{\citenamefont {Gyhm}\ \emph {et~al.}(2022)\citenamefont {Gyhm},
  \citenamefont {{\v{S}}afr{\'a}nek},\ and\ \citenamefont
  {Rosa}}]{gyhm2022quantum}%
  \BibitemOpen
  \bibfield  {author} {\bibinfo {author} {\bibfnamefont {J.-Y.}\ \bibnamefont
  {Gyhm}}, \bibinfo {author} {\bibfnamefont {D.}~\bibnamefont
  {{\v{S}}afr{\'a}nek}},\ and\ \bibinfo {author} {\bibfnamefont
  {D.}~\bibnamefont {Rosa}},\ }\bibfield  {title} {\bibinfo {title} {Quantum
  charging advantage cannot be extensive without global operations},\ }\href
  {https://doi.org/https://doi.org/10.1103/PhysRevLett.128.140501} {\bibfield
  {journal} {\bibinfo  {journal} {Physical Review Letters}\ }\textbf {\bibinfo
  {volume} {128}},\ \bibinfo {pages} {140501} (\bibinfo {year}
  {2022})}\BibitemShut {NoStop}%
\bibitem [{\citenamefont {Salvia}\ \emph {et~al.}(2023)\citenamefont {Salvia},
  \citenamefont {Perarnau-Llobet}, \citenamefont {Haack}, \citenamefont
  {Brunner},\ and\ \citenamefont {Nimmrichter}}]{Salvia2023}%
  \BibitemOpen
  \bibfield  {author} {\bibinfo {author} {\bibfnamefont {R.}~\bibnamefont
  {Salvia}}, \bibinfo {author} {\bibfnamefont {M.}~\bibnamefont
  {Perarnau-Llobet}}, \bibinfo {author} {\bibfnamefont {G.}~\bibnamefont
  {Haack}}, \bibinfo {author} {\bibfnamefont {N.}~\bibnamefont {Brunner}},\
  and\ \bibinfo {author} {\bibfnamefont {S.}~\bibnamefont {Nimmrichter}},\
  }\bibfield  {title} {\bibinfo {title} {Quantum advantage in charging cavity
  and spin batteries by repeated interactions},\ }\href
  {https://doi.org/10.1103/PhysRevResearch.5.013155} {\bibfield  {journal}
  {\bibinfo  {journal} {Physical Review Research}\ }\textbf {\bibinfo {volume}
  {5}},\ \bibinfo {pages} {013155} (\bibinfo {year} {2023})}\BibitemShut
  {NoStop}%
\bibitem [{\citenamefont {Rodr\'{\i}guez}\ \emph {et~al.}(2023)\citenamefont
  {Rodr\'{\i}guez}, \citenamefont {Ahmadi}, \citenamefont {Mazurek},
  \citenamefont {Barzanjeh}, \citenamefont {Alicki},\ and\ \citenamefont
  {Horodecki}}]{Rodriguez2023}%
  \BibitemOpen
  \bibfield  {author} {\bibinfo {author} {\bibfnamefont {R.~R.}\ \bibnamefont
  {Rodr\'{\i}guez}}, \bibinfo {author} {\bibfnamefont {B.}~\bibnamefont
  {Ahmadi}}, \bibinfo {author} {\bibfnamefont {P.}~\bibnamefont {Mazurek}},
  \bibinfo {author} {\bibfnamefont {S.}~\bibnamefont {Barzanjeh}}, \bibinfo
  {author} {\bibfnamefont {R.}~\bibnamefont {Alicki}},\ and\ \bibinfo {author}
  {\bibfnamefont {P.}~\bibnamefont {Horodecki}},\ }\bibfield  {title} {\bibinfo
  {title} {Catalysis in charging quantum batteries},\ }\href
  {https://doi.org/10.1103/PhysRevA.107.042419} {\bibfield  {journal} {\bibinfo
   {journal} {Physical Review A}\ }\textbf {\bibinfo {volume} {107}},\ \bibinfo
  {pages} {042419} (\bibinfo {year} {2023})}\BibitemShut {NoStop}%
\bibitem [{\citenamefont {Ferraro}\ \emph {et~al.}(2018)\citenamefont
  {Ferraro}, \citenamefont {Campisi}, \citenamefont {Andolina}, \citenamefont
  {Pellegrini},\ and\ \citenamefont {Polini}}]{ferraro2018high}%
  \BibitemOpen
  \bibfield  {author} {\bibinfo {author} {\bibfnamefont {D.}~\bibnamefont
  {Ferraro}}, \bibinfo {author} {\bibfnamefont {M.}~\bibnamefont {Campisi}},
  \bibinfo {author} {\bibfnamefont {G.~M.}\ \bibnamefont {Andolina}}, \bibinfo
  {author} {\bibfnamefont {V.}~\bibnamefont {Pellegrini}},\ and\ \bibinfo
  {author} {\bibfnamefont {M.}~\bibnamefont {Polini}},\ }\bibfield  {title}
  {\bibinfo {title} {High-power collective charging of a solid-state quantum
  battery},\ }\href
  {https://doi.org/https://doi.org/10.1103/PhysRevLett.120.117702} {\bibfield
  {journal} {\bibinfo  {journal} {Physical Review Letters}\ }\textbf {\bibinfo
  {volume} {120}},\ \bibinfo {pages} {117702} (\bibinfo {year}
  {2018})}\BibitemShut {NoStop}%
\bibitem [{\citenamefont {Le}\ \emph {et~al.}(2018)\citenamefont {Le},
  \citenamefont {Levinsen}, \citenamefont {Modi}, \citenamefont {Parish},\ and\
  \citenamefont {Pollock}}]{le2018spin}%
  \BibitemOpen
  \bibfield  {author} {\bibinfo {author} {\bibfnamefont {T.~P.}\ \bibnamefont
  {Le}}, \bibinfo {author} {\bibfnamefont {J.}~\bibnamefont {Levinsen}},
  \bibinfo {author} {\bibfnamefont {K.}~\bibnamefont {Modi}}, \bibinfo {author}
  {\bibfnamefont {M.~M.}\ \bibnamefont {Parish}},\ and\ \bibinfo {author}
  {\bibfnamefont {F.~A.}\ \bibnamefont {Pollock}},\ }\bibfield  {title}
  {\bibinfo {title} {Spin-chain model of a many-body quantum battery},\ }\href
  {https://doi.org/https://doi.org/10.1103/PhysRevA.97.022106} {\bibfield
  {journal} {\bibinfo  {journal} {Physical Review A}\ }\textbf {\bibinfo
  {volume} {97}},\ \bibinfo {pages} {022106} (\bibinfo {year}
  {2018})}\BibitemShut {NoStop}%
\bibitem [{\citenamefont {Allahverdyan}\ \emph {et~al.}(2004)\citenamefont
  {Allahverdyan}, \citenamefont {Balian},\ and\ \citenamefont
  {Nieuwenhuizen}}]{allahverdyan2004maximal}%
  \BibitemOpen
  \bibfield  {author} {\bibinfo {author} {\bibfnamefont {A.~E.}\ \bibnamefont
  {Allahverdyan}}, \bibinfo {author} {\bibfnamefont {R.}~\bibnamefont
  {Balian}},\ and\ \bibinfo {author} {\bibfnamefont {T.~M.}\ \bibnamefont
  {Nieuwenhuizen}},\ }\bibfield  {title} {\bibinfo {title} {Maximal work
  extraction from finite quantum systems},\ }\href
  {https://doi.org/10.1209/epl/i2004-10101-2} {\bibfield  {journal} {\bibinfo
  {journal} {Europhysics Letters}\ }\textbf {\bibinfo {volume} {67}},\ \bibinfo
  {pages} {565} (\bibinfo {year} {2004})}\BibitemShut {NoStop}%
\bibitem [{\citenamefont {Ro\ss{}nagel}\ \emph {et~al.}(2014)\citenamefont
  {Ro\ss{}nagel}, \citenamefont {Abah}, \citenamefont {Schmidt-Kaler},
  \citenamefont {Singer},\ and\ \citenamefont {Lutz}}]{Rossnagel2014}%
  \BibitemOpen
  \bibfield  {author} {\bibinfo {author} {\bibfnamefont {J.}~\bibnamefont
  {Ro\ss{}nagel}}, \bibinfo {author} {\bibfnamefont {O.}~\bibnamefont {Abah}},
  \bibinfo {author} {\bibfnamefont {F.}~\bibnamefont {Schmidt-Kaler}}, \bibinfo
  {author} {\bibfnamefont {K.}~\bibnamefont {Singer}},\ and\ \bibinfo {author}
  {\bibfnamefont {E.}~\bibnamefont {Lutz}},\ }\bibfield  {title} {\bibinfo
  {title} {Nanoscale heat engine beyond the carnot limit},\ }\href
  {https://doi.org/10.1103/PhysRevLett.112.030602} {\bibfield  {journal}
  {\bibinfo  {journal} {Physical Review Letters}\ }\textbf {\bibinfo {volume}
  {112}},\ \bibinfo {pages} {030602} (\bibinfo {year} {2014})}\BibitemShut
  {NoStop}%
\bibitem [{\citenamefont {Klaers}\ \emph {et~al.}(2017)\citenamefont {Klaers},
  \citenamefont {Faelt}, \citenamefont {Imamoglu},\ and\ \citenamefont
  {Togan}}]{Klaers2017}%
  \BibitemOpen
  \bibfield  {author} {\bibinfo {author} {\bibfnamefont {J.}~\bibnamefont
  {Klaers}}, \bibinfo {author} {\bibfnamefont {S.}~\bibnamefont {Faelt}},
  \bibinfo {author} {\bibfnamefont {A.}~\bibnamefont {Imamoglu}},\ and\
  \bibinfo {author} {\bibfnamefont {E.}~\bibnamefont {Togan}},\ }\bibfield
  {title} {\bibinfo {title} {Squeezed thermal reservoirs as a resource for a
  nanomechanical engine beyond the carnot limit},\ }\href
  {https://doi.org/10.1103/PhysRevX.7.031044} {\bibfield  {journal} {\bibinfo
  {journal} {Physical Review X}\ }\textbf {\bibinfo {volume} {7}},\ \bibinfo
  {pages} {031044} (\bibinfo {year} {2017})}\BibitemShut {NoStop}%
\bibitem [{\citenamefont {Niedenzu}\ \emph {et~al.}(2018)\citenamefont
  {Niedenzu}, \citenamefont {Mukherjee}, \citenamefont {Ghosh}, \citenamefont
  {Kofman},\ and\ \citenamefont {Kurizki}}]{Niedenzu2018}%
  \BibitemOpen
  \bibfield  {author} {\bibinfo {author} {\bibfnamefont {W.}~\bibnamefont
  {Niedenzu}}, \bibinfo {author} {\bibfnamefont {V.}~\bibnamefont {Mukherjee}},
  \bibinfo {author} {\bibfnamefont {A.}~\bibnamefont {Ghosh}}, \bibinfo
  {author} {\bibfnamefont {A.~G.}\ \bibnamefont {Kofman}},\ and\ \bibinfo
  {author} {\bibfnamefont {G.}~\bibnamefont {Kurizki}},\ }\bibfield  {title}
  {\bibinfo {title} {Quantum engine efficiency bound beyond the second law of
  thermodynamics},\ }\href {https://doi.org/10.1038/s41467-017-01991-6}
  {\bibfield  {journal} {\bibinfo  {journal} {Nature Communications}\ }\textbf
  {\bibinfo {volume} {9}},\ \bibinfo {pages} {165} (\bibinfo {year}
  {2018})}\BibitemShut {NoStop}%
\bibitem [{\citenamefont {Biswas}\ \emph {et~al.}(2022)\citenamefont {Biswas},
  \citenamefont {{\L{}}obejko}, \citenamefont {Mazurek}, \citenamefont
  {Ja{\l{}}owiecki},\ and\ \citenamefont {Horodecki}}]{Biswas2022extractionof}%
  \BibitemOpen
  \bibfield  {author} {\bibinfo {author} {\bibfnamefont {T.}~\bibnamefont
  {Biswas}}, \bibinfo {author} {\bibfnamefont {M.}~\bibnamefont
  {{\L{}}obejko}}, \bibinfo {author} {\bibfnamefont {P.}~\bibnamefont
  {Mazurek}}, \bibinfo {author} {\bibfnamefont {K.}~\bibnamefont
  {Ja{\l{}}owiecki}},\ and\ \bibinfo {author} {\bibfnamefont {M.}~\bibnamefont
  {Horodecki}},\ }\bibfield  {title} {\bibinfo {title} {Extraction of
  ergotropy: free energy bound and application to open cycle engines},\ }\href
  {https://doi.org/10.22331/q-2022-10-17-841} {\bibfield  {journal} {\bibinfo
  {journal} {{Quantum}}\ }\textbf {\bibinfo {volume} {6}},\ \bibinfo {pages}
  {841} (\bibinfo {year} {2022})}\BibitemShut {NoStop}%
\bibitem [{\citenamefont {Goold}\ \emph {et~al.}(2016)\citenamefont {Goold},
  \citenamefont {Huber}, \citenamefont {Riera}, \citenamefont {del Rio},\ and\
  \citenamefont {Skrzypczyk}}]{Goold_2016}%
  \BibitemOpen
  \bibfield  {author} {\bibinfo {author} {\bibfnamefont {J.}~\bibnamefont
  {Goold}}, \bibinfo {author} {\bibfnamefont {M.}~\bibnamefont {Huber}},
  \bibinfo {author} {\bibfnamefont {A.}~\bibnamefont {Riera}}, \bibinfo
  {author} {\bibfnamefont {L.}~\bibnamefont {del Rio}},\ and\ \bibinfo {author}
  {\bibfnamefont {P.}~\bibnamefont {Skrzypczyk}},\ }\bibfield  {title}
  {\bibinfo {title} {The role of quantum information in thermodynamics—a
  topical review},\ }\href {https://doi.org/10.1088/1751-8113/49/14/143001}
  {\bibfield  {journal} {\bibinfo  {journal} {Journal of Physics A:
  Mathematical and Theoretical}\ }\textbf {\bibinfo {volume} {49}},\ \bibinfo
  {pages} {143001} (\bibinfo {year} {2016})}\BibitemShut {NoStop}%
\bibitem [{\citenamefont {Bera}\ \emph {et~al.}(2017)\citenamefont {Bera},
  \citenamefont {Riera}, \citenamefont {Lewenstein},\ and\ \citenamefont
  {Winter}}]{Bera2017}%
  \BibitemOpen
  \bibfield  {author} {\bibinfo {author} {\bibfnamefont {M.~N.}\ \bibnamefont
  {Bera}}, \bibinfo {author} {\bibfnamefont {A.}~\bibnamefont {Riera}},
  \bibinfo {author} {\bibfnamefont {M.}~\bibnamefont {Lewenstein}},\ and\
  \bibinfo {author} {\bibfnamefont {A.}~\bibnamefont {Winter}},\ }\bibfield
  {title} {\bibinfo {title} {Generalized laws of thermodynamics in the presence
  of correlations},\ }\href {https://doi.org/10.1038/s41467-017-02370-x}
  {\bibfield  {journal} {\bibinfo  {journal} {Nature Communications}\ }\textbf
  {\bibinfo {volume} {8}},\ \bibinfo {pages} {2180} (\bibinfo {year}
  {2017})}\BibitemShut {NoStop}%
\bibitem [{\citenamefont {Manzano}\ \emph {et~al.}(2018)\citenamefont
  {Manzano}, \citenamefont {Plastina},\ and\ \citenamefont
  {Zambrini}}]{Manzano2018}%
  \BibitemOpen
  \bibfield  {author} {\bibinfo {author} {\bibfnamefont {G.}~\bibnamefont
  {Manzano}}, \bibinfo {author} {\bibfnamefont {F.}~\bibnamefont {Plastina}},\
  and\ \bibinfo {author} {\bibfnamefont {R.}~\bibnamefont {Zambrini}},\
  }\bibfield  {title} {\bibinfo {title} {Optimal work extraction and
  thermodynamics of quantum measurements and correlations},\ }\href
  {https://doi.org/10.1103/PhysRevLett.121.120602} {\bibfield  {journal}
  {\bibinfo  {journal} {Physical Review Letters}\ }\textbf {\bibinfo {volume}
  {121}},\ \bibinfo {pages} {120602} (\bibinfo {year} {2018})}\BibitemShut
  {NoStop}%
\bibitem [{\citenamefont {Vitagliano}\ \emph {et~al.}(2018)\citenamefont
  {Vitagliano}, \citenamefont {Kl{\"o}ckl}, \citenamefont {Huber},\ and\
  \citenamefont {Friis}}]{Vitagliano2018}%
  \BibitemOpen
  \bibfield  {author} {\bibinfo {author} {\bibfnamefont {G.}~\bibnamefont
  {Vitagliano}}, \bibinfo {author} {\bibfnamefont {C.}~\bibnamefont
  {Kl{\"o}ckl}}, \bibinfo {author} {\bibfnamefont {M.}~\bibnamefont {Huber}},\
  and\ \bibinfo {author} {\bibfnamefont {N.}~\bibnamefont {Friis}},\ }\bibinfo
  {title} {Trade-off between work and correlations in quantum thermodynamics},\
  in\ \href {https://doi.org/10.1007/978-3-319-99046-0_30} {\emph {\bibinfo
  {booktitle} {Thermodynamics in the Quantum Regime: Fundamental Aspects and
  New Directions}}},\ \bibinfo {editor} {edited by\ \bibinfo {editor}
  {\bibfnamefont {F.}~\bibnamefont {Binder}}, \bibinfo {editor} {\bibfnamefont
  {L.~A.}\ \bibnamefont {Correa}}, \bibinfo {editor} {\bibfnamefont
  {C.}~\bibnamefont {Gogolin}}, \bibinfo {editor} {\bibfnamefont
  {J.}~\bibnamefont {Anders}},\ and\ \bibinfo {editor} {\bibfnamefont
  {G.}~\bibnamefont {Adesso}}}\ (\bibinfo  {publisher} {Springer International
  Publishing},\ \bibinfo {address} {Cham},\ \bibinfo {year} {2018})\ pp.\
  \bibinfo {pages} {731--750}\BibitemShut {NoStop}%
\bibitem [{\citenamefont {Dou}\ \emph {et~al.}(2021)\citenamefont {Dou},
  \citenamefont {Wang},\ and\ \citenamefont {Sun}}]{Dou2021}%
  \BibitemOpen
  \bibfield  {author} {\bibinfo {author} {\bibfnamefont {F.-Q.}\ \bibnamefont
  {Dou}}, \bibinfo {author} {\bibfnamefont {Y.-J.}\ \bibnamefont {Wang}},\ and\
  \bibinfo {author} {\bibfnamefont {J.-A.}\ \bibnamefont {Sun}},\ }\bibfield
  {title} {\bibinfo {title} {Highly efficient charging and discharging of
  three-level quantum batteries through shortcuts to adiabaticity},\ }\href
  {https://doi.org/10.1007/s11467-021-1130-5} {\bibfield  {journal} {\bibinfo
  {journal} {Frontiers of Physics}\ }\textbf {\bibinfo {volume} {17}},\
  \bibinfo {pages} {31503} (\bibinfo {year} {2021})}\BibitemShut {NoStop}%
\bibitem [{\citenamefont {Gemme}\ \emph {et~al.}(2022)\citenamefont {Gemme},
  \citenamefont {Grossi}, \citenamefont {Ferraro}, \citenamefont {Vallecorsa},\
  and\ \citenamefont {Sassetti}}]{Gemme2022}%
  \BibitemOpen
  \bibfield  {author} {\bibinfo {author} {\bibfnamefont {G.}~\bibnamefont
  {Gemme}}, \bibinfo {author} {\bibfnamefont {M.}~\bibnamefont {Grossi}},
  \bibinfo {author} {\bibfnamefont {D.}~\bibnamefont {Ferraro}}, \bibinfo
  {author} {\bibfnamefont {S.}~\bibnamefont {Vallecorsa}},\ and\ \bibinfo
  {author} {\bibfnamefont {M.}~\bibnamefont {Sassetti}},\ }\bibfield  {title}
  {\bibinfo {title} {Ibm quantum platforms: A quantum battery perspective},\
  }\bibfield  {journal} {\bibinfo  {journal} {Batteries}\ }\textbf {\bibinfo
  {volume} {8}},\ \href {https://doi.org/10.3390/batteries8050043}
  {10.3390/batteries8050043} (\bibinfo {year} {2022})\BibitemShut {NoStop}%
\bibitem [{\citenamefont {Dou}\ \emph {et~al.}(2022{\natexlab{a}})\citenamefont
  {Dou}, \citenamefont {Zhou},\ and\ \citenamefont {Sun}}]{Dou2022}%
  \BibitemOpen
  \bibfield  {author} {\bibinfo {author} {\bibfnamefont {F.-Q.}\ \bibnamefont
  {Dou}}, \bibinfo {author} {\bibfnamefont {H.}~\bibnamefont {Zhou}},\ and\
  \bibinfo {author} {\bibfnamefont {J.-A.}\ \bibnamefont {Sun}},\ }\bibfield
  {title} {\bibinfo {title} {Cavity heisenberg-spin-chain quantum battery},\
  }\href {https://doi.org/10.1103/PhysRevA.106.032212} {\bibfield  {journal}
  {\bibinfo  {journal} {Phys. Rev. A}\ }\textbf {\bibinfo {volume} {106}},\
  \bibinfo {pages} {032212} (\bibinfo {year} {2022}{\natexlab{a}})}\BibitemShut
  {NoStop}%
\bibitem [{\citenamefont {Dou}\ \emph {et~al.}(2022{\natexlab{b}})\citenamefont
  {Dou}, \citenamefont {Lu}, \citenamefont {Wang},\ and\ \citenamefont
  {Sun}}]{Dou2022b}%
  \BibitemOpen
  \bibfield  {author} {\bibinfo {author} {\bibfnamefont {F.-Q.}\ \bibnamefont
  {Dou}}, \bibinfo {author} {\bibfnamefont {Y.-Q.}\ \bibnamefont {Lu}},
  \bibinfo {author} {\bibfnamefont {Y.-J.}\ \bibnamefont {Wang}},\ and\
  \bibinfo {author} {\bibfnamefont {J.-A.}\ \bibnamefont {Sun}},\ }\bibfield
  {title} {\bibinfo {title} {Extended dicke quantum battery with interatomic
  interactions and driving field},\ }\href
  {https://doi.org/10.1103/PhysRevB.105.115405} {\bibfield  {journal} {\bibinfo
   {journal} {Phys. Rev. B}\ }\textbf {\bibinfo {volume} {105}},\ \bibinfo
  {pages} {115405} (\bibinfo {year} {2022}{\natexlab{b}})}\BibitemShut
  {NoStop}%
\bibitem [{\citenamefont {Dou}\ and\ \citenamefont {Yang}(2023)}]{Dou2023}%
  \BibitemOpen
  \bibfield  {author} {\bibinfo {author} {\bibfnamefont {F.-Q.}\ \bibnamefont
  {Dou}}\ and\ \bibinfo {author} {\bibfnamefont {F.-M.}\ \bibnamefont {Yang}},\
  }\bibfield  {title} {\bibinfo {title} {Superconducting transmon
  qubit-resonator quantum battery},\ }\href
  {https://doi.org/10.1103/PhysRevA.107.023725} {\bibfield  {journal} {\bibinfo
   {journal} {Phys. Rev. A}\ }\textbf {\bibinfo {volume} {107}},\ \bibinfo
  {pages} {023725} (\bibinfo {year} {2023})}\BibitemShut {NoStop}%
\bibitem [{\citenamefont {Nielsen}\ and\ \citenamefont
  {Chuang}(2002)}]{nielsen2002quantum}%
  \BibitemOpen
  \bibfield  {author} {\bibinfo {author} {\bibfnamefont {M.~A.}\ \bibnamefont
  {Nielsen}}\ and\ \bibinfo {author} {\bibfnamefont {I.}~\bibnamefont
  {Chuang}},\ }\href@noop {} {\bibinfo {title} {Quantum computation and quantum
  information}} (\bibinfo {year} {2002})\BibitemShut {NoStop}%
\bibitem [{\citenamefont {Lloyd}\ \emph {et~al.}(2014)\citenamefont {Lloyd},
  \citenamefont {Mohseni},\ and\ \citenamefont
  {Rebentrost}}]{lloyd2014quantum}%
  \BibitemOpen
  \bibfield  {author} {\bibinfo {author} {\bibfnamefont {S.}~\bibnamefont
  {Lloyd}}, \bibinfo {author} {\bibfnamefont {M.}~\bibnamefont {Mohseni}},\
  and\ \bibinfo {author} {\bibfnamefont {P.}~\bibnamefont {Rebentrost}},\
  }\bibfield  {title} {\bibinfo {title} {Quantum principal component
  analysis},\ }\href {https://doi.org/https://doi.org/10.1038/nphys3029}
  {\bibfield  {journal} {\bibinfo  {journal} {Nature Physics}\ }\textbf
  {\bibinfo {volume} {10}},\ \bibinfo {pages} {631} (\bibinfo {year}
  {2014})}\BibitemShut {NoStop}%
\bibitem [{\citenamefont {Arute}\ \emph {et~al.}(2019)\citenamefont {Arute},
  \citenamefont {Arya}, \citenamefont {Babbush}, \citenamefont {Bacon},
  \citenamefont {Bardin}, \citenamefont {Barends}, \citenamefont {Biswas},
  \citenamefont {Boixo}, \citenamefont {Brandao}, \citenamefont {Buell} \emph
  {et~al.}}]{arute2019quantum}%
  \BibitemOpen
  \bibfield  {author} {\bibinfo {author} {\bibfnamefont {F.}~\bibnamefont
  {Arute}}, \bibinfo {author} {\bibfnamefont {K.}~\bibnamefont {Arya}},
  \bibinfo {author} {\bibfnamefont {R.}~\bibnamefont {Babbush}}, \bibinfo
  {author} {\bibfnamefont {D.}~\bibnamefont {Bacon}}, \bibinfo {author}
  {\bibfnamefont {J.~C.}\ \bibnamefont {Bardin}}, \bibinfo {author}
  {\bibfnamefont {R.}~\bibnamefont {Barends}}, \bibinfo {author} {\bibfnamefont
  {R.}~\bibnamefont {Biswas}}, \bibinfo {author} {\bibfnamefont
  {S.}~\bibnamefont {Boixo}}, \bibinfo {author} {\bibfnamefont {F.~G.}\
  \bibnamefont {Brandao}}, \bibinfo {author} {\bibfnamefont {D.~A.}\
  \bibnamefont {Buell}}, \emph {et~al.},\ }\bibfield  {title} {\bibinfo {title}
  {Quantum supremacy using a programmable superconducting processor},\ }\href
  {https://doi.org/https://doi.org/10.1038/s41586-019-1666-5} {\bibfield
  {journal} {\bibinfo  {journal} {Nature}\ }\textbf {\bibinfo {volume} {574}},\
  \bibinfo {pages} {505} (\bibinfo {year} {2019})}\BibitemShut {NoStop}%
\bibitem [{\citenamefont {Sakurai}\ \emph {et~al.}(2022)\citenamefont
  {Sakurai}, \citenamefont {Mizukami},\ and\ \citenamefont
  {Shinaoka}}]{sakurai2022hybrid}%
  \BibitemOpen
  \bibfield  {author} {\bibinfo {author} {\bibfnamefont {R.}~\bibnamefont
  {Sakurai}}, \bibinfo {author} {\bibfnamefont {W.}~\bibnamefont {Mizukami}},\
  and\ \bibinfo {author} {\bibfnamefont {H.}~\bibnamefont {Shinaoka}},\
  }\bibfield  {title} {\bibinfo {title} {Hybrid quantum-classical algorithm for
  computing imaginary-time correlation functions},\ }\href
  {https://doi.org/https://doi.org/10.1103/PhysRevResearch.4.023219} {\bibfield
   {journal} {\bibinfo  {journal} {Physical Review Research}\ }\textbf
  {\bibinfo {volume} {4}},\ \bibinfo {pages} {023219} (\bibinfo {year}
  {2022})}\BibitemShut {NoStop}%
\bibitem [{\citenamefont {Cain}\ \emph {et~al.}(2023)\citenamefont {Cain},
  \citenamefont {Chattopadhyay}, \citenamefont {Liu}, \citenamefont {Samajdar},
  \citenamefont {Pichler},\ and\ \citenamefont {Lukin}}]{cain2023quantum}%
  \BibitemOpen
  \bibfield  {author} {\bibinfo {author} {\bibfnamefont {M.}~\bibnamefont
  {Cain}}, \bibinfo {author} {\bibfnamefont {S.}~\bibnamefont {Chattopadhyay}},
  \bibinfo {author} {\bibfnamefont {J.-G.}\ \bibnamefont {Liu}}, \bibinfo
  {author} {\bibfnamefont {R.}~\bibnamefont {Samajdar}}, \bibinfo {author}
  {\bibfnamefont {H.}~\bibnamefont {Pichler}},\ and\ \bibinfo {author}
  {\bibfnamefont {M.~D.}\ \bibnamefont {Lukin}},\ }\bibfield  {title} {\bibinfo
  {title} {Quantum speedup for combinatorial optimization with flat energy
  landscapes},\ }\bibfield  {journal} {\bibinfo  {journal} {arXiv preprint
  arXiv:2306.13123}\ }\href
  {https://doi.org/https://doi.org/10.48550/arXiv.2306.13123}
  {https://doi.org/10.48550/arXiv.2306.13123} (\bibinfo {year}
  {2023})\BibitemShut {NoStop}%
\bibitem [{\citenamefont {Yonezu}\ \emph {et~al.}(2023)\citenamefont {Yonezu},
  \citenamefont {Enomoto}, \citenamefont {Yoshida},\ and\ \citenamefont
  {Takeda}}]{yonezu2023time}%
  \BibitemOpen
  \bibfield  {author} {\bibinfo {author} {\bibfnamefont {K.}~\bibnamefont
  {Yonezu}}, \bibinfo {author} {\bibfnamefont {Y.}~\bibnamefont {Enomoto}},
  \bibinfo {author} {\bibfnamefont {T.}~\bibnamefont {Yoshida}},\ and\ \bibinfo
  {author} {\bibfnamefont {S.}~\bibnamefont {Takeda}},\ }\bibfield  {title}
  {\bibinfo {title} {Time-domain universal linear-optical operations for
  universal quantum information processing},\ }\href
  {https://doi.org/https://doi.org/10.1103/PhysRevLett.131.040601} {\bibfield
  {journal} {\bibinfo  {journal} {Physical Review Letters}\ }\textbf {\bibinfo
  {volume} {131}},\ \bibinfo {pages} {040601} (\bibinfo {year}
  {2023})}\BibitemShut {NoStop}%
\bibitem [{\citenamefont {Shtanko}\ \emph {et~al.}(2023)\citenamefont
  {Shtanko}, \citenamefont {Wang}, \citenamefont {Zhang}, \citenamefont
  {Harle}, \citenamefont {Seif}, \citenamefont {Movassagh},\ and\ \citenamefont
  {Minev}}]{shtanko2023uncovering}%
  \BibitemOpen
  \bibfield  {author} {\bibinfo {author} {\bibfnamefont {O.}~\bibnamefont
  {Shtanko}}, \bibinfo {author} {\bibfnamefont {D.~S.}\ \bibnamefont {Wang}},
  \bibinfo {author} {\bibfnamefont {H.}~\bibnamefont {Zhang}}, \bibinfo
  {author} {\bibfnamefont {N.}~\bibnamefont {Harle}}, \bibinfo {author}
  {\bibfnamefont {A.}~\bibnamefont {Seif}}, \bibinfo {author} {\bibfnamefont
  {R.}~\bibnamefont {Movassagh}},\ and\ \bibinfo {author} {\bibfnamefont
  {Z.}~\bibnamefont {Minev}},\ }\bibfield  {title} {\bibinfo {title}
  {Uncovering local integrability in quantum many-body dynamics},\ }\bibfield
  {journal} {\bibinfo  {journal} {arXiv preprint arXiv:2307.07552}\ }\href
  {https://doi.org/https://doi.org/10.48550/arXiv.2307.07552}
  {https://doi.org/10.48550/arXiv.2307.07552} (\bibinfo {year}
  {2023})\BibitemShut {NoStop}%
\bibitem [{\citenamefont {O’Brien}\ \emph {et~al.}(2019)\citenamefont
  {O’Brien}, \citenamefont {Senjean}, \citenamefont {Sagastizabal},
  \citenamefont {Bonet-Monroig}, \citenamefont {Dutkiewicz}, \citenamefont
  {Buda}, \citenamefont {DiCarlo},\ and\ \citenamefont
  {Visscher}}]{o2019calculating}%
  \BibitemOpen
  \bibfield  {author} {\bibinfo {author} {\bibfnamefont {T.~E.}\ \bibnamefont
  {O’Brien}}, \bibinfo {author} {\bibfnamefont {B.}~\bibnamefont {Senjean}},
  \bibinfo {author} {\bibfnamefont {R.}~\bibnamefont {Sagastizabal}}, \bibinfo
  {author} {\bibfnamefont {X.}~\bibnamefont {Bonet-Monroig}}, \bibinfo {author}
  {\bibfnamefont {A.}~\bibnamefont {Dutkiewicz}}, \bibinfo {author}
  {\bibfnamefont {F.}~\bibnamefont {Buda}}, \bibinfo {author} {\bibfnamefont
  {L.}~\bibnamefont {DiCarlo}},\ and\ \bibinfo {author} {\bibfnamefont
  {L.}~\bibnamefont {Visscher}},\ }\bibfield  {title} {\bibinfo {title}
  {Calculating energy derivatives for quantum chemistry on a quantum
  computer},\ }\href
  {https://doi.org/https://doi.org/10.1038/s41534-019-0213-4} {\bibfield
  {journal} {\bibinfo  {journal} {npj Quantum Information}\ }\textbf {\bibinfo
  {volume} {5}},\ \bibinfo {pages} {113} (\bibinfo {year} {2019})}\BibitemShut
  {NoStop}%
\bibitem [{\citenamefont {Mizukami}\ \emph {et~al.}(2020)\citenamefont
  {Mizukami}, \citenamefont {Mitarai}, \citenamefont {Nakagawa}, \citenamefont
  {Yamamoto}, \citenamefont {Yan},\ and\ \citenamefont
  {Ohnishi}}]{mizukami2020orbital}%
  \BibitemOpen
  \bibfield  {author} {\bibinfo {author} {\bibfnamefont {W.}~\bibnamefont
  {Mizukami}}, \bibinfo {author} {\bibfnamefont {K.}~\bibnamefont {Mitarai}},
  \bibinfo {author} {\bibfnamefont {Y.~O.}\ \bibnamefont {Nakagawa}}, \bibinfo
  {author} {\bibfnamefont {T.}~\bibnamefont {Yamamoto}}, \bibinfo {author}
  {\bibfnamefont {T.}~\bibnamefont {Yan}},\ and\ \bibinfo {author}
  {\bibfnamefont {Y.-y.}\ \bibnamefont {Ohnishi}},\ }\bibfield  {title}
  {\bibinfo {title} {Orbital optimized unitary coupled cluster theory for
  quantum computer},\ }\href
  {https://doi.org/https://doi.org/10.1103/PhysRevResearch.2.033421} {\bibfield
   {journal} {\bibinfo  {journal} {Physical Review Research}\ }\textbf
  {\bibinfo {volume} {2}},\ \bibinfo {pages} {033421} (\bibinfo {year}
  {2020})}\BibitemShut {NoStop}%
\bibitem [{\citenamefont {Fujii}\ \emph {et~al.}(2022)\citenamefont {Fujii},
  \citenamefont {Mizuta}, \citenamefont {Ueda}, \citenamefont {Mitarai},
  \citenamefont {Mizukami},\ and\ \citenamefont {Nakagawa}}]{fujii2022deep}%
  \BibitemOpen
  \bibfield  {author} {\bibinfo {author} {\bibfnamefont {K.}~\bibnamefont
  {Fujii}}, \bibinfo {author} {\bibfnamefont {K.}~\bibnamefont {Mizuta}},
  \bibinfo {author} {\bibfnamefont {H.}~\bibnamefont {Ueda}}, \bibinfo {author}
  {\bibfnamefont {K.}~\bibnamefont {Mitarai}}, \bibinfo {author} {\bibfnamefont
  {W.}~\bibnamefont {Mizukami}},\ and\ \bibinfo {author} {\bibfnamefont
  {Y.~O.}\ \bibnamefont {Nakagawa}},\ }\bibfield  {title} {\bibinfo {title}
  {Deep variational quantum eigensolver: a divide-and-conquer method for
  solving a larger problem with smaller size quantum computers},\ }\href
  {https://doi.org/https://doi.org/10.1103/PRXQuantum.3.010346} {\bibfield
  {journal} {\bibinfo  {journal} {PRX Quantum}\ }\textbf {\bibinfo {volume}
  {3}},\ \bibinfo {pages} {010346} (\bibinfo {year} {2022})}\BibitemShut
  {NoStop}%
\bibitem [{\citenamefont {Simon}\ \emph {et~al.}(2023)\citenamefont {Simon},
  \citenamefont {Santagati}, \citenamefont {Degroote}, \citenamefont {Moll},
  \citenamefont {Streif},\ and\ \citenamefont {Wiebe}}]{simon2023improved}%
  \BibitemOpen
  \bibfield  {author} {\bibinfo {author} {\bibfnamefont {S.}~\bibnamefont
  {Simon}}, \bibinfo {author} {\bibfnamefont {R.}~\bibnamefont {Santagati}},
  \bibinfo {author} {\bibfnamefont {M.}~\bibnamefont {Degroote}}, \bibinfo
  {author} {\bibfnamefont {N.}~\bibnamefont {Moll}}, \bibinfo {author}
  {\bibfnamefont {M.}~\bibnamefont {Streif}},\ and\ \bibinfo {author}
  {\bibfnamefont {N.}~\bibnamefont {Wiebe}},\ }\bibfield  {title} {\bibinfo
  {title} {Improved precision scaling for simulating coupled quantum-classical
  dynamics},\ }\bibfield  {journal} {\bibinfo  {journal} {arXiv preprint
  arXiv:2307.13033}\ }\href
  {https://doi.org/https://doi.org/10.48550/arXiv.2307.13033}
  {https://doi.org/10.48550/arXiv.2307.13033} (\bibinfo {year}
  {2023})\BibitemShut {NoStop}%
\bibitem [{\citenamefont {Ko}\ \emph {et~al.}(2023)\citenamefont {Ko},
  \citenamefont {Li},\ and\ \citenamefont {Wang}}]{ko2023implementation}%
  \BibitemOpen
  \bibfield  {author} {\bibinfo {author} {\bibfnamefont {T.}~\bibnamefont
  {Ko}}, \bibinfo {author} {\bibfnamefont {X.}~\bibnamefont {Li}},\ and\
  \bibinfo {author} {\bibfnamefont {C.}~\bibnamefont {Wang}},\ }\bibfield
  {title} {\bibinfo {title} {Implementation of the density-functional theory on
  quantum computers with linear scaling with respect to the number of atoms},\
  }\bibfield  {journal} {\bibinfo  {journal} {arXiv preprint arXiv:2307.07067}\
  }\href {https://doi.org/https://doi.org/10.48550/arXiv.2307.07067}
  {https://doi.org/10.48550/arXiv.2307.07067} (\bibinfo {year}
  {2023})\BibitemShut {NoStop}%
\bibitem [{\citenamefont {Senjean}\ \emph {et~al.}(2023)\citenamefont
  {Senjean}, \citenamefont {Yalouz},\ and\ \citenamefont
  {Sauban{\`e}re}}]{senjean2023toward}%
  \BibitemOpen
  \bibfield  {author} {\bibinfo {author} {\bibfnamefont {B.}~\bibnamefont
  {Senjean}}, \bibinfo {author} {\bibfnamefont {S.}~\bibnamefont {Yalouz}},\
  and\ \bibinfo {author} {\bibfnamefont {M.}~\bibnamefont {Sauban{\`e}re}},\
  }\bibfield  {title} {\bibinfo {title} {Toward density functional theory on
  quantum computers?},\ }\href {https://doi.org/10.21468/SciPostPhys.14.3.055}
  {\bibfield  {journal} {\bibinfo  {journal} {SciPost Physics}\ }\textbf
  {\bibinfo {volume} {14}},\ \bibinfo {pages} {055} (\bibinfo {year}
  {2023})}\BibitemShut {NoStop}%
\bibitem [{\citenamefont {Ma}\ \emph {et~al.}(2020)\citenamefont {Ma},
  \citenamefont {Govoni},\ and\ \citenamefont {Galli}}]{ma2020quantum}%
  \BibitemOpen
  \bibfield  {author} {\bibinfo {author} {\bibfnamefont {H.}~\bibnamefont
  {Ma}}, \bibinfo {author} {\bibfnamefont {M.}~\bibnamefont {Govoni}},\ and\
  \bibinfo {author} {\bibfnamefont {G.}~\bibnamefont {Galli}},\ }\bibfield
  {title} {\bibinfo {title} {Quantum simulations of materials on near-term
  quantum computers},\ }\href
  {https://doi.org/https://doi.org/10.1038/s41524-020-00353-z} {\bibfield
  {journal} {\bibinfo  {journal} {npj Computational Materials}\ }\textbf
  {\bibinfo {volume} {6}},\ \bibinfo {pages} {85} (\bibinfo {year}
  {2020})}\BibitemShut {NoStop}%
\bibitem [{\citenamefont {Kanno}\ \emph {et~al.}(2022)\citenamefont {Kanno},
  \citenamefont {Endo}, \citenamefont {Utsumi},\ and\ \citenamefont
  {Tada}}]{kanno2022resource}%
  \BibitemOpen
  \bibfield  {author} {\bibinfo {author} {\bibfnamefont {S.}~\bibnamefont
  {Kanno}}, \bibinfo {author} {\bibfnamefont {S.}~\bibnamefont {Endo}},
  \bibinfo {author} {\bibfnamefont {T.}~\bibnamefont {Utsumi}},\ and\ \bibinfo
  {author} {\bibfnamefont {T.}~\bibnamefont {Tada}},\ }\bibfield  {title}
  {\bibinfo {title} {Resource estimations for the hamiltonian simulation in
  correlated electron materials},\ }\href
  {https://doi.org/https://doi.org/10.1103/PhysRevA.106.012612} {\bibfield
  {journal} {\bibinfo  {journal} {Physical Review A}\ }\textbf {\bibinfo
  {volume} {106}},\ \bibinfo {pages} {012612} (\bibinfo {year}
  {2022})}\BibitemShut {NoStop}%
\bibitem [{\citenamefont {Zini}\ \emph {et~al.}(2023)\citenamefont {Zini},
  \citenamefont {Delgado}, \citenamefont {dos Reis}, \citenamefont {Casares},
  \citenamefont {Mueller}, \citenamefont {Voigt},\ and\ \citenamefont
  {Arrazola}}]{zini2023quantum}%
  \BibitemOpen
  \bibfield  {author} {\bibinfo {author} {\bibfnamefont {M.~S.}\ \bibnamefont
  {Zini}}, \bibinfo {author} {\bibfnamefont {A.}~\bibnamefont {Delgado}},
  \bibinfo {author} {\bibfnamefont {R.}~\bibnamefont {dos Reis}}, \bibinfo
  {author} {\bibfnamefont {P.~A.~M.}\ \bibnamefont {Casares}}, \bibinfo
  {author} {\bibfnamefont {J.~E.}\ \bibnamefont {Mueller}}, \bibinfo {author}
  {\bibfnamefont {A.-C.}\ \bibnamefont {Voigt}},\ and\ \bibinfo {author}
  {\bibfnamefont {J.~M.}\ \bibnamefont {Arrazola}},\ }\bibfield  {title}
  {\bibinfo {title} {Quantum simulation of battery materials using ionic
  pseudopotentials},\ }\href
  {https://doi.org/https://doi.org/10.22331/q-2023-07-10-1049} {\bibfield
  {journal} {\bibinfo  {journal} {Quantum}\ }\textbf {\bibinfo {volume} {7}},\
  \bibinfo {pages} {1049} (\bibinfo {year} {2023})}\BibitemShut {NoStop}%
\bibitem [{\citenamefont {Rubin}\ \emph {et~al.}(2023)\citenamefont {Rubin},
  \citenamefont {Berry}, \citenamefont {Malone}, \citenamefont {White},
  \citenamefont {Khattar}, \citenamefont {DePrince~III}, \citenamefont
  {Sicolo}, \citenamefont {K{\"u}ehn}, \citenamefont {Kaicher}, \citenamefont
  {Lee} \emph {et~al.}}]{rubin2023fault}%
  \BibitemOpen
  \bibfield  {author} {\bibinfo {author} {\bibfnamefont {N.~C.}\ \bibnamefont
  {Rubin}}, \bibinfo {author} {\bibfnamefont {D.~W.}\ \bibnamefont {Berry}},
  \bibinfo {author} {\bibfnamefont {F.~D.}\ \bibnamefont {Malone}}, \bibinfo
  {author} {\bibfnamefont {A.~F.}\ \bibnamefont {White}}, \bibinfo {author}
  {\bibfnamefont {T.}~\bibnamefont {Khattar}}, \bibinfo {author} {\bibfnamefont
  {A.~E.}\ \bibnamefont {DePrince~III}}, \bibinfo {author} {\bibfnamefont
  {S.}~\bibnamefont {Sicolo}}, \bibinfo {author} {\bibfnamefont
  {M.}~\bibnamefont {K{\"u}ehn}}, \bibinfo {author} {\bibfnamefont
  {M.}~\bibnamefont {Kaicher}}, \bibinfo {author} {\bibfnamefont
  {J.}~\bibnamefont {Lee}}, \emph {et~al.},\ }\bibfield  {title} {\bibinfo
  {title} {Fault-tolerant quantum simulation of materials using bloch
  orbitals},\ }\href
  {https://doi.org/https://doi.org/10.1103/PRXQuantum.4.040303} {\bibfield
  {journal} {\bibinfo  {journal} {PRX Quantum}\ }\textbf {\bibinfo {volume}
  {4}},\ \bibinfo {pages} {040303} (\bibinfo {year} {2023})}\BibitemShut
  {NoStop}%
\bibitem [{\citenamefont {Westermayr}\ \emph {et~al.}(2023)\citenamefont
  {Westermayr}, \citenamefont {Gilkes}, \citenamefont {Barrett},\ and\
  \citenamefont {Maurer}}]{westermayr2023high}%
  \BibitemOpen
  \bibfield  {author} {\bibinfo {author} {\bibfnamefont {J.}~\bibnamefont
  {Westermayr}}, \bibinfo {author} {\bibfnamefont {J.}~\bibnamefont {Gilkes}},
  \bibinfo {author} {\bibfnamefont {R.}~\bibnamefont {Barrett}},\ and\ \bibinfo
  {author} {\bibfnamefont {R.~J.}\ \bibnamefont {Maurer}},\ }\bibfield  {title}
  {\bibinfo {title} {High-throughput property-driven generative design of
  functional organic molecules},\ }\href
  {https://doi.org/https://doi.org/10.1038/s43588-022-00391-1} {\bibfield
  {journal} {\bibinfo  {journal} {Nature Computational Science}\ }\textbf
  {\bibinfo {volume} {3}},\ \bibinfo {pages} {139} (\bibinfo {year}
  {2023})}\BibitemShut {NoStop}%
\bibitem [{\citenamefont {Kim}\ \emph {et~al.}(2023)\citenamefont {Kim},
  \citenamefont {Eddins}, \citenamefont {Anand}, \citenamefont {Wei},
  \citenamefont {Berg}, \citenamefont {Rosenblatt}, \citenamefont {Nayfeh},
  \citenamefont {Wu}, \citenamefont {Zaletel}, \citenamefont {Temme},\ and\
  \citenamefont {Kandala}}]{Kim2023}%
  \BibitemOpen
  \bibfield  {author} {\bibinfo {author} {\bibfnamefont {Y.}~\bibnamefont
  {Kim}}, \bibinfo {author} {\bibfnamefont {A.}~\bibnamefont {Eddins}},
  \bibinfo {author} {\bibfnamefont {S.}~\bibnamefont {Anand}}, \bibinfo
  {author} {\bibfnamefont {K.}~\bibnamefont {Wei}}, \bibinfo {author}
  {\bibfnamefont {E.}~\bibnamefont {Berg}}, \bibinfo {author} {\bibfnamefont
  {S.}~\bibnamefont {Rosenblatt}}, \bibinfo {author} {\bibfnamefont
  {H.}~\bibnamefont {Nayfeh}}, \bibinfo {author} {\bibfnamefont
  {Y.}~\bibnamefont {Wu}}, \bibinfo {author} {\bibfnamefont {M.}~\bibnamefont
  {Zaletel}}, \bibinfo {author} {\bibfnamefont {K.}~\bibnamefont {Temme}},\
  and\ \bibinfo {author} {\bibfnamefont {A.}~\bibnamefont {Kandala}},\
  }\bibfield  {title} {\bibinfo {title} {Evidence for the utility of quantum
  computing before fault tolerance},\ }\href
  {https://doi.org/10.1038/s41586-023-06096-3} {\bibfield  {journal} {\bibinfo
  {journal} {Nature}\ }\textbf {\bibinfo {volume} {618}},\ \bibinfo {pages}
  {500} (\bibinfo {year} {2023})}\BibitemShut {NoStop}%
\bibitem [{\citenamefont {Layden}\ \emph {et~al.}(2023)\citenamefont {Layden},
  \citenamefont {Mazzola}, \citenamefont {Mishmash}, \citenamefont {Motta},
  \citenamefont {Wocjan}, \citenamefont {Kim},\ and\ \citenamefont
  {Sheldon}}]{layden2023quantum}%
  \BibitemOpen
  \bibfield  {author} {\bibinfo {author} {\bibfnamefont {D.}~\bibnamefont
  {Layden}}, \bibinfo {author} {\bibfnamefont {G.}~\bibnamefont {Mazzola}},
  \bibinfo {author} {\bibfnamefont {R.~V.}\ \bibnamefont {Mishmash}}, \bibinfo
  {author} {\bibfnamefont {M.}~\bibnamefont {Motta}}, \bibinfo {author}
  {\bibfnamefont {P.}~\bibnamefont {Wocjan}}, \bibinfo {author} {\bibfnamefont
  {J.-S.}\ \bibnamefont {Kim}},\ and\ \bibinfo {author} {\bibfnamefont
  {S.}~\bibnamefont {Sheldon}},\ }\bibfield  {title} {\bibinfo {title}
  {Quantum-enhanced markov chain monte carlo},\ }\href
  {https://doi.org/https://doi.org/10.1038/s41586-023-06095-4} {\bibfield
  {journal} {\bibinfo  {journal} {Nature}\ }\textbf {\bibinfo {volume} {619}},\
  \bibinfo {pages} {282} (\bibinfo {year} {2023})}\BibitemShut {NoStop}%
\bibitem [{\citenamefont {Cao}\ \emph {et~al.}(2023)\citenamefont {Cao},
  \citenamefont {Wu}, \citenamefont {Chen}, \citenamefont {Gong}, \citenamefont
  {Wu}, \citenamefont {Ye}, \citenamefont {Zha}, \citenamefont {Qian},
  \citenamefont {Ying}, \citenamefont {Guo} \emph
  {et~al.}}]{cao2023generation}%
  \BibitemOpen
  \bibfield  {author} {\bibinfo {author} {\bibfnamefont {S.}~\bibnamefont
  {Cao}}, \bibinfo {author} {\bibfnamefont {B.}~\bibnamefont {Wu}}, \bibinfo
  {author} {\bibfnamefont {F.}~\bibnamefont {Chen}}, \bibinfo {author}
  {\bibfnamefont {M.}~\bibnamefont {Gong}}, \bibinfo {author} {\bibfnamefont
  {Y.}~\bibnamefont {Wu}}, \bibinfo {author} {\bibfnamefont {Y.}~\bibnamefont
  {Ye}}, \bibinfo {author} {\bibfnamefont {C.}~\bibnamefont {Zha}}, \bibinfo
  {author} {\bibfnamefont {H.}~\bibnamefont {Qian}}, \bibinfo {author}
  {\bibfnamefont {C.}~\bibnamefont {Ying}}, \bibinfo {author} {\bibfnamefont
  {S.}~\bibnamefont {Guo}}, \emph {et~al.},\ }\bibfield  {title} {\bibinfo
  {title} {Generation of genuine entanglement up to 51 superconducting
  qubits},\ }\href {https://doi.org/https://doi.org/10.1038/s41586-023-06195-1}
  {\bibfield  {journal} {\bibinfo  {journal} {Nature}\ ,\ \bibinfo {pages} {1}}
  (\bibinfo {year} {2023})}\BibitemShut {NoStop}%
\bibitem [{\citenamefont {Preskill}(2018)}]{preskill2018quantum}%
  \BibitemOpen
  \bibfield  {author} {\bibinfo {author} {\bibfnamefont {J.}~\bibnamefont
  {Preskill}},\ }\bibfield  {title} {\bibinfo {title} {Quantum computing in the
  nisq era and beyond},\ }\href
  {https://doi.org/https://doi.org/10.22331/q-2018-08-06-79} {\bibfield
  {journal} {\bibinfo  {journal} {Quantum}\ }\textbf {\bibinfo {volume} {2}},\
  \bibinfo {pages} {79} (\bibinfo {year} {2018})}\BibitemShut {NoStop}%
\bibitem [{\citenamefont {Bauer}\ \emph {et~al.}(2020)\citenamefont {Bauer},
  \citenamefont {Bravyi}, \citenamefont {Motta},\ and\ \citenamefont
  {Chan}}]{bauer2020quantum}%
  \BibitemOpen
  \bibfield  {author} {\bibinfo {author} {\bibfnamefont {B.}~\bibnamefont
  {Bauer}}, \bibinfo {author} {\bibfnamefont {S.}~\bibnamefont {Bravyi}},
  \bibinfo {author} {\bibfnamefont {M.}~\bibnamefont {Motta}},\ and\ \bibinfo
  {author} {\bibfnamefont {G.~K.-L.}\ \bibnamefont {Chan}},\ }\bibfield
  {title} {\bibinfo {title} {Quantum algorithms for quantum chemistry and
  quantum materials science},\ }\href
  {https://doi.org/https://doi.org/10.1021/acs.chemrev.9b00829} {\bibfield
  {journal} {\bibinfo  {journal} {Chemical Reviews}\ }\textbf {\bibinfo
  {volume} {120}},\ \bibinfo {pages} {12685} (\bibinfo {year}
  {2020})}\BibitemShut {NoStop}%
\bibitem [{\citenamefont {Wurtz}\ \emph {et~al.}(2023)\citenamefont {Wurtz},
  \citenamefont {Bylinskii}, \citenamefont {Braverman}, \citenamefont
  {Amato-Grill}, \citenamefont {Cantu}, \citenamefont {Huber}, \citenamefont
  {Lukin}, \citenamefont {Liu}, \citenamefont {Weinberg}, \citenamefont {Long}
  \emph {et~al.}}]{wurtz2023aquila}%
  \BibitemOpen
  \bibfield  {author} {\bibinfo {author} {\bibfnamefont {J.}~\bibnamefont
  {Wurtz}}, \bibinfo {author} {\bibfnamefont {A.}~\bibnamefont {Bylinskii}},
  \bibinfo {author} {\bibfnamefont {B.}~\bibnamefont {Braverman}}, \bibinfo
  {author} {\bibfnamefont {J.}~\bibnamefont {Amato-Grill}}, \bibinfo {author}
  {\bibfnamefont {S.~H.}\ \bibnamefont {Cantu}}, \bibinfo {author}
  {\bibfnamefont {F.}~\bibnamefont {Huber}}, \bibinfo {author} {\bibfnamefont
  {A.}~\bibnamefont {Lukin}}, \bibinfo {author} {\bibfnamefont
  {F.}~\bibnamefont {Liu}}, \bibinfo {author} {\bibfnamefont {P.}~\bibnamefont
  {Weinberg}}, \bibinfo {author} {\bibfnamefont {J.}~\bibnamefont {Long}},
  \emph {et~al.},\ }\bibfield  {title} {\bibinfo {title} {Aquila: Quera's
  256-qubit neutral-atom quantum computer},\ }\bibfield  {journal} {\bibinfo
  {journal} {arXiv preprint arXiv:2306.11727}\ }\href
  {https://doi.org/https://doi.org/10.48550/arXiv.2306.11727}
  {https://doi.org/10.48550/arXiv.2306.11727} (\bibinfo {year}
  {2023})\BibitemShut {NoStop}%
\bibitem [{\citenamefont {Peruzzo}\ \emph {et~al.}(2014)\citenamefont
  {Peruzzo}, \citenamefont {McClean}, \citenamefont {Shadbolt}, \citenamefont
  {Yung}, \citenamefont {Zhou}, \citenamefont {Love}, \citenamefont
  {Aspuru-Guzik},\ and\ \citenamefont {O'Brien}}]{peruzzo2014variational}%
  \BibitemOpen
  \bibfield  {author} {\bibinfo {author} {\bibfnamefont {A.}~\bibnamefont
  {Peruzzo}}, \bibinfo {author} {\bibfnamefont {J.}~\bibnamefont {McClean}},
  \bibinfo {author} {\bibfnamefont {P.}~\bibnamefont {Shadbolt}}, \bibinfo
  {author} {\bibfnamefont {M.-H.}\ \bibnamefont {Yung}}, \bibinfo {author}
  {\bibfnamefont {X.-Q.}\ \bibnamefont {Zhou}}, \bibinfo {author}
  {\bibfnamefont {P.~J.}\ \bibnamefont {Love}}, \bibinfo {author}
  {\bibfnamefont {A.}~\bibnamefont {Aspuru-Guzik}},\ and\ \bibinfo {author}
  {\bibfnamefont {J.~L.}\ \bibnamefont {O'Brien}},\ }\bibfield  {title}
  {\bibinfo {title} {A variational eigenvalue solver on a photonic quantum
  processor},\ }\bibfield  {journal} {\bibinfo  {journal} {Nature
  Communications}\ }\textbf {\bibinfo {volume} {5}},\ \href
  {https://doi.org/10.1038/ncomms5213} {10.1038/ncomms5213} (\bibinfo {year}
  {2014})\BibitemShut {NoStop}%
\bibitem [{\citenamefont {Kandala}\ \emph {et~al.}(2017)\citenamefont
  {Kandala}, \citenamefont {Mezzacapo}, \citenamefont {Temme}, \citenamefont
  {Takita}, \citenamefont {Brink}, \citenamefont {Chow},\ and\ \citenamefont
  {Gambetta}}]{kandala2017hardware}%
  \BibitemOpen
  \bibfield  {author} {\bibinfo {author} {\bibfnamefont {A.}~\bibnamefont
  {Kandala}}, \bibinfo {author} {\bibfnamefont {A.}~\bibnamefont {Mezzacapo}},
  \bibinfo {author} {\bibfnamefont {K.}~\bibnamefont {Temme}}, \bibinfo
  {author} {\bibfnamefont {M.}~\bibnamefont {Takita}}, \bibinfo {author}
  {\bibfnamefont {M.}~\bibnamefont {Brink}}, \bibinfo {author} {\bibfnamefont
  {J.~M.}\ \bibnamefont {Chow}},\ and\ \bibinfo {author} {\bibfnamefont
  {J.~M.}\ \bibnamefont {Gambetta}},\ }\bibfield  {title} {\bibinfo {title}
  {Hardware-efficient variational quantum eigensolver for small molecules and
  quantum magnets},\ }\href
  {https://doi.org/https://doi.org/10.1038/nature23879} {\bibfield  {journal}
  {\bibinfo  {journal} {Nature}\ }\textbf {\bibinfo {volume} {549}},\ \bibinfo
  {pages} {242} (\bibinfo {year} {2017})}\BibitemShut {NoStop}%
\bibitem [{\citenamefont {McClean}\ \emph {et~al.}(2016)\citenamefont
  {McClean}, \citenamefont {Romero}, \citenamefont {Babbush},\ and\
  \citenamefont {Aspuru-Guzik}}]{mcclean2016theory}%
  \BibitemOpen
  \bibfield  {author} {\bibinfo {author} {\bibfnamefont {J.~R.}\ \bibnamefont
  {McClean}}, \bibinfo {author} {\bibfnamefont {J.}~\bibnamefont {Romero}},
  \bibinfo {author} {\bibfnamefont {R.}~\bibnamefont {Babbush}},\ and\ \bibinfo
  {author} {\bibfnamefont {A.}~\bibnamefont {Aspuru-Guzik}},\ }\bibfield
  {title} {\bibinfo {title} {The theory of variational hybrid quantum-classical
  algorithms},\ }\href {https://doi.org/10.1088/1367-2630/18/2/023023}
  {\bibfield  {journal} {\bibinfo  {journal} {New Journal of Physics}\ }\textbf
  {\bibinfo {volume} {18}},\ \bibinfo {pages} {023023} (\bibinfo {year}
  {2016})}\BibitemShut {NoStop}%
\bibitem [{\citenamefont {Cerezo}\ \emph
  {et~al.}(2021{\natexlab{a}})\citenamefont {Cerezo}, \citenamefont
  {Arrasmith}, \citenamefont {Babbush}, \citenamefont {Benjamin}, \citenamefont
  {Endo}, \citenamefont {Fujii}, \citenamefont {McClean}, \citenamefont
  {Mitarai}, \citenamefont {Yuan}, \citenamefont {Cincio},\ and\ \citenamefont
  {Coles}}]{Cerezo2021}%
  \BibitemOpen
  \bibfield  {author} {\bibinfo {author} {\bibfnamefont {M.}~\bibnamefont
  {Cerezo}}, \bibinfo {author} {\bibfnamefont {A.}~\bibnamefont {Arrasmith}},
  \bibinfo {author} {\bibfnamefont {R.}~\bibnamefont {Babbush}}, \bibinfo
  {author} {\bibfnamefont {S.~C.}\ \bibnamefont {Benjamin}}, \bibinfo {author}
  {\bibfnamefont {S.}~\bibnamefont {Endo}}, \bibinfo {author} {\bibfnamefont
  {K.}~\bibnamefont {Fujii}}, \bibinfo {author} {\bibfnamefont {J.~R.}\
  \bibnamefont {McClean}}, \bibinfo {author} {\bibfnamefont {K.}~\bibnamefont
  {Mitarai}}, \bibinfo {author} {\bibfnamefont {X.}~\bibnamefont {Yuan}},
  \bibinfo {author} {\bibfnamefont {L.}~\bibnamefont {Cincio}},\ and\ \bibinfo
  {author} {\bibfnamefont {P.~J.}\ \bibnamefont {Coles}},\ }\bibfield  {title}
  {\bibinfo {title} {Variational quantum algorithms},\ }\href
  {https://doi.org/10.1038/s42254-021-00348-9} {\bibfield  {journal} {\bibinfo
  {journal} {Nature Reviews Physics}\ }\textbf {\bibinfo {volume} {3}},\
  \bibinfo {pages} {625} (\bibinfo {year} {2021}{\natexlab{a}})}\BibitemShut
  {NoStop}%
\bibitem [{\citenamefont {Bharti}\ \emph {et~al.}(2022)\citenamefont {Bharti},
  \citenamefont {Cervera-Lierta}, \citenamefont {Kyaw}, \citenamefont {Haug},
  \citenamefont {Alperin-Lea}, \citenamefont {Anand}, \citenamefont {Degroote},
  \citenamefont {Heimonen}, \citenamefont {Kottmann}, \citenamefont {Menke},
  \citenamefont {Mok}, \citenamefont {Sim}, \citenamefont {Kwek},\ and\
  \citenamefont {Aspuru-Guzik}}]{Bharti2022}%
  \BibitemOpen
  \bibfield  {author} {\bibinfo {author} {\bibfnamefont {K.}~\bibnamefont
  {Bharti}}, \bibinfo {author} {\bibfnamefont {A.}~\bibnamefont
  {Cervera-Lierta}}, \bibinfo {author} {\bibfnamefont {T.~H.}\ \bibnamefont
  {Kyaw}}, \bibinfo {author} {\bibfnamefont {T.}~\bibnamefont {Haug}}, \bibinfo
  {author} {\bibfnamefont {S.}~\bibnamefont {Alperin-Lea}}, \bibinfo {author}
  {\bibfnamefont {A.}~\bibnamefont {Anand}}, \bibinfo {author} {\bibfnamefont
  {M.}~\bibnamefont {Degroote}}, \bibinfo {author} {\bibfnamefont
  {H.}~\bibnamefont {Heimonen}}, \bibinfo {author} {\bibfnamefont {J.~S.}\
  \bibnamefont {Kottmann}}, \bibinfo {author} {\bibfnamefont {T.}~\bibnamefont
  {Menke}}, \bibinfo {author} {\bibfnamefont {W.-K.}\ \bibnamefont {Mok}},
  \bibinfo {author} {\bibfnamefont {S.}~\bibnamefont {Sim}}, \bibinfo {author}
  {\bibfnamefont {L.-C.}\ \bibnamefont {Kwek}},\ and\ \bibinfo {author}
  {\bibfnamefont {A.}~\bibnamefont {Aspuru-Guzik}},\ }\bibfield  {title}
  {\bibinfo {title} {Noisy intermediate-scale quantum algorithms},\ }\href
  {https://doi.org/10.1103/RevModPhys.94.015004} {\bibfield  {journal}
  {\bibinfo  {journal} {Review of Modern Physics}\ }\textbf {\bibinfo {volume}
  {94}},\ \bibinfo {pages} {015004} (\bibinfo {year} {2022})}\BibitemShut
  {NoStop}%
\bibitem [{\citenamefont {Farhi}\ \emph {et~al.}(2014)\citenamefont {Farhi},
  \citenamefont {Goldstone},\ and\ \citenamefont {Gutmann}}]{farhi2014quantum}%
  \BibitemOpen
  \bibfield  {author} {\bibinfo {author} {\bibfnamefont {E.}~\bibnamefont
  {Farhi}}, \bibinfo {author} {\bibfnamefont {J.}~\bibnamefont {Goldstone}},\
  and\ \bibinfo {author} {\bibfnamefont {S.}~\bibnamefont {Gutmann}},\
  }\bibfield  {title} {\bibinfo {title} {A quantum approximate optimization
  algorithm},\ }\bibfield  {journal} {\bibinfo  {journal} {arXiv preprint
  arXiv:1411.4028}\ }\href
  {https://doi.org/https://doi.org/10.48550/arXiv.1411.4028}
  {https://doi.org/10.48550/arXiv.1411.4028} (\bibinfo {year}
  {2014})\BibitemShut {NoStop}%
\bibitem [{\citenamefont {Omiya}\ \emph {et~al.}(2022)\citenamefont {Omiya},
  \citenamefont {Nakagawa}, \citenamefont {Koh}, \citenamefont {Mizukami},
  \citenamefont {Gao},\ and\ \citenamefont {Kobayashi}}]{omiya2022analytical}%
  \BibitemOpen
  \bibfield  {author} {\bibinfo {author} {\bibfnamefont {K.}~\bibnamefont
  {Omiya}}, \bibinfo {author} {\bibfnamefont {Y.~O.}\ \bibnamefont {Nakagawa}},
  \bibinfo {author} {\bibfnamefont {S.}~\bibnamefont {Koh}}, \bibinfo {author}
  {\bibfnamefont {W.}~\bibnamefont {Mizukami}}, \bibinfo {author}
  {\bibfnamefont {Q.}~\bibnamefont {Gao}},\ and\ \bibinfo {author}
  {\bibfnamefont {T.}~\bibnamefont {Kobayashi}},\ }\bibfield  {title} {\bibinfo
  {title} {Analytical energy gradient for state-averaged orbital-optimized
  variational quantum eigensolvers and its application to a photochemical
  reaction},\ }\href {https://doi.org/https://doi.org/10.1021/acs.jctc.1c00877}
  {\bibfield  {journal} {\bibinfo  {journal} {Journal of Chemical Theory and
  Computation}\ }\textbf {\bibinfo {volume} {18}},\ \bibinfo {pages} {741}
  (\bibinfo {year} {2022})}\BibitemShut {NoStop}%
\bibitem [{\citenamefont {Ibe}\ \emph {et~al.}(2022)\citenamefont {Ibe},
  \citenamefont {Nakagawa}, \citenamefont {Earnest}, \citenamefont {Yamamoto},
  \citenamefont {Mitarai}, \citenamefont {Gao},\ and\ \citenamefont
  {Kobayashi}}]{ibe2022calculating}%
  \BibitemOpen
  \bibfield  {author} {\bibinfo {author} {\bibfnamefont {Y.}~\bibnamefont
  {Ibe}}, \bibinfo {author} {\bibfnamefont {Y.~O.}\ \bibnamefont {Nakagawa}},
  \bibinfo {author} {\bibfnamefont {N.}~\bibnamefont {Earnest}}, \bibinfo
  {author} {\bibfnamefont {T.}~\bibnamefont {Yamamoto}}, \bibinfo {author}
  {\bibfnamefont {K.}~\bibnamefont {Mitarai}}, \bibinfo {author} {\bibfnamefont
  {Q.}~\bibnamefont {Gao}},\ and\ \bibinfo {author} {\bibfnamefont
  {T.}~\bibnamefont {Kobayashi}},\ }\bibfield  {title} {\bibinfo {title}
  {Calculating transition amplitudes by variational quantum deflation},\ }\href
  {https://doi.org/https://doi.org/10.1103/PhysRevResearch.4.013173} {\bibfield
   {journal} {\bibinfo  {journal} {Physical Review Research}\ }\textbf
  {\bibinfo {volume} {4}},\ \bibinfo {pages} {013173} (\bibinfo {year}
  {2022})}\BibitemShut {NoStop}%
\bibitem [{\citenamefont {Nakagawa}\ \emph {et~al.}(2023)\citenamefont
  {Nakagawa}, \citenamefont {Chen}, \citenamefont {Sudo}, \citenamefont
  {Ohnishi},\ and\ \citenamefont {Mizukami}}]{nakagawa2023analytical}%
  \BibitemOpen
  \bibfield  {author} {\bibinfo {author} {\bibfnamefont {Y.~O.}\ \bibnamefont
  {Nakagawa}}, \bibinfo {author} {\bibfnamefont {J.}~\bibnamefont {Chen}},
  \bibinfo {author} {\bibfnamefont {S.}~\bibnamefont {Sudo}}, \bibinfo {author}
  {\bibfnamefont {Y.-y.}\ \bibnamefont {Ohnishi}},\ and\ \bibinfo {author}
  {\bibfnamefont {W.}~\bibnamefont {Mizukami}},\ }\bibfield  {title} {\bibinfo
  {title} {Analytical formulation of the second-order derivative of energy for
  the orbital-optimized variational quantum eigensolver: Application to
  polarizability},\ }\href
  {https://doi.org/https://doi.org/10.1021/acs.jctc.2c01176} {\bibfield
  {journal} {\bibinfo  {journal} {Journal of Chemical Theory and Computation}\
  }\textbf {\bibinfo {volume} {19}},\ \bibinfo {pages} {1998} (\bibinfo {year}
  {2023})}\BibitemShut {NoStop}%
\bibitem [{\citenamefont {Cong}\ \emph {et~al.}(2019)\citenamefont {Cong},
  \citenamefont {Choi},\ and\ \citenamefont {Lukin}}]{cong2019quantum}%
  \BibitemOpen
  \bibfield  {author} {\bibinfo {author} {\bibfnamefont {I.}~\bibnamefont
  {Cong}}, \bibinfo {author} {\bibfnamefont {S.}~\bibnamefont {Choi}},\ and\
  \bibinfo {author} {\bibfnamefont {M.~D.}\ \bibnamefont {Lukin}},\ }\bibfield
  {title} {\bibinfo {title} {Quantum convolutional neural networks},\ }\href
  {https://doi.org/https://doi.org/10.1038/s41567-019-0648-8} {\bibfield
  {journal} {\bibinfo  {journal} {Nature Physics}\ }\textbf {\bibinfo {volume}
  {15}},\ \bibinfo {pages} {1273} (\bibinfo {year} {2019})}\BibitemShut
  {NoStop}%
\bibitem [{\citenamefont {McClean}\ \emph {et~al.}(2018)\citenamefont
  {McClean}, \citenamefont {Boixo}, \citenamefont {Smelyanskiy}, \citenamefont
  {Babbush},\ and\ \citenamefont {Neven}}]{McClean2018}%
  \BibitemOpen
  \bibfield  {author} {\bibinfo {author} {\bibfnamefont {J.~R.}\ \bibnamefont
  {McClean}}, \bibinfo {author} {\bibfnamefont {S.}~\bibnamefont {Boixo}},
  \bibinfo {author} {\bibfnamefont {V.~N.}\ \bibnamefont {Smelyanskiy}},
  \bibinfo {author} {\bibfnamefont {R.}~\bibnamefont {Babbush}},\ and\ \bibinfo
  {author} {\bibfnamefont {H.}~\bibnamefont {Neven}},\ }\bibfield  {title}
  {\bibinfo {title} {Barren plateaus in quantum neural network training
  landscapes},\ }\bibfield  {journal} {\bibinfo  {journal} {Nature
  Communications}\ }\textbf {\bibinfo {volume} {9}},\ \href
  {https://doi.org/10.1038/s41467-018-07090-4} {10.1038/s41467-018-07090-4}
  (\bibinfo {year} {2018})\BibitemShut {NoStop}%
\bibitem [{\citenamefont {Zhao}\ and\ \citenamefont
  {Gao}(2021)}]{zhao2021analyzing}%
  \BibitemOpen
  \bibfield  {author} {\bibinfo {author} {\bibfnamefont {C.}~\bibnamefont
  {Zhao}}\ and\ \bibinfo {author} {\bibfnamefont {X.-S.}\ \bibnamefont {Gao}},\
  }\bibfield  {title} {\bibinfo {title} {Analyzing the barren plateau
  phenomenon in training quantum neural networks with the zx-calculus},\ }\href
  {https://doi.org/https://doi.org/10.22331/q-2021-06-04-466} {\bibfield
  {journal} {\bibinfo  {journal} {Quantum}\ }\textbf {\bibinfo {volume} {5}},\
  \bibinfo {pages} {466} (\bibinfo {year} {2021})}\BibitemShut {NoStop}%
\bibitem [{\citenamefont {Sack}\ \emph {et~al.}(2022)\citenamefont {Sack},
  \citenamefont {Medina}, \citenamefont {Michailidis}, \citenamefont {Kueng},\
  and\ \citenamefont {Serbyn}}]{sack2022avoiding}%
  \BibitemOpen
  \bibfield  {author} {\bibinfo {author} {\bibfnamefont {S.~H.}\ \bibnamefont
  {Sack}}, \bibinfo {author} {\bibfnamefont {R.~A.}\ \bibnamefont {Medina}},
  \bibinfo {author} {\bibfnamefont {A.~A.}\ \bibnamefont {Michailidis}},
  \bibinfo {author} {\bibfnamefont {R.}~\bibnamefont {Kueng}},\ and\ \bibinfo
  {author} {\bibfnamefont {M.}~\bibnamefont {Serbyn}},\ }\bibfield  {title}
  {\bibinfo {title} {Avoiding barren plateaus using classical shadows},\ }\href
  {https://doi.org/https://doi.org/10.1103/PRXQuantum.3.020365} {\bibfield
  {journal} {\bibinfo  {journal} {PRX Quantum}\ }\textbf {\bibinfo {volume}
  {3}},\ \bibinfo {pages} {020365} (\bibinfo {year} {2022})}\BibitemShut
  {NoStop}%
\bibitem [{\citenamefont {Endo}\ \emph {et~al.}(2021)\citenamefont {Endo},
  \citenamefont {Cai}, \citenamefont {Benjamin},\ and\ \citenamefont
  {Yuan}}]{endo2021hybrid}%
  \BibitemOpen
  \bibfield  {author} {\bibinfo {author} {\bibfnamefont {S.}~\bibnamefont
  {Endo}}, \bibinfo {author} {\bibfnamefont {Z.}~\bibnamefont {Cai}}, \bibinfo
  {author} {\bibfnamefont {S.~C.}\ \bibnamefont {Benjamin}},\ and\ \bibinfo
  {author} {\bibfnamefont {X.}~\bibnamefont {Yuan}},\ }\bibfield  {title}
  {\bibinfo {title} {Hybrid quantum-classical algorithms and quantum error
  mitigation},\ }\href {https://doi.org/https://doi.org/10.7566/JPSJ.90.032001}
  {\bibfield  {journal} {\bibinfo  {journal} {Journal of the Physical Society
  of Japan}\ }\textbf {\bibinfo {volume} {90}},\ \bibinfo {pages} {032001}
  (\bibinfo {year} {2021})}\BibitemShut {NoStop}%
\bibitem [{\citenamefont {Suzuki}\ \emph {et~al.}(2022)\citenamefont {Suzuki},
  \citenamefont {Endo}, \citenamefont {Fujii},\ and\ \citenamefont
  {Tokunaga}}]{suzuki2022quantum}%
  \BibitemOpen
  \bibfield  {author} {\bibinfo {author} {\bibfnamefont {Y.}~\bibnamefont
  {Suzuki}}, \bibinfo {author} {\bibfnamefont {S.}~\bibnamefont {Endo}},
  \bibinfo {author} {\bibfnamefont {K.}~\bibnamefont {Fujii}},\ and\ \bibinfo
  {author} {\bibfnamefont {Y.}~\bibnamefont {Tokunaga}},\ }\bibfield  {title}
  {\bibinfo {title} {Quantum error mitigation as a universal error reduction
  technique: Applications from the nisq to the fault-tolerant quantum computing
  eras},\ }\href {https://doi.org/https://doi.org/10.1103/PRXQuantum.3.010345}
  {\bibfield  {journal} {\bibinfo  {journal} {PRX Quantum}\ }\textbf {\bibinfo
  {volume} {3}},\ \bibinfo {pages} {010345} (\bibinfo {year}
  {2022})}\BibitemShut {NoStop}%
\bibitem [{\citenamefont {Cai}\ \emph {et~al.}(2022)\citenamefont {Cai},
  \citenamefont {Babbush}, \citenamefont {Benjamin}, \citenamefont {Endo},
  \citenamefont {Huggins}, \citenamefont {Li}, \citenamefont {McClean},\ and\
  \citenamefont {O'Brien}}]{cai2022quantum}%
  \BibitemOpen
  \bibfield  {author} {\bibinfo {author} {\bibfnamefont {Z.}~\bibnamefont
  {Cai}}, \bibinfo {author} {\bibfnamefont {R.}~\bibnamefont {Babbush}},
  \bibinfo {author} {\bibfnamefont {S.~C.}\ \bibnamefont {Benjamin}}, \bibinfo
  {author} {\bibfnamefont {S.}~\bibnamefont {Endo}}, \bibinfo {author}
  {\bibfnamefont {W.~J.}\ \bibnamefont {Huggins}}, \bibinfo {author}
  {\bibfnamefont {Y.}~\bibnamefont {Li}}, \bibinfo {author} {\bibfnamefont
  {J.~R.}\ \bibnamefont {McClean}},\ and\ \bibinfo {author} {\bibfnamefont
  {T.~E.}\ \bibnamefont {O'Brien}},\ }\bibfield  {title} {\bibinfo {title}
  {Quantum error mitigation},\ }\bibfield  {journal} {\bibinfo  {journal}
  {arXiv preprint arXiv:2210.00921}\ }\href
  {https://doi.org/https://doi.org/10.48550/arXiv.2210.00921}
  {https://doi.org/10.48550/arXiv.2210.00921} (\bibinfo {year}
  {2022})\BibitemShut {NoStop}%
\bibitem [{\citenamefont {Akahoshi}\ \emph {et~al.}(2023)\citenamefont
  {Akahoshi}, \citenamefont {Maruyama}, \citenamefont {Oshima}, \citenamefont
  {Sato},\ and\ \citenamefont {Fujii}}]{akahoshi2023partially}%
  \BibitemOpen
  \bibfield  {author} {\bibinfo {author} {\bibfnamefont {Y.}~\bibnamefont
  {Akahoshi}}, \bibinfo {author} {\bibfnamefont {K.}~\bibnamefont {Maruyama}},
  \bibinfo {author} {\bibfnamefont {H.}~\bibnamefont {Oshima}}, \bibinfo
  {author} {\bibfnamefont {S.}~\bibnamefont {Sato}},\ and\ \bibinfo {author}
  {\bibfnamefont {K.}~\bibnamefont {Fujii}},\ }\bibfield  {title} {\bibinfo
  {title} {Partially fault-tolerant quantum computing architecture with
  error-corrected clifford gates and space-time efficient analog rotations},\
  }\bibfield  {journal} {\bibinfo  {journal} {arXiv preprint arXiv:2303.13181}\
  }\href {https://doi.org/https://doi.org/10.48550/arXiv.2303.13181}
  {https://doi.org/10.48550/arXiv.2303.13181} (\bibinfo {year}
  {2023})\BibitemShut {NoStop}%
\bibitem [{\citenamefont {Bultrini}\ \emph {et~al.}(2023)\citenamefont
  {Bultrini}, \citenamefont {Wang}, \citenamefont {Czarnik}, \citenamefont
  {Gordon}, \citenamefont {Cerezo}, \citenamefont {Coles},\ and\ \citenamefont
  {Cincio}}]{bultrini2023battle}%
  \BibitemOpen
  \bibfield  {author} {\bibinfo {author} {\bibfnamefont {D.}~\bibnamefont
  {Bultrini}}, \bibinfo {author} {\bibfnamefont {S.}~\bibnamefont {Wang}},
  \bibinfo {author} {\bibfnamefont {P.}~\bibnamefont {Czarnik}}, \bibinfo
  {author} {\bibfnamefont {M.~H.}\ \bibnamefont {Gordon}}, \bibinfo {author}
  {\bibfnamefont {M.}~\bibnamefont {Cerezo}}, \bibinfo {author} {\bibfnamefont
  {P.~J.}\ \bibnamefont {Coles}},\ and\ \bibinfo {author} {\bibfnamefont
  {L.}~\bibnamefont {Cincio}},\ }\bibfield  {title} {\bibinfo {title} {The
  battle of clean and dirty qubits in the era of partial error correction},\
  }\href {https://doi.org/https://doi.org/10.22331/q-2023-07-13-1060}
  {\bibfield  {journal} {\bibinfo  {journal} {Quantum}\ }\textbf {\bibinfo
  {volume} {7}},\ \bibinfo {pages} {1060} (\bibinfo {year} {2023})}\BibitemShut
  {NoStop}%
\bibitem [{\citenamefont {Barison}\ \emph {et~al.}(2021)\citenamefont
  {Barison}, \citenamefont {Vicentini},\ and\ \citenamefont
  {Carleo}}]{barison2021efficient}%
  \BibitemOpen
  \bibfield  {author} {\bibinfo {author} {\bibfnamefont {S.}~\bibnamefont
  {Barison}}, \bibinfo {author} {\bibfnamefont {F.}~\bibnamefont {Vicentini}},\
  and\ \bibinfo {author} {\bibfnamefont {G.}~\bibnamefont {Carleo}},\
  }\bibfield  {title} {\bibinfo {title} {An efficient quantum algorithm for the
  time evolution of parameterized circuits},\ }\href
  {https://doi.org/https://doi.org/10.22331/q-2021-07-28-512} {\bibfield
  {journal} {\bibinfo  {journal} {Quantum}\ }\textbf {\bibinfo {volume} {5}},\
  \bibinfo {pages} {512} (\bibinfo {year} {2021})}\BibitemShut {NoStop}%
\bibitem [{\citenamefont {Solfanelli}\ \emph {et~al.}(2021)\citenamefont
  {Solfanelli}, \citenamefont {Santini},\ and\ \citenamefont
  {Campisi}}]{Solfanelli2021}%
  \BibitemOpen
  \bibfield  {author} {\bibinfo {author} {\bibfnamefont {A.}~\bibnamefont
  {Solfanelli}}, \bibinfo {author} {\bibfnamefont {A.}~\bibnamefont
  {Santini}},\ and\ \bibinfo {author} {\bibfnamefont {M.}~\bibnamefont
  {Campisi}},\ }\bibfield  {title} {\bibinfo {title} {Experimental verification
  of fluctuation relations with a quantum computer},\ }\href
  {https://doi.org/10.1103/PRXQuantum.2.030353} {\bibfield  {journal} {\bibinfo
   {journal} {PRX Quantum}\ }\textbf {\bibinfo {volume} {2}},\ \bibinfo {pages}
  {030353} (\bibinfo {year} {2021})}\BibitemShut {NoStop}%
\bibitem [{\citenamefont {Melo}\ \emph {et~al.}(2022)\citenamefont {Melo},
  \citenamefont {S\'a}, \citenamefont {Roditi}, \citenamefont {Souza},
  \citenamefont {Oliveira}, \citenamefont {Sarthour},\ and\ \citenamefont
  {Landi}}]{Melo2022}%
  \BibitemOpen
  \bibfield  {author} {\bibinfo {author} {\bibfnamefont {F.~V.}\ \bibnamefont
  {Melo}}, \bibinfo {author} {\bibfnamefont {N.}~\bibnamefont {S\'a}}, \bibinfo
  {author} {\bibfnamefont {I.}~\bibnamefont {Roditi}}, \bibinfo {author}
  {\bibfnamefont {A.~M.}\ \bibnamefont {Souza}}, \bibinfo {author}
  {\bibfnamefont {I.~S.}\ \bibnamefont {Oliveira}}, \bibinfo {author}
  {\bibfnamefont {R.~S.}\ \bibnamefont {Sarthour}},\ and\ \bibinfo {author}
  {\bibfnamefont {G.~T.}\ \bibnamefont {Landi}},\ }\bibfield  {title} {\bibinfo
  {title} {Implementation of a two-stroke quantum heat engine with a
  collisional model},\ }\href {https://doi.org/10.1103/PhysRevA.106.032410}
  {\bibfield  {journal} {\bibinfo  {journal} {Physical Review A}\ }\textbf
  {\bibinfo {volume} {106}},\ \bibinfo {pages} {032410} (\bibinfo {year}
  {2022})}\BibitemShut {NoStop}%
\bibitem [{\citenamefont {Solfanelli}\ \emph {et~al.}(2022)\citenamefont
  {Solfanelli}, \citenamefont {Santini},\ and\ \citenamefont
  {Campisi}}]{Solfanelli2022}%
  \BibitemOpen
  \bibfield  {author} {\bibinfo {author} {\bibfnamefont {A.}~\bibnamefont
  {Solfanelli}}, \bibinfo {author} {\bibfnamefont {A.}~\bibnamefont
  {Santini}},\ and\ \bibinfo {author} {\bibfnamefont {M.}~\bibnamefont
  {Campisi}},\ }\bibfield  {title} {\bibinfo {title} {{Quantum thermodynamic
  methods to purify a qubit on a quantum processing unit}},\ }\href
  {https://doi.org/10.1116/5.0091121} {\bibfield  {journal} {\bibinfo
  {journal} {AVS Quantum Science}\ }\textbf {\bibinfo {volume} {4}},\ \bibinfo
  {pages} {026802} (\bibinfo {year} {2022})}\BibitemShut {NoStop}%
\bibitem [{\citenamefont {Consiglio}\ \emph {et~al.}(2023)\citenamefont
  {Consiglio}, \citenamefont {Settino}, \citenamefont {Giordano}, \citenamefont
  {Mastroianni}, \citenamefont {Plastina}, \citenamefont {Lorenzo},
  \citenamefont {Maniscalco}, \citenamefont {Goold},\ and\ \citenamefont
  {Apollaro}}]{Consiglio2023}%
  \BibitemOpen
  \bibfield  {author} {\bibinfo {author} {\bibfnamefont {M.}~\bibnamefont
  {Consiglio}}, \bibinfo {author} {\bibfnamefont {J.}~\bibnamefont {Settino}},
  \bibinfo {author} {\bibfnamefont {A.}~\bibnamefont {Giordano}}, \bibinfo
  {author} {\bibfnamefont {C.}~\bibnamefont {Mastroianni}}, \bibinfo {author}
  {\bibfnamefont {F.}~\bibnamefont {Plastina}}, \bibinfo {author}
  {\bibfnamefont {S.}~\bibnamefont {Lorenzo}}, \bibinfo {author} {\bibfnamefont
  {S.}~\bibnamefont {Maniscalco}}, \bibinfo {author} {\bibfnamefont
  {J.}~\bibnamefont {Goold}},\ and\ \bibinfo {author} {\bibfnamefont
  {T.~J.~G.}\ \bibnamefont {Apollaro}},\ }\href@noop {} {\bibinfo {title}
  {Variational gibbs state preparation on nisq devices}} (\bibinfo {year}
  {2023}),\ \Eprint {https://arxiv.org/abs/arXiv:2303.11276} {arXiv:2303.11276}
  \BibitemShut {NoStop}%
\bibitem [{IBM()}]{IBMQuantum}%
  \BibitemOpen
  \href@noop {} {\bibinfo {title} {{{IBM Quantum}}}},\ \bibinfo {howpublished}
  {https://quantum-computing.ibm.com/services/resources}\BibitemShut {NoStop}%
\bibitem [{\citenamefont {Nakanishi}\ \emph {et~al.}(2019)\citenamefont
  {Nakanishi}, \citenamefont {Mitarai},\ and\ \citenamefont
  {Fujii}}]{Nakanishi2019}%
  \BibitemOpen
  \bibfield  {author} {\bibinfo {author} {\bibfnamefont {K.~M.}\ \bibnamefont
  {Nakanishi}}, \bibinfo {author} {\bibfnamefont {K.}~\bibnamefont {Mitarai}},\
  and\ \bibinfo {author} {\bibfnamefont {K.}~\bibnamefont {Fujii}},\ }\bibfield
   {title} {\bibinfo {title} {Subspace-search variational quantum eigensolver
  for excited states},\ }\href
  {https://doi.org/10.1103/PhysRevResearch.1.033062} {\bibfield  {journal}
  {\bibinfo  {journal} {Physical Review Research}\ }\textbf {\bibinfo {volume}
  {1}},\ \bibinfo {pages} {033062} (\bibinfo {year} {2019})}\BibitemShut
  {NoStop}%
\bibitem [{\citenamefont {Higgott}\ \emph {et~al.}(2019)\citenamefont
  {Higgott}, \citenamefont {Wang},\ and\ \citenamefont
  {Brierley}}]{Higgott2019}%
  \BibitemOpen
  \bibfield  {author} {\bibinfo {author} {\bibfnamefont {O.}~\bibnamefont
  {Higgott}}, \bibinfo {author} {\bibfnamefont {D.}~\bibnamefont {Wang}},\ and\
  \bibinfo {author} {\bibfnamefont {S.}~\bibnamefont {Brierley}},\ }\bibfield
  {title} {\bibinfo {title} {Variational {Q}uantum {C}omputation of {E}xcited
  {S}tates},\ }\href {https://doi.org/10.22331/q-2019-07-01-156} {\bibfield
  {journal} {\bibinfo  {journal} {{Quantum}}\ }\textbf {\bibinfo {volume}
  {3}},\ \bibinfo {pages} {156} (\bibinfo {year} {2019})}\BibitemShut {NoStop}%
\bibitem [{\citenamefont {Jones}\ \emph {et~al.}(2019)\citenamefont {Jones},
  \citenamefont {Endo}, \citenamefont {McArdle}, \citenamefont {Yuan},\ and\
  \citenamefont {Benjamin}}]{Jones2019}%
  \BibitemOpen
  \bibfield  {author} {\bibinfo {author} {\bibfnamefont {T.}~\bibnamefont
  {Jones}}, \bibinfo {author} {\bibfnamefont {S.}~\bibnamefont {Endo}},
  \bibinfo {author} {\bibfnamefont {S.}~\bibnamefont {McArdle}}, \bibinfo
  {author} {\bibfnamefont {X.}~\bibnamefont {Yuan}},\ and\ \bibinfo {author}
  {\bibfnamefont {S.~C.}\ \bibnamefont {Benjamin}},\ }\bibfield  {title}
  {\bibinfo {title} {Variational quantum algorithms for discovering hamiltonian
  spectra},\ }\href {https://doi.org/10.1103/PhysRevA.99.062304} {\bibfield
  {journal} {\bibinfo  {journal} {Physical Review A}\ }\textbf {\bibinfo
  {volume} {99}},\ \bibinfo {pages} {062304} (\bibinfo {year}
  {2019})}\BibitemShut {NoStop}%
\bibitem [{\citenamefont {C{\^{\i}}rstoiu}\ \emph {et~al.}(2020)\citenamefont
  {C{\^{\i}}rstoiu}, \citenamefont {Holmes}, \citenamefont {Iosue},
  \citenamefont {Cincio}, \citenamefont {Coles},\ and\ \citenamefont
  {Sornborger}}]{Crstoiu2020}%
  \BibitemOpen
  \bibfield  {author} {\bibinfo {author} {\bibfnamefont {C.}~\bibnamefont
  {C{\^{\i}}rstoiu}}, \bibinfo {author} {\bibfnamefont {Z.}~\bibnamefont
  {Holmes}}, \bibinfo {author} {\bibfnamefont {J.}~\bibnamefont {Iosue}},
  \bibinfo {author} {\bibfnamefont {L.}~\bibnamefont {Cincio}}, \bibinfo
  {author} {\bibfnamefont {P.~J.}\ \bibnamefont {Coles}},\ and\ \bibinfo
  {author} {\bibfnamefont {A.}~\bibnamefont {Sornborger}},\ }\bibfield  {title}
  {\bibinfo {title} {Variational fast forwarding for quantum simulation beyond
  the coherence time},\ }\bibfield  {journal} {\bibinfo  {journal} {npj Quantum
  Information}\ }\textbf {\bibinfo {volume} {6}},\ \href
  {https://doi.org/10.1038/s41534-020-00302-0} {10.1038/s41534-020-00302-0}
  (\bibinfo {year} {2020})\BibitemShut {NoStop}%
\bibitem [{\citenamefont {Yuan}\ \emph {et~al.}(2019)\citenamefont {Yuan},
  \citenamefont {Endo}, \citenamefont {Zhao}, \citenamefont {Li},\ and\
  \citenamefont {Benjamin}}]{Yuan2019}%
  \BibitemOpen
  \bibfield  {author} {\bibinfo {author} {\bibfnamefont {X.}~\bibnamefont
  {Yuan}}, \bibinfo {author} {\bibfnamefont {S.}~\bibnamefont {Endo}}, \bibinfo
  {author} {\bibfnamefont {Q.}~\bibnamefont {Zhao}}, \bibinfo {author}
  {\bibfnamefont {Y.}~\bibnamefont {Li}},\ and\ \bibinfo {author}
  {\bibfnamefont {S.~C.}\ \bibnamefont {Benjamin}},\ }\bibfield  {title}
  {\bibinfo {title} {Theory of variational quantum simulation},\ }\href
  {https://doi.org/10.22331/q-2019-10-07-191} {\bibfield  {journal} {\bibinfo
  {journal} {{Quantum}}\ }\textbf {\bibinfo {volume} {3}},\ \bibinfo {pages}
  {191} (\bibinfo {year} {2019})}\BibitemShut {NoStop}%
\bibitem [{\citenamefont {Li}\ and\ \citenamefont
  {Benjamin}(2017)}]{li2017efficient}%
  \BibitemOpen
  \bibfield  {author} {\bibinfo {author} {\bibfnamefont {Y.}~\bibnamefont
  {Li}}\ and\ \bibinfo {author} {\bibfnamefont {S.~C.}\ \bibnamefont
  {Benjamin}},\ }\bibfield  {title} {\bibinfo {title} {Efficient variational
  quantum simulator incorporating active error minimization},\ }\href
  {https://doi.org/https://doi.org/10.1103/PhysRevX.7.021050} {\bibfield
  {journal} {\bibinfo  {journal} {Physical Review X}\ }\textbf {\bibinfo
  {volume} {7}},\ \bibinfo {pages} {021050} (\bibinfo {year}
  {2017})}\BibitemShut {NoStop}%
\bibitem [{\citenamefont {Bharti}\ and\ \citenamefont
  {Haug}(2021)}]{Bharti2021}%
  \BibitemOpen
  \bibfield  {author} {\bibinfo {author} {\bibfnamefont {K.}~\bibnamefont
  {Bharti}}\ and\ \bibinfo {author} {\bibfnamefont {T.}~\bibnamefont {Haug}},\
  }\bibfield  {title} {\bibinfo {title} {Quantum-assisted simulator},\ }\href
  {https://doi.org/10.1103/PhysRevA.104.042418} {\bibfield  {journal} {\bibinfo
   {journal} {Physical Review A}\ }\textbf {\bibinfo {volume} {104}},\ \bibinfo
  {pages} {042418} (\bibinfo {year} {2021})}\BibitemShut {NoStop}%
\bibitem [{\citenamefont {Heya}\ \emph {et~al.}(2023)\citenamefont {Heya},
  \citenamefont {Nakanishi}, \citenamefont {Mitarai}, \citenamefont {Yan},
  \citenamefont {Zuo}, \citenamefont {Suzuki}, \citenamefont {Sugiyama},
  \citenamefont {Tamate}, \citenamefont {Tabuchi}, \citenamefont {Fujii} \emph
  {et~al.}}]{heya2019subspace}%
  \BibitemOpen
  \bibfield  {author} {\bibinfo {author} {\bibfnamefont {K.}~\bibnamefont
  {Heya}}, \bibinfo {author} {\bibfnamefont {K.~M.}\ \bibnamefont {Nakanishi}},
  \bibinfo {author} {\bibfnamefont {K.}~\bibnamefont {Mitarai}}, \bibinfo
  {author} {\bibfnamefont {Z.}~\bibnamefont {Yan}}, \bibinfo {author}
  {\bibfnamefont {K.}~\bibnamefont {Zuo}}, \bibinfo {author} {\bibfnamefont
  {Y.}~\bibnamefont {Suzuki}}, \bibinfo {author} {\bibfnamefont
  {T.}~\bibnamefont {Sugiyama}}, \bibinfo {author} {\bibfnamefont
  {S.}~\bibnamefont {Tamate}}, \bibinfo {author} {\bibfnamefont
  {Y.}~\bibnamefont {Tabuchi}}, \bibinfo {author} {\bibfnamefont
  {K.}~\bibnamefont {Fujii}}, \emph {et~al.},\ }\bibfield  {title} {\bibinfo
  {title} {Subspace variational quantum simulator},\ }\href@noop {} {\bibfield
  {journal} {\bibinfo  {journal} {Physical Review Research}\ }\textbf {\bibinfo
  {volume} {5}},\ \bibinfo {pages} {023078} (\bibinfo {year}
  {2023})}\BibitemShut {NoStop}%
\bibitem [{\citenamefont {Bernien}\ \emph {et~al.}(2017)\citenamefont
  {Bernien}, \citenamefont {Schwartz}, \citenamefont {Keesling}, \citenamefont
  {Levine}, \citenamefont {Omran}, \citenamefont {Pichler}, \citenamefont
  {Choi}, \citenamefont {Zibrov}, \citenamefont {Endres}, \citenamefont
  {Greiner}, \citenamefont {Vuleti{\'{c}}},\ and\ \citenamefont
  {Lukin}}]{Bernien2017}%
  \BibitemOpen
  \bibfield  {author} {\bibinfo {author} {\bibfnamefont {H.}~\bibnamefont
  {Bernien}}, \bibinfo {author} {\bibfnamefont {S.}~\bibnamefont {Schwartz}},
  \bibinfo {author} {\bibfnamefont {A.}~\bibnamefont {Keesling}}, \bibinfo
  {author} {\bibfnamefont {H.}~\bibnamefont {Levine}}, \bibinfo {author}
  {\bibfnamefont {A.}~\bibnamefont {Omran}}, \bibinfo {author} {\bibfnamefont
  {H.}~\bibnamefont {Pichler}}, \bibinfo {author} {\bibfnamefont
  {S.}~\bibnamefont {Choi}}, \bibinfo {author} {\bibfnamefont {A.~S.}\
  \bibnamefont {Zibrov}}, \bibinfo {author} {\bibfnamefont {M.}~\bibnamefont
  {Endres}}, \bibinfo {author} {\bibfnamefont {M.}~\bibnamefont {Greiner}},
  \bibinfo {author} {\bibfnamefont {V.}~\bibnamefont {Vuleti{\'{c}}}},\ and\
  \bibinfo {author} {\bibfnamefont {M.~D.}\ \bibnamefont {Lukin}},\ }\bibfield
  {title} {\bibinfo {title} {Probing many-body dynamics on a 51-atom quantum
  simulator},\ }\href {https://doi.org/10.1038/nature24622} {\bibfield
  {journal} {\bibinfo  {journal} {Nature}\ }\textbf {\bibinfo {volume} {551}},\
  \bibinfo {pages} {579} (\bibinfo {year} {2017})}\BibitemShut {NoStop}%
\bibitem [{\citenamefont {Labuhn}\ \emph {et~al.}(2016)\citenamefont {Labuhn},
  \citenamefont {Barredo}, \citenamefont {Ravets}, \citenamefont
  {de~L{\'{e}}s{\'{e}}leuc}, \citenamefont {Macr{\`{\i}}}, \citenamefont
  {Lahaye},\ and\ \citenamefont {Browaeys}}]{Labuhn2016}%
  \BibitemOpen
  \bibfield  {author} {\bibinfo {author} {\bibfnamefont {H.}~\bibnamefont
  {Labuhn}}, \bibinfo {author} {\bibfnamefont {D.}~\bibnamefont {Barredo}},
  \bibinfo {author} {\bibfnamefont {S.}~\bibnamefont {Ravets}}, \bibinfo
  {author} {\bibfnamefont {S.}~\bibnamefont {de~L{\'{e}}s{\'{e}}leuc}},
  \bibinfo {author} {\bibfnamefont {T.}~\bibnamefont {Macr{\`{\i}}}}, \bibinfo
  {author} {\bibfnamefont {T.}~\bibnamefont {Lahaye}},\ and\ \bibinfo {author}
  {\bibfnamefont {A.}~\bibnamefont {Browaeys}},\ }\bibfield  {title} {\bibinfo
  {title} {Tunable two-dimensional arrays of single rydberg atoms for realizing
  quantum ising models},\ }\href {https://doi.org/10.1038/nature18274}
  {\bibfield  {journal} {\bibinfo  {journal} {Nature}\ }\textbf {\bibinfo
  {volume} {534}},\ \bibinfo {pages} {667} (\bibinfo {year}
  {2016})}\BibitemShut {NoStop}%
\bibitem [{\citenamefont {Mitarai}\ \emph {et~al.}(2018)\citenamefont
  {Mitarai}, \citenamefont {Negoro}, \citenamefont {Kitagawa},\ and\
  \citenamefont {Fujii}}]{mitarai2018}%
  \BibitemOpen
  \bibfield  {author} {\bibinfo {author} {\bibfnamefont {K.}~\bibnamefont
  {Mitarai}}, \bibinfo {author} {\bibfnamefont {M.}~\bibnamefont {Negoro}},
  \bibinfo {author} {\bibfnamefont {M.}~\bibnamefont {Kitagawa}},\ and\
  \bibinfo {author} {\bibfnamefont {K.}~\bibnamefont {Fujii}},\ }\bibfield
  {title} {\bibinfo {title} {Quantum circuit learning},\ }\href
  {https://doi.org/10.1103/PhysRevA.98.032309} {\bibfield  {journal} {\bibinfo
  {journal} {Physical Review A}\ }\textbf {\bibinfo {volume} {98}},\ \bibinfo
  {pages} {032309} (\bibinfo {year} {2018})}\BibitemShut {NoStop}%
\bibitem [{\citenamefont {Schuld}\ \emph {et~al.}(2019)\citenamefont {Schuld},
  \citenamefont {Bergholm}, \citenamefont {Gogolin}, \citenamefont {Izaac},\
  and\ \citenamefont {Killoran}}]{schuld2019}%
  \BibitemOpen
  \bibfield  {author} {\bibinfo {author} {\bibfnamefont {M.}~\bibnamefont
  {Schuld}}, \bibinfo {author} {\bibfnamefont {V.}~\bibnamefont {Bergholm}},
  \bibinfo {author} {\bibfnamefont {C.}~\bibnamefont {Gogolin}}, \bibinfo
  {author} {\bibfnamefont {J.}~\bibnamefont {Izaac}},\ and\ \bibinfo {author}
  {\bibfnamefont {N.}~\bibnamefont {Killoran}},\ }\bibfield  {title} {\bibinfo
  {title} {Evaluating analytic gradients on quantum hardware},\ }\href
  {https://doi.org/10.1103/PhysRevA.99.032331} {\bibfield  {journal} {\bibinfo
  {journal} {Physical Review A}\ }\textbf {\bibinfo {volume} {99}},\ \bibinfo
  {pages} {032331} (\bibinfo {year} {2019})}\BibitemShut {NoStop}%
\bibitem [{\citenamefont {Mondal}\ and\ \citenamefont
  {Bhattacharjee}(2022)}]{Mondal2022}%
  \BibitemOpen
  \bibfield  {author} {\bibinfo {author} {\bibfnamefont {S.}~\bibnamefont
  {Mondal}}\ and\ \bibinfo {author} {\bibfnamefont {S.}~\bibnamefont
  {Bhattacharjee}},\ }\bibfield  {title} {\bibinfo {title} {Periodically driven
  many-body quantum battery},\ }\href
  {https://doi.org/10.1103/PhysRevE.105.044125} {\bibfield  {journal} {\bibinfo
   {journal} {Physical Review E}\ }\textbf {\bibinfo {volume} {105}},\ \bibinfo
  {pages} {044125} (\bibinfo {year} {2022})}\BibitemShut {NoStop}%
\bibitem [{\citenamefont {Catalano}\ \emph {et~al.}(2023)\citenamefont
  {Catalano}, \citenamefont {Giampaolo}, \citenamefont {Morsch}, \citenamefont
  {Giovannetti},\ and\ \citenamefont {Franchini}}]{Catalano2023}%
  \BibitemOpen
  \bibfield  {author} {\bibinfo {author} {\bibfnamefont {A.~G.}\ \bibnamefont
  {Catalano}}, \bibinfo {author} {\bibfnamefont {S.~M.}\ \bibnamefont
  {Giampaolo}}, \bibinfo {author} {\bibfnamefont {O.}~\bibnamefont {Morsch}},
  \bibinfo {author} {\bibfnamefont {V.}~\bibnamefont {Giovannetti}},\ and\
  \bibinfo {author} {\bibfnamefont {F.}~\bibnamefont {Franchini}},\ }\href
  {https://doi.org/https://doi.org/10.48550/arXiv.2307.02529} {\bibinfo {title}
  {Frustrating quantum batteries}} (\bibinfo {year} {2023}),\ \Eprint
  {https://arxiv.org/abs/arXiv:2307.02529} {arXiv:2307.02529} \BibitemShut
  {NoStop}%
\bibitem [{\citenamefont {Cerezo}\ \emph
  {et~al.}(2021{\natexlab{b}})\citenamefont {Cerezo}, \citenamefont {Sone},
  \citenamefont {Volkoff}, \citenamefont {Cincio},\ and\ \citenamefont
  {Coles}}]{Cerezo2021b}%
  \BibitemOpen
  \bibfield  {author} {\bibinfo {author} {\bibfnamefont {M.}~\bibnamefont
  {Cerezo}}, \bibinfo {author} {\bibfnamefont {A.}~\bibnamefont {Sone}},
  \bibinfo {author} {\bibfnamefont {T.}~\bibnamefont {Volkoff}}, \bibinfo
  {author} {\bibfnamefont {L.}~\bibnamefont {Cincio}},\ and\ \bibinfo {author}
  {\bibfnamefont {P.~J.}\ \bibnamefont {Coles}},\ }\bibfield  {title} {\bibinfo
  {title} {Cost function dependent barren plateaus in shallow parametrized
  quantum circuits},\ }\bibfield  {journal} {\bibinfo  {journal} {Nature
  Communications}\ }\textbf {\bibinfo {volume} {12}},\ \href
  {https://doi.org/10.1038/s41467-021-21728-w} {10.1038/s41467-021-21728-w}
  (\bibinfo {year} {2021}{\natexlab{b}})\BibitemShut {NoStop}%
\bibitem [{\citenamefont {Temme}\ \emph {et~al.}(2017)\citenamefont {Temme},
  \citenamefont {Bravyi},\ and\ \citenamefont {Gambetta}}]{Temme2017}%
  \BibitemOpen
  \bibfield  {author} {\bibinfo {author} {\bibfnamefont {K.}~\bibnamefont
  {Temme}}, \bibinfo {author} {\bibfnamefont {S.}~\bibnamefont {Bravyi}},\ and\
  \bibinfo {author} {\bibfnamefont {J.~M.}\ \bibnamefont {Gambetta}},\
  }\bibfield  {title} {\bibinfo {title} {Error mitigation for short-depth
  quantum circuits},\ }\href {https://doi.org/10.1103/PhysRevLett.119.180509}
  {\bibfield  {journal} {\bibinfo  {journal} {Physical Review Letters}\
  }\textbf {\bibinfo {volume} {119}},\ \bibinfo {pages} {180509} (\bibinfo
  {year} {2017})}\BibitemShut {NoStop}%
\bibitem [{\citenamefont {Kandala}\ \emph {et~al.}(2019)\citenamefont
  {Kandala}, \citenamefont {Temme}, \citenamefont {C{\'{o}}rcoles},
  \citenamefont {Mezzacapo}, \citenamefont {Chow},\ and\ \citenamefont
  {Gambetta}}]{Kandala2019}%
  \BibitemOpen
  \bibfield  {author} {\bibinfo {author} {\bibfnamefont {A.}~\bibnamefont
  {Kandala}}, \bibinfo {author} {\bibfnamefont {K.}~\bibnamefont {Temme}},
  \bibinfo {author} {\bibfnamefont {A.~D.}\ \bibnamefont {C{\'{o}}rcoles}},
  \bibinfo {author} {\bibfnamefont {A.}~\bibnamefont {Mezzacapo}}, \bibinfo
  {author} {\bibfnamefont {J.~M.}\ \bibnamefont {Chow}},\ and\ \bibinfo
  {author} {\bibfnamefont {J.~M.}\ \bibnamefont {Gambetta}},\ }\bibfield
  {title} {\bibinfo {title} {Error mitigation extends the computational reach
  of a noisy quantum processor},\ }\href
  {https://doi.org/10.1038/s41586-019-1040-7} {\bibfield  {journal} {\bibinfo
  {journal} {Nature}\ }\textbf {\bibinfo {volume} {567}},\ \bibinfo {pages}
  {491} (\bibinfo {year} {2019})}\BibitemShut {NoStop}%
\bibitem [{\citenamefont {Majumdar}\ \emph {et~al.}(2023)\citenamefont
  {Majumdar}, \citenamefont {Rivero}, \citenamefont {Metz}, \citenamefont
  {Hasan},\ and\ \citenamefont {Wang}}]{majumdar2023best}%
  \BibitemOpen
  \bibfield  {author} {\bibinfo {author} {\bibfnamefont {R.}~\bibnamefont
  {Majumdar}}, \bibinfo {author} {\bibfnamefont {P.}~\bibnamefont {Rivero}},
  \bibinfo {author} {\bibfnamefont {F.}~\bibnamefont {Metz}}, \bibinfo {author}
  {\bibfnamefont {A.}~\bibnamefont {Hasan}},\ and\ \bibinfo {author}
  {\bibfnamefont {D.~S.}\ \bibnamefont {Wang}},\ }\href
  {https://doi.org/https://doi.org/10.48550/arXiv.2307.05203} {\bibinfo {title}
  {Best practices for quantum error mitigation with digital zero-noise
  extrapolation}} (\bibinfo {year} {2023}),\ \Eprint
  {https://arxiv.org/abs/2307.05203} {arXiv:2307.05203 [quant-ph]} \BibitemShut
  {NoStop}%
\bibitem [{\citenamefont {Levy}\ \emph {et~al.}(2016)\citenamefont {Levy},
  \citenamefont {Di\'osi},\ and\ \citenamefont {Kosloff}}]{Levy2016}%
  \BibitemOpen
  \bibfield  {author} {\bibinfo {author} {\bibfnamefont {A.}~\bibnamefont
  {Levy}}, \bibinfo {author} {\bibfnamefont {L.}~\bibnamefont {Di\'osi}},\ and\
  \bibinfo {author} {\bibfnamefont {R.}~\bibnamefont {Kosloff}},\ }\bibfield
  {title} {\bibinfo {title} {Quantum flywheel},\ }\href
  {https://doi.org/10.1103/PhysRevA.93.052119} {\bibfield  {journal} {\bibinfo
  {journal} {Phys. Rev. A}\ }\textbf {\bibinfo {volume} {93}},\ \bibinfo
  {pages} {052119} (\bibinfo {year} {2016})}\BibitemShut {NoStop}%
\bibitem [{\citenamefont {von Lindenfels}\ \emph {et~al.}(2019)\citenamefont
  {von Lindenfels}, \citenamefont {Gr\"ab}, \citenamefont {Schmiegelow},
  \citenamefont {Kaushal}, \citenamefont {Schulz}, \citenamefont {Mitchison},
  \citenamefont {Goold}, \citenamefont {Schmidt-Kaler},\ and\ \citenamefont
  {Poschinger}}]{flywheel2019}%
  \BibitemOpen
  \bibfield  {author} {\bibinfo {author} {\bibfnamefont {D.}~\bibnamefont {von
  Lindenfels}}, \bibinfo {author} {\bibfnamefont {O.}~\bibnamefont {Gr\"ab}},
  \bibinfo {author} {\bibfnamefont {C.~T.}\ \bibnamefont {Schmiegelow}},
  \bibinfo {author} {\bibfnamefont {V.}~\bibnamefont {Kaushal}}, \bibinfo
  {author} {\bibfnamefont {J.}~\bibnamefont {Schulz}}, \bibinfo {author}
  {\bibfnamefont {M.~T.}\ \bibnamefont {Mitchison}}, \bibinfo {author}
  {\bibfnamefont {J.}~\bibnamefont {Goold}}, \bibinfo {author} {\bibfnamefont
  {F.}~\bibnamefont {Schmidt-Kaler}},\ and\ \bibinfo {author} {\bibfnamefont
  {U.~G.}\ \bibnamefont {Poschinger}},\ }\bibfield  {title} {\bibinfo {title}
  {Spin heat engine coupled to a harmonic-oscillator flywheel},\ }\href
  {https://doi.org/10.1103/PhysRevLett.123.080602} {\bibfield  {journal}
  {\bibinfo  {journal} {Phys. Rev. Lett.}\ }\textbf {\bibinfo {volume} {123}},\
  \bibinfo {pages} {080602} (\bibinfo {year} {2019})}\BibitemShut {NoStop}%
\bibitem [{\citenamefont {Puliyil}\ \emph {et~al.}(2022)\citenamefont
  {Puliyil}, \citenamefont {Banik},\ and\ \citenamefont
  {Alimuddin}}]{Puliyil2022}%
  \BibitemOpen
  \bibfield  {author} {\bibinfo {author} {\bibfnamefont {S.}~\bibnamefont
  {Puliyil}}, \bibinfo {author} {\bibfnamefont {M.}~\bibnamefont {Banik}},\
  and\ \bibinfo {author} {\bibfnamefont {M.}~\bibnamefont {Alimuddin}},\
  }\bibfield  {title} {\bibinfo {title} {Thermodynamic signatures of genuinely
  multipartite entanglement},\ }\href
  {https://doi.org/10.1103/PhysRevLett.129.070601} {\bibfield  {journal}
  {\bibinfo  {journal} {Physical Review Letters}\ }\textbf {\bibinfo {volume}
  {129}},\ \bibinfo {pages} {070601} (\bibinfo {year} {2022})}\BibitemShut
  {NoStop}%
\bibitem [{\citenamefont {Cerezo}\ \emph {et~al.}(2022)\citenamefont {Cerezo},
  \citenamefont {Sharma}, \citenamefont {Arrasmith},\ and\ \citenamefont
  {Coles}}]{cerezo2022variational}%
  \BibitemOpen
  \bibfield  {author} {\bibinfo {author} {\bibfnamefont {M.}~\bibnamefont
  {Cerezo}}, \bibinfo {author} {\bibfnamefont {K.}~\bibnamefont {Sharma}},
  \bibinfo {author} {\bibfnamefont {A.}~\bibnamefont {Arrasmith}},\ and\
  \bibinfo {author} {\bibfnamefont {P.~J.}\ \bibnamefont {Coles}},\ }\bibfield
  {title} {\bibinfo {title} {Variational quantum state eigensolver},\
  }\href@noop {} {\bibfield  {journal} {\bibinfo  {journal} {npj Quantum
  Information}\ }\textbf {\bibinfo {volume} {8}},\ \bibinfo {pages} {113}
  (\bibinfo {year} {2022})}\BibitemShut {NoStop}%
\bibitem [{\citenamefont {Cross}\ \emph {et~al.}(2019)\citenamefont {Cross},
  \citenamefont {Bishop}, \citenamefont {Sheldon}, \citenamefont {Nation},\
  and\ \citenamefont {Gambetta}}]{cross2019validating}%
  \BibitemOpen
  \bibfield  {author} {\bibinfo {author} {\bibfnamefont {A.~W.}\ \bibnamefont
  {Cross}}, \bibinfo {author} {\bibfnamefont {L.~S.}\ \bibnamefont {Bishop}},
  \bibinfo {author} {\bibfnamefont {S.}~\bibnamefont {Sheldon}}, \bibinfo
  {author} {\bibfnamefont {P.~D.}\ \bibnamefont {Nation}},\ and\ \bibinfo
  {author} {\bibfnamefont {J.~M.}\ \bibnamefont {Gambetta}},\ }\bibfield
  {title} {\bibinfo {title} {Validating quantum computers using randomized
  model circuits},\ }\href
  {https://doi.org/https://doi.org/10.1103/PhysRevA.100.032328} {\bibfield
  {journal} {\bibinfo  {journal} {Physical Review A}\ }\textbf {\bibinfo
  {volume} {100}},\ \bibinfo {pages} {032328} (\bibinfo {year}
  {2019})}\BibitemShut {NoStop}%
\bibitem [{IBM(2021)}]{IBMQuantumHas2021}%
  \BibitemOpen
  \href@noop {} {\bibinfo {title} {{{IBM Quantum}} has achieved its highest
  {{Quantum Volume}} yet}},\ \bibinfo {howpublished}
  {https://research.ibm.com/blog/quantum-volume-256} (\bibinfo {year}
  {2021})\BibitemShut {NoStop}%
\bibitem [{\citenamefont {{Qiskit contributors}}(2023)}]{Qiskit}%
  \BibitemOpen
  \bibfield  {author} {\bibinfo {author} {\bibnamefont {{Qiskit
  contributors}}},\ }\href {https://doi.org/10.5281/zenodo.2573505} {\bibinfo
  {title} {Qiskit: An open-source framework for quantum computing}} (\bibinfo
  {year} {2023})\BibitemShut {NoStop}%
\bibitem [{\citenamefont {Spall}(1992)}]{spall1992multivariate}%
  \BibitemOpen
  \bibfield  {author} {\bibinfo {author} {\bibfnamefont {J.~C.}\ \bibnamefont
  {Spall}},\ }\bibfield  {title} {\bibinfo {title} {Multivariate stochastic
  approximation using a simultaneous perturbation gradient approximation},\
  }\href {https://doi.org/10.1109/9.119632} {\bibfield  {journal} {\bibinfo
  {journal} {IEEE transactions on automatic control}\ }\textbf {\bibinfo
  {volume} {37}},\ \bibinfo {pages} {332} (\bibinfo {year} {1992})}\BibitemShut
  {NoStop}%
\bibitem [{\citenamefont {Weinberg}\ and\ \citenamefont
  {Bukov}(2017)}]{quspin}%
  \BibitemOpen
  \bibfield  {author} {\bibinfo {author} {\bibfnamefont {P.}~\bibnamefont
  {Weinberg}}\ and\ \bibinfo {author} {\bibfnamefont {M.}~\bibnamefont
  {Bukov}},\ }\bibfield  {title} {\bibinfo {title} {{QuSpin: a Python Package
  for Dynamics and Exact Diagonalisation of Quantum Many Body Systems part I:
  spin chains}},\ }\href {https://doi.org/10.21468/SciPostPhys.2.1.003}
  {\bibfield  {journal} {\bibinfo  {journal} {SciPost Physics}\ }\textbf
  {\bibinfo {volume} {2}},\ \bibinfo {pages} {003} (\bibinfo {year}
  {2017})}\BibitemShut {NoStop}%
\end{thebibliography}%
\end{document}